\documentclass[12pt]{article}

\pdfoutput=1

\usepackage{graphics}
\usepackage{epsfig}

\usepackage{amsfonts}
\usepackage{amssymb}
\usepackage{amsmath}
\usepackage{dsfont}
\usepackage{pifont}
\usepackage{bbm}
\usepackage{multirow}
\usepackage{latexsym}
\usepackage{verbatim}
\usepackage{mcite}
\usepackage{cite}
\usepackage{units}
\usepackage[all]{xy}
\usepackage{arydshln}
\usepackage{cancel}
\usepackage{slashed}

\def\beq{\begin{equation}}
\def\eeq{\end{equation}}
\def\beqa{\begin{eqnarray}}
\def\eeqa{\end{eqnarray}}

\def\a{{\alpha}}
\def\b{{\beta}}

\def\g{{\gamma}}
\def\bg{{\bar{\gamma}}}

\def\d{{\delta}}
\def\eps{{\epsilon}}

\def\m{{\mu}}
\def\n{{\nu}}
\def\r{{\rho}}
\def\s{{\sigma}}

\def\dm{{\dot{\mu}}}
\def\dn{{\dot{\nu}}}

\def\bfone{\relax{\rm 1\kern-.35em 1}}

\def\hi{{\hat{i}}}
\def\hj{{\hat{j}}}
\def\hk{{\hat{k}}}
\def\hl{{\hat{l}}}
\def\hm{{\hat{m}}}

\newcommand{\cC}{{\cal C}}
\newcommand{\cM}{{\cal M}}
\newcommand{\cN}{{\cal N}}

\newcommand{\bbM}{{\mathbb{M}}}
\newcommand{\bbN}{{\mathbb{N}}}
\newcommand{\bbP}{{\mathbb{P}}}
\newcommand{\bbQ}{{\mathbb{Q}}}
\newcommand{\bbR}{{\mathbb{R}}}
\newcommand{\bbS}{{\mathbb{S}}}

\newcommand{\be}{\begin{equation}}
\newcommand{\ee}{\end{equation}}
\newcommand{\ben}{\begin{displaymath}}
\newcommand{\een}{\end{displaymath}}
\newcommand{\bea}{\begin{eqnarray}}
\newcommand{\eea}{\end{eqnarray}}

\newcommand{\bean}{\begin{eqnarray*}}
\newcommand{\eean}{\end{eqnarray*}}

\DeclareMathAlphabet{\mathpzc}{OT1}{pzc}{m}{it}

\begin{document}
\pagestyle{plain}


\makeatletter
\@addtoreset{equation}{section}
\makeatother
\renewcommand{\thesection}{\arabic{section}}
\renewcommand{\theequation}{\thesection.\arabic{equation}}
\renewcommand{\thefootnote}{\arabic{footnote}}


\setcounter{page}{1}
\setcounter{footnote}{0}


\begin{titlepage}
\begin{flushright}
\small ~~
\end{flushright}

\bigskip

\begin{center}

\vskip 0cm

{\LARGE \bf Exceptional Flux Compactifications} \\[6mm]

\vskip 0.5cm

{\bf Giuseppe Dibitetto,\, Adolfo Guarino \,and\, Diederik Roest}\\

\vskip 25pt

{\em Centre for Theoretical Physics,\\
University of Groningen, \\
Nijenborgh 4, 9747 AG Groningen, The Netherlands\\
{\small {\tt \{g.dibitetto , j.a.guarino , d.roest\}@rug.nl}}} \\

\vskip 0.8cm

\end{center}

\vskip 1cm

\begin{center}

{\bf ABSTRACT}\\[3ex]

\begin{minipage}{13cm}
\small

We consider type II (non-)geometric flux backgrounds in the absence of brane sources, and construct their explicit embedding into maximal gauged $D=4$ supergravity. This enables one to investigate the critical points, mass spectra and gauge groups of such backgrounds. We focus on a class of type IIA geometric vacua and find a novel, non-supersymmetric and stable AdS vacuum in maximal supergravity with a non-semisimple gauge group. Our construction relies on a non-trivial mapping between $\textrm{SL}(2) \times \textrm{SO}(6,6)$ fluxes, $\textrm{SU}(8)$ mass spectra and gaugings of $\textrm{E}_{7(7)}$ subgroups. 

\end{minipage}

\end{center}

\vfill

\end{titlepage}


\tableofcontents

\section{Introduction}
\label{sec:introduction}

In the last decade a lot of progress has been made in understanding flux compactifications in type II string theory \cite{Grana:2005jc}. The introduction of background fluxes is needed in order to perturbatively generate a potential for the moduli fields and stabilise them into a minimum. Generically, a given flux background contains a number of local sources such as branes and orientifold planes that can break different amounts of supersymmetry, thus yielding an effective four-dimensional (4D) description with lower supersymmetry.

Flux backgrounds compatible with minimal supersymmetry have received a lot of attention in the literature, with a particular focus on the mechanism of inducing an effective superpotential from fluxes \cite{Giddings:2001yu}, e.g. in the context of the so-called $STU$-models in four dimensions \cite{STU_models(1),STU_models(2),STU_models(3)}. However, the task of finding de Sitter (dS) solutions for cosmological purposes in this set-up turns out to be quite hard; the only known dS solutions in geometric backgrounds are presented in refs~\cite{dS_solutions}. The situation gets dramatically worse if one specialises to backgrounds preserving extended supersymmetry; in particular, in $\,\cN=4\,$ \cite{Aldazabal:2008zza,Dall'Agata:2009gv}, geometric fluxes have been found to be insufficient to obtain dS solutions \cite{Dibitetto:2010rg,Dibitetto:2011gm}. 

A crucial ingredient for dS extrema is therefore given by non-geometric fluxes \cite{Shelton:2005cf}, whose existence was first conjectured in order for the low energy effective theory to be duality covariant. This duality is correctly encoded in the global symmetry of the underlying gauged supergravity in four dimensions \cite{Samtleben:2008pe}. In this sense, T-duality singles out the important role of half-maximal supergravities, whereas, in order to supplement it with non-perturbative dualities to generate the full U-duality group, one has to consider maximal supergravity.

Concentrating in particular on T-duality, different ways have been investigated in the literature in order to implement T-duality covariance at a more fundamental level in order to gain a better understanding of how non-geometric fluxes change the compactification prescription, thus appearing in the effective theory \cite{Andriot:2011uh}. One direction to follow is generalised geometry \cite{GG_list}, in which the internal manifold is given a particular bundle structure in which the gauge fields now span the full $\,\textrm{O}(6,6)\,$ group. Another possibility is that of doubling the internal coordinates \cite{Hull:2006va,Hull&Edwards} by supplementing them with the corresponding duals to winding modes and viewing a non-geometric flux background as something created by means of a twisted double torus compactification \cite{Dall'Agata:2007sr}.

Recently, this second approach has been further developed into the so-called Double Field Theory (DFT) \cite{DFT_list}, which aims to promote T-duality to a fundamental symmetry even independently of whether spacetime directions are compact or not \cite{Duff_duality}. This theory in 10+10 dimensions is formulated in terms of a generalised metric, whose action can be constructed to be fully $\,\textrm{O}(10,10)\,$ invariant. Since it has been shown how to obtain gaugings of $\,\cN=4\,$ supergravity from DFT \cite{DFT_N=4}, this might provide a higher-dimensional origin for non-geometric flux backgrounds, even though the concrete construction leading to the most general background still needs to be accomplished.

In a parallel development, progress has also been made on classifying the landscape of such theories. In ref.~\cite{Dibitetto:2011gm}, an exhaustive analysis of the landscape of isotropic type II geometric vacua has been carried out within the context of half-maximal gauged supergravity. By making use of the dictionary between fluxes and embedding tensor components, the most general geometric flux background in type IIA and IIB has been studied, all the vacua found and the underlying gaugings identified. The crucial ingredient to obtain these results turned out to be the fact of dealing with sets of embedding tensor components which were invariant under non-compact duality transformations. This allows one to restrict the search for critical points to the origin without loss of generality, where all the equations of motion for the scalars simply reduce to quadratic conditions in the fluxes, which can be treated algebraically.

The set of geometric vacua in type IIB, where only gauge fluxes are allowed, turned out to only contain the Minkowski solutions found in ref.~\cite{Giddings:2001yu}. Much more interesting is the vacua structure of type IIA $\,\cN=4\,$ compactifications, where metric flux is allowed as well as gauge fluxes. There turn out to be two inequivalent theories (related by a $\mathbb{Z}_{2}$) with gauge group $\,\textrm{ISO}(3)\,\ltimes\,\textrm{U}(1)^{6}\,$, having 4 inequivalent Anti-de Sitter (AdS) critical points each. Surprisingly, two of these break all supersymmetry and nevertheless has non-negative masses, indicating perturbative stability. Furthermore, all critical points of this theory turned out to be compatible with the absence of supersymmetry-breaking sources, hence admitting an uplift to maximal supergravity. Again, we would like to stress that these solutions were obtained by moving to the origin all the $\,\textrm{SL}(2)\,\times\,\textrm{SO}(6,6)\,$ scalars; for this procedure to be fully general, our set of fluxes needed to be invariant under non-compact S- and T-duality transformations. Indeed this is the case for geometric fluxes in these theories.

The combination of these recent developments makes it interesting to further investigate the structure of extended gauged supergravities in order to better understand which role T- and U-dualities play in the context of flux compactifications. The embedding of half-maximal into maximal supergravity \cite{Aldazabal:2011yz, Dibitetto:2011eu} allows one to study flux backgrounds that preserve maximal supersymmetry. An interesting fact is that the completion of half-maximal supergravity deformations to maximal is given by objects which behave as spinors
under T-duality. Following the generalised geometry approach, a further extension has been proposed called exceptional generalised geometry \cite{Aldazabal:2010ef,EGG} in which the internal manifold gains more structure by including R-R gauge fields which extend $\,\textrm{O}(6,6)\,$ to $\,\textrm{E}_{7(7)}$.

From a purely supergravity point of view, a number of critical points of maximal theories have been found. In particular, the $\textrm{SO}(8)$-gauged theory has been studied in detail and a classification of AdS critical points based on their residual symmetry group $G$ has been carried out \cite{Warner_SO8}. These critical points correspond to configurations in which only $G$-invariant scalars can acquire a vacuum expectation value and hence they are to be found as solutions of the $G$-truncated theory, in which only the $G$-invariant sector of the theory is retained. Amongst the solutions with $\textrm{SO}(3) \times \textrm{SO}(3)$ invariance, there is the first example of a non-supersymmetric and nevertheless stable vacuum in a theory of gauged supergravity with the maximal amount of supersymmetry \cite{Fischbacher:2010ec}. Note that this critical point relies on the Breitenlohner-Freedman bound to be perturbatively stable. More recently, this classification has been extended with a number of critical points with smaller or trivial invariance groups, which have been obtained with a numerical procedure \cite{Fischbacher_numerical}. 

Moving beyond $\textrm{SO}(8)$, analytic continuations such as $\textrm{SO}(p,q)$ and their contracted versions have been investigated for critical points \cite{Hull:1988jw}. Moreover, by restricting to a subset of all embedding tensor irrep's and by applying the method developed in ref.~\cite{Dibitetto:2011gm}, critical points were found in theories with non-semisimple gaugings \cite{DallAgata:2011aa}. In terms of the Levi decomposition these gauge groups can be written as the semi-direct product of a semisimple group with an Abelian ideal. In this case, the subset of embedding tensor irrep's is not closed under non-compact duality transformations. Therefore one cannot employ all such transformations; in other words, one has to restrict to a submanifold of the full moduli space.

The aforementioned results provide a strong indication that the issue of stability without supersymmetry in extended supergravity is much more involved than what was generally expected
in the past. In particular, the presence of non-supersymmetric yet stable Anti-de Sitter critical points came somewhat as a surprise. Whether this is also possible for Minkowski and/or De Sitter remains an open question. The sGoldstino analysis recently carried out in ref.~\cite{Borghese:2011en} does not provide a final statement concerning this issue, but does indicate increasing restrictions as one goes towards a positive scalar potential. From here we can thus conclude that the study of vacua in $\,\cN=8\,$ flux compactifications and their stability
analysis is also motivated from the pure supergravity point of view in that it can provide new critical points and new examples of non-supersymmetric and nevertheless stable solutions.

Our goal in this work will be to elaborate on the results of \cite{Dibitetto:2011gm} and explicitly show how these IIA geometric flux backgrounds and any other type II background can be embedded in maximal supergravity. To this end we will therefore need to relate different formulations of $\,\cN=8\,$ gauged supergravity. The embedding tensor formalism provides an $\,\textrm{E}_{7(7)}\,$ covariant formulation of maximal gauged supergravity in $D=4$. However, in order to make contact with flux compactifications, we need a rewriting of this theory in terms of irrep's of the S- and T-duality groups, i.e. $\,\textrm{SL}(2)\,\times\,\textrm{SO}(6,6)$, following the philosophy of ref.~\cite{Aldazabal:2010ef}. Finally, in order to study the physical properties of scalars, such as equations of motions and the mass matrix, $\,\textrm{SU}(8)\,$is the correct group rearranging all the 70 scalar physical degrees of freedom into an irrep. This can be summarised as

\begin{center}
\scalebox{0.9}[0.9]{\xymatrix{
*+[F-,]{\begin{array}{c}\boldsymbol{\cN=8\textbf{ SUGRA}}\\
\textrm{E}_{7(7)} \end{array} } & \leftrightarrow &
*+[F-,]{\begin{array}{c}\textbf{Fluxes}\\
\textrm{SL}(2)\,\times\,\textrm{SO}(6,6) \end{array} } &
\leftrightarrow &
*+[F-,]{\begin{array}{c}\textbf{Mass spectra}\\
\textrm{SU}(8) \end{array} }  }}
\end{center}
\vspace{2mm}

\noindent Employing this mapping, we will derive the mass spectrum and the gauge group of such ``exceptional'' cases of flux backgrounds without branes. 
\\

Interestingly, we will find that the mass spectrum of one of the geometric IIA solutions remains non-negative in the maximal theory, providing the first example of a non-supersymmetric critical point with this property. Similarly, the gauge group will turn out to be a semi-direct product between a semisimple group and a nilpotent ideal. Again this provides the first such example in maximal supergravity. A point to note is that once this set of $\,\cN=4\,$ fluxes will be embedded in $\,\cN=8\,$, it will not be a closed set under all non-compact $\,E_{7(7)}\,$ transformations; these fluxes will be part of a bigger U-duality orbit of configurations. Therefore our results leave open the possibility of including fluxes which are odd under the orientifold involution (i.e. the spinorial embedding tensor components), which would fill out the U-duality orbits. In a further line of research we hope to investigate backgrounds with non-zero spinorial fluxes in order to study the landscape of geometric type II compactifications within $\,\cN=8\,$.

The paper is organised as follows. In section~\ref{sec:maximal_supergravity}, we first review maximal supergravity in four dimensions in its formulation in terms of $\,\textrm{SU}(8)\,$ irrep's. Secondly we rewrite it in terms of $\,\textrm{SL}(2)\,\times\,\textrm{SO}(6,6)$ irrep's, and finally we show how to relate these two different formulations. In section~\ref{sec:truncations}, we present a web of consistent truncations of $\,{\cN=8}\,$ supergravity yielding four-dimensional theories with lower amounts of supersymmetry and a smaller content in terms of fields and deformations which, nevertheless, capture interesting information related to flux compactifications in type II strings. In section~\ref{sec:typeII_examples}, we explicitly show how to embed type II flux compactifications in maximal gauged supergravity in $D=4$ and we work out the examples of particular non-geometric backgrounds in type IIB as well as the general isotropic geometric backgrounds of type IIA. In the latter case we compute the mass spectrum for the scalars and we identify the gauging. Finally, our conclusions are presented in section~\ref{sec:conclusions}. We defer a number of technical details to the appendices.

\section{Maximal supergravity in four dimensions}
\label{sec:maximal_supergravity}

Maximal supergravity appears when reducing type II ten-dimensional supergravities on a torus down to four dimensions. In recent years, a unified framework where to systematically investigate maximal supergravities has been developed, the so-called embedding tensor formalism \cite{deWit:2007mt}. This formalism relies on the \textit{gauging procedure}, i.e., promoting to local a part of the $\textrm{E}_{7(7)}$ global symmetry of the 4D theory. After applying a gauging, a non-Abelian gauge symmetry is realised in a way compatible with still keeping $\,\mathcal{N}=8\,$ supersymmetry in four dimensions. Moreover, a non-trivial potential $\,V\,$ for the $70$ (physical) scalar fields in the lower-dimensional theory (a.k.a moduli fields) is also generated, hence opening the possibility for them to get stabilised, i.e., to acquire a mass, due to the gauging. The aim of this paper is to explore the interplay between gaugings and moduli stabilisation in the context of maximal supergravity.

\subsection{Embedding tensor and the $\,\textrm{E}_{7(7)}\,$ formulation}

A gauging is totally encoded inside the embedding tensor ${\Theta_{\mathbb{M}}}^{\mathpzc{A}}$, where $\mathbb{M}=1,...,56$ and ${\mathpzc{A}=1,...,133}$ respectively denote indices in the fundamental $\textbf{56}$ and adjoint $\textbf{133}$ representations of $\textrm{E}_{7(7)}$. The tensor ${\Theta_{\mathbb{M}}}^{\mathpzc{A}}$ lives in the $\,\textbf{56} \, \times \, \textbf{133} = \textbf{56} \,+\, \textbf{912} \,+\, \textbf{6480}\,$ irreducible representations (irrep's) of $\textrm{E}_{7(7)}$ and specifies which subset of the $\textrm{E}_{7(7)}$ generators $\left\lbrace t_{\mathpzc{A}=1,...,133} \right\rbrace$ become gauge symmetries after the gauging procedure and hence have an associated gauge boson $V_{\mathbb{M}}$ in four dimensions. As in standard gauge theories, the ordinary derivative is replaced by a covariant one, $\nabla \rightarrow \nabla -g\, V^{\mathbb{M}} \,{\Theta_{\mathbb{M}}}^{\mathpzc{A}} \, t_{\mathpzc{A}} $, and a non-Abelian gauge algebra
\beq
\label{gauge_algebra}
\left[ X_{\mathbb{M}} , X_{\mathbb{N}} \right] = - {X_{\mathbb{M} \mathbb{N}}}^{\mathbb{P}} \, X_{\mathbb{P}} \hspace{15mm} \textrm{with} \hspace{15mm} {X_{\mathbb{M} \mathbb{N}}}^{\mathbb{P}} = {\Theta_{\mathbb{M}}}^{\mathpzc{A}} \, {[t_{\mathpzc{A}}]_{\mathbb{N}}}^{\mathbb{P}} \ ,
\eeq
is spanned by the generators $X_{\mathbb{M}}$. Since $\textrm{E}_{7(7)} \subset \textrm{Sp}(56,\mathbb{R})$, and even though $\textrm{E}_{7(7)}$ does not have any invariant metric, one can still use the $\textrm{Sp}(56,\mathbb{R})$ invariant matrix $\Omega_{\mathbb{M}\mathbb{N}}$ (skew-symmetric) in order to raise and lower $\textrm{E}_{7(7)}$ fundamental indices. In what follows, we adopt the SouthWest-NorthEast (SW-NE) convention, e.g., $X^{\mathbb{M}}=X_{\mathbb{N}} \, \Omega^{\mathbb{N} \mathbb{M}}$, together with $\,\Omega_{\mathbb{M}\mathbb{P}}\,\Omega^{\mathbb{N}\mathbb{P}}=\delta_{\mathbb{M}}^{\mathbb{N}}$.

Maximal supersymmetry requires the tensor $X_{\mathbb{M} \mathbb{N} \mathbb{P}}=X_{\mathbb{M} (\mathbb{N} \mathbb{P})}=-{X_{\mathbb{M} \mathbb{N}}}^{\mathbb{Q}} \, \Omega_{\mathbb{Q} \mathbb{P}}$ to live in the $\textbf{912}$ irrep of $\textrm{E}_{7(7)}$. This translates into the following set of linear constraints
\beq
\label{linear_const}
X_{(\mathbb{M} \mathbb{N} \mathbb{P})}=0 
\hspace{10mm},\hspace{10mm}
{X_{\mathbb{P} \mathbb{M}}}^{\mathbb{P}}=0 \ .
\eeq
On the other hand, the tensor $X_{\mathbb{M} \mathbb{N} \mathbb{P}}$ must also satisfy a set of quadratic constraints coming from the consistency of the gauge algebra in (\ref{gauge_algebra}). These quadratic constraints sit in the $(\textbf{133} \,\times\, \textbf{133})_{a} = \textbf{133} \,+\, \textbf{8645}$ irrep's of $\textrm{E}_{7(7)}$ and are given by
\beq
\label{quadratic_const}
\Omega^{\mathbb{R} \mathbb{S}} \, X_{\mathbb{R} \mathbb{M} \mathbb{N}} \, X_{\mathbb{S} \mathbb{P} \mathbb{Q}}=0 \ .  
\eeq
The above set of linear (\ref{linear_const}) and quadratic (\ref{quadratic_const}) constraints guarantee the consistency of the maximal gauged supergravity in four dimensions \cite{deWit:2007mt}. 

Switching on a gauging has strong implications for the scalar sector of the four-dimensional theory. It consists of $133$ scalars out of which only $70$ are physical degrees of freedom -- the remaining $63$ can be removed from the theory after gauge fixing -- and parameterise an $\,\,\textrm{E}_{7(7)}/\textrm{SU}(8)$ coset element $\mathcal{M}_{\mathbb{M}\mathbb{N}}=\mathcal{M}_{(\mathbb{M}\mathbb{N})}$. Because of the gauging, a non-trivial scalar potential appears
\beq
\begin{array}{ccc}
\label{V_N=8}
V &=& \dfrac{g^{2}}{672} \, X_{\bbM \bbN \bbP}  \, X_{\bbQ \bbR \bbS} \left( \mathcal{M}^{\bbM \bbQ} \, \mathcal{M}^{\bbN \bbR} \, \mathcal{M}^{\bbP \bbS}   +   7 \, \mathcal{M}^{\bbM \bbQ} \, \Omega^{\bbN \bbR} \, \Omega^{\bbP \bbS}    \right) \ ,
\end{array}
\eeq
which is invariant under the linear action of $\textrm{E}_{7(7)}$ transformations. This scalar potential might contain a reach structure of critical points where to stabilise all the moduli fields in the four-dimensional theory.

\subsection{Fermionic mass terms and the $\,\textrm{SU}(8)\,$ formulation}
\label{sec:fermi_mass_A7}

The Lagrangian of maximal supergravity in four dimensions can be unambiguously written in terms of $\textrm{SU}(8)$ tensors, since $\textrm{SU}(8)$ is one of the maximal subgroups of $\textrm{E}_{7(7)}$. More concretely, it is its maximal compact subgroup and is identified with the R-symmetry group under which the eight gravitini of the theory get rotated amongst themselves.

\subsubsection*{Bosonic field content}

Under its $\textrm{SU}(8)$ maximal subgroup, we have the following branching for some relevant $\textrm{E}_{7(7)}$ representations
\beq
\begin{array}{cclcl}
\textrm{E}_{7(7)} & \supset &  \textrm{SU}(8) & \hspace{5mm} &   \hspace{40mm} \textrm{\textsc{fields}}\\[2mm]
\textbf{56} & \rightarrow & \textbf{28} + \overline{\textbf{28}} & &  \textrm{vectors: } \,\,\,\,\,\,V_{\mathcal{IJ}} \oplus V^{\mathcal{IJ}}\\[2mm]
\textbf{133} & \rightarrow & \textbf{63} + \textbf{70} & & \textrm{scalars: } \,\,\,\,\,\,{\phi_{\mathcal{I}}}^{\mathcal{J}} \,\,(\textrm{\footnotesize{unphysical}})\oplus \phi_{\mathcal{I} \mathcal{J} \mathcal{K} \mathcal{L}}  \,\,(\textrm{\footnotesize{physical}}) \\[2mm]
\textbf{912} & \rightarrow & \textbf{36} + \textbf{420} + \overline{\textbf{36}} + \overline{\textbf{420}} &  & \textrm{emb tens: } \mathcal{A}^{\mathcal{IJ}} \oplus {\mathcal{A}_{\mathcal{I}}}^{\mathcal{JKL}} \oplus \mathcal{A}_{\mathcal{IJ}} \oplus {\mathcal{A}^{\mathcal{I}}}_{\mathcal{JKL}}
\end{array}
\nonumber
\eeq
related to the vectors, scalars and embedding tensor\footnote{Strictly speaking, the $\mathcal{A}^{\mathcal{IJ}}$ and ${\mathcal{A}_{\mathcal{I}}}^{\mathcal{JKL}}$ fermionic mass terms as well as their complex conjugates correspond to the embedding tensor dressed up with the scalar fields of the theory.} in the four-dimensional theory. When expressed in terms of $\textrm{SU}(8)$ fundamental indices $\mathcal{I}=1,...,8$, the above fields have the following symmetry properties according to the irrep's they are associated to: 
\begin{itemize}

\item[$i)$] $V_{\mathcal{IJ}}=V_{\mathcal{[IJ]}}\,$ and $\,V^{\mathcal{IJ}}=(V_{\mathcal{IJ}})^{*}\,$ for the complex vector fields.

\item[$ii)$] ${\phi_{\mathcal{I}}}^{\mathcal{I}}=0\,$
and $\,\phi_{\mathcal{I} \mathcal{J} \mathcal{K} \mathcal{L}} = \phi_{[\mathcal{I} \mathcal{J} \mathcal{K} \mathcal{L}]}\,$ for the scalar fields which are further restricted by the pseudo-reality condition
\beq
\phi_{\mathcal{I} \mathcal{J} \mathcal{K} \mathcal{L}} = \frac{1}{24} \, \epsilon_{\mathcal{IJKL MNPQ}} \,  \phi^{\mathcal{MNPQ}}  \hspace{10mm} \textrm{with} \hspace{10mm} \phi^{\mathcal{MNPQ}} = (\phi_{\mathcal{MNPQ}})^{*} \ .
\eeq
It is worth noticing that the physical scalars $\,\phi_{\mathcal{I} \mathcal{J} \mathcal{K} \mathcal{L}}\,$ in the theory fit an irrep, namely the $\textbf{70}$ of $\textrm{SU}(8)$. This will no longer be the case when using another formulation of the theory, as we will see in the next section.

\item[$iii)$] $\mathcal{A}^{\mathcal{IJ}}=\mathcal{A}^{(\mathcal{IJ})}$, $\,{\mathcal{A}_{\mathcal{I}}}^{\mathcal{J} \mathcal{K} \mathcal{L}} = {\mathcal{A}_{\mathcal{I}}}^{[\mathcal{J} \mathcal{K} \mathcal{L}]}\,$ and $\,{\mathcal{A}_{\mathcal{Q}}}^{\mathcal{Q} \mathcal{I} \mathcal{J}} =0\,$ and equivalently for their complex conjugate counterparts $\mathcal{A}_{\mathcal{IJ}}=(\mathcal{A}^{\mathcal{IJ}})^{*}\,$ and $\,{\mathcal{A}^{\mathcal{I}}}_{\mathcal{J} \mathcal{K} \mathcal{L}} =({\mathcal{A}_{\mathcal{I}}}^{\mathcal{J} \mathcal{K} \mathcal{L}} )^{*}\,$.

\end{itemize}

In the $\,\textrm{SU}(8)\,$ formulation, the $\,\textrm{Sp}(56,\mathbb{R})$ invariant (skew-symmetric) matrix $\,\Omega_{\mathbb{M} \mathbb{N}}\,$ takes the form
\beq
\Omega_{\mathbb{M} \mathbb{N}} = - i
\left( 
\begin{array}{c|c}
0 & \delta_{\mathcal{IJ}}^{\mathcal{KL}} \\[1mm] 
\hline
\\[-4mm]
-\delta^{\mathcal{IJ}}_{\mathcal{KL}} & 0 
\end{array}
\right) \ .
\eeq

\subsubsection*{Fermionic mass terms and scalar potential}

The $\mathcal{A}^{\mathcal{IJ}}$ and ${\mathcal{A}_{\mathcal{I}}}^{\mathcal{JKL}}$ tensors play a central role in the $\textrm{SU}(8)$ formulation of maximal supergravity. They determine the fermionic mass terms for the gravitini $\psi^{\,\,\mathcal{I}}_{\mu}$ and the dilatini $\chi_{\mathcal{IJK}}$ in the four-dimensional Lagrangian \cite{deWit:2007mt} (where in this formula $\mu, \nu$ are understood as space-time indices)
\beq
\label{Fermi_Lagrangian}
e^{-1} \, g^{-1} \, \mathcal{L}_{\textrm{fermi}} = \frac{\sqrt{2}}{2} \, \mathcal{A}_{\mathcal{I} \mathcal{J}} \,\overline{\psi}^{\,\,\mathcal{I}}_{\mu} \, \gamma^{\mu \nu} \,\psi^{\,\,\mathcal{J}}_{\nu} + \frac{1}{6} \, {\mathcal{A}_{\mathcal{I}}}^{\mathcal{JKL}} \,\overline{\psi}^{\,\,\mathcal{I}}_{\mu} \, \gamma^{\mu} \, \chi_{\mathcal{JKL}} + \mathcal{A}^{\mathcal{IJK},\mathcal{LMN}} \, \overline{\chi}_{\mathcal{IJK}} \, \chi_{\mathcal{LMN}} + \textrm{h.c.} \ ,
\eeq
where $\,\mathcal{A}^{\mathcal{IJK},\mathcal{LMN}} \equiv \frac{\sqrt{2}}{144} \, \epsilon^{\mathcal{IJKPQR[LM}} \, {\mathcal{A}^{\mathcal{N}]}}_{\mathcal{PQR}}\,$. The number of supersymmetries preserved by an AdS ($V_{0}<0$) or Minkowski ($V_{0}=0$) solution of the theory is related to the number of spinors satisfying the Killing equations
\beq
\label{Killing_equations}
g \, \mathcal{A}_{\mathcal{IJ}} \, \epsilon^{\mathcal{J}} \, = \, \sqrt{-\frac{1}{6}\,V_{0}} \,\, \epsilon_{\mathcal{I}} \ .
\eeq

The scalar potential in (\ref{V_N=8}) can also be rewritten in terms of the fermionic mass terms as 
\beq
\label{V_SU8}
g^{-2} \,V = -\frac{3}{4} \, |\mathcal{A}_1|^{2} + \frac{1}{24} \, |\mathcal{A}_2|^{2} \ ,
\eeq
where $\,|\mathcal{A}_1|^{2}=\mathcal{A}_{\mathcal{IJ}} \, \, \mathcal{A}^{\mathcal{IJ}}\,$ and $\,|\mathcal{A}_2|^{2}={\mathcal{A}_{\mathcal{I}}}^{\mathcal{JKL}} \, {\mathcal{A}^{\mathcal{I}}}_{\mathcal{JKL}}\,$. This potential will possess a structure of critical points satisfying
\beq
\left. \frac{\partial V}{\partial \phi_{\mathcal{IJKL}}} \right|_{\left< \phi_{\mathcal{IJKL}} \right>}= 0 \ ,
\eeq
where $\left< \phi_{\mathcal{IJKL}} \right>$ denotes the VEV for the $70$ physical scalar fields. Provided $\,\left< V_{\mathcal{IJ}}\right>=0\,$ for the vector fields, maximally symmetric solutions of the theory are obtained by solving the equations of motion \cite{Diffon:2011wt} of the physical scalars
\beq
\label{scalars_eom}
\begin{array}{ccc}
\mathcal{C}_{\mathcal{IJKL}} \,+\, \dfrac{1}{24} \, \epsilon_{\mathcal{IJKLMNPQ}} \, \mathcal{C}^{\mathcal{MNPQ}} &=& 0 \ ,
\end{array}
\eeq
where $\,\mathcal{C}_{\mathcal{IJKL}}={\mathcal{A}^{\mathcal{M}}}_{[\mathcal{IJK}}\,\mathcal{A}_{\mathcal{L}]\mathcal{M}} \, +\, \frac{3}{4} \, {\mathcal{A}^{\mathcal{M}}}_{\mathcal{N}[\mathcal{IJ}} \, {\mathcal{A}^{\mathcal{N}}}_{\mathcal{KL}]\mathcal{M}}\,$. At these solutions, the mass matrix for the physical scalars \cite{Diffon:2011wt,Borghese:2011en} reads
\beq
\label{Mass-matrix}
\begin{array}{ccl}
g^{-2} \, {\left(\textrm{mass}^{2}\right)_{\mathcal{IJKL}}}^{\mathcal{MNPQ}}  & =  &  \delta_{\mathcal{IJKL}}^{\mathcal{MNPQ}} \, \left( \frac{5}{24} \, \mathcal{A}^{\mathcal{R}}{}_{\mathcal{STU}} \, \mathcal{A}_{\mathcal{R}}{}^{\mathcal{STU}} - \frac{1}{2} \, \mathcal{A}_{\mathcal{RS}} \, \mathcal{A}^{\mathcal{RS}} \right) \\[2mm]
& + & 6 \, \delta_{[\mathcal{IJ}}^{[\mathcal{MN}} \, \left( \mathcal{A}_{\mathcal{K}}{}^{\mathcal{RS} |\mathcal{P}} \, \mathcal{A}^{\mathcal{Q}]}{}_{\mathcal{L}]\mathcal{RS}} - \frac{1}{4} \, \mathcal{A}_{\mathcal{R}}{}^{\mathcal{S} |\mathcal{PQ}]} \, \mathcal{A}^{\mathcal{R}}{}_{\mathcal{S}|\mathcal{KL}]} \right) \\[2mm] 
&-& \frac{2}{3} \, A_{[\mathcal{I}}{}^{[\mathcal{MNP}} \, \mathcal{A}^{\mathcal{Q}]}{}_{\mathcal{JKL}]} \ .
\end{array}
\eeq
Defining the normalised mass as $\,\left(\textrm{mass}^{2}\right)_{\textrm{norm}}= \frac{1}{|V_{0}|}\,\left(\textrm{mass}^{2}\right)\,$, then the Breitenlohner-Freedman (B.F.) bound for the stability of an AdS solution is given by
\beq 
\label{BF_bound}
m^2\geq-\frac{3}{4}\ , 
\eeq 
where $m^2$ denotes the lowest eigenvalue of the normalised mass matrix at the AdS extremum with energy $\,V_{0}<0\,$. We will make extensive use of (\ref{Mass-matrix}) and (\ref{BF_bound}) in the last part of the work when discussing stability of solutions in specific maximal supergravity models arising from flux compactifications of type II strings.

\subsubsection*{Quadratic constraints}

The set of quadratic constraints in (\ref{quadratic_const}) can also be expressed in terms of the $\mathcal{A}^{\mathcal{IJ}}$ and ${\mathcal{A}_{\mathcal{I}}}^{\mathcal{JKL}}$ tensors. Using the branching relations
\beqa
\textrm{E}_{7(7)} & \supset &  \textrm{SU}(8) \\[2mm]
\label{irrep_133}
\textbf{133} & \longrightarrow  & \textbf{63}\,+\,\textbf{70} \ , \\[2mm]
\label{irrep_8640}
\textbf{8645} & \longrightarrow  & \textbf{63}\,+ \, \textbf{378} \,+ \,\overline{\textbf{378}} \,+\,\textbf{945}\,+\,\overline{\textbf{945}} \,+ \,\textbf{2352}\, + \,\textbf{3584} \ , 
\eeqa
as an organising principle, one gets the following quadratic constraints \cite{deWit:2007mt}
\beq
\label{QC_SU8}
\begin{array}{crcl}
\hspace{-5mm} & 9 \, \mathcal{A}_{\mathcal{R}}{}^{\mathcal{STM}} \mathcal{A}^{\mathcal{R}}{}_{\mathcal{STI}} - \mathcal{A}_{\mathcal{I}}{}^{\mathcal{RST}} \mathcal{A}^{\mathcal{M}}{}_{\mathcal{RST}} - \delta_{\mathcal{I}}^{\mathcal{M}} \, |\mathcal{A}_2|^{2} & = & 0 \ , \\[3mm]
\hspace{-5mm} &3 \, \mathcal{A}_{\mathcal{R}}{}^{\mathcal{STM}} \mathcal{A}^{\mathcal{R}}{}_{\mathcal{STI}} - \mathcal{A}_{\mathcal{I}}{}^{\mathcal{RST}} \mathcal{A}^{\mathcal{M}}{}_{\mathcal{RST}} + 12 \, \mathcal{A}_{\mathcal{IR}} \mathcal{A}^{\mathcal{MR}} - \frac{1}{4} \, \delta_{\mathcal{I}}^{\mathcal{M}} \, |\mathcal{A}_2|^{2} - \frac{3}{2} \, \delta_{\mathcal{I}}^{\mathcal{M}} \, |\mathcal{A}_1|^{2} & = & 0 \ , \\[3mm]
\hspace{-5mm} &\mathcal{A}^{\mathcal{I}}{}_{\mathcal{JV}[\mathcal{M}} \, \mathcal{A}^{\mathcal{V}}{}_{\mathcal{NPQ}]} + \mathcal{A}_{\mathcal{JV}} \, \delta^{\mathcal{I}}_{[\mathcal{M}} \, \mathcal{A}^{\mathcal{V}}{}_{\mathcal{NPQ}]} - \mathcal{A}_{\mathcal{J}[\mathcal{M}} \, \mathcal{A}^{\mathcal{I}}{}_{\mathcal{NPQ}]}  &  & \\[1mm] 
\hspace{-5mm} &+ \, \frac{1}{24} \, \epsilon_{\mathcal{MNPQRSTU}} \, \left( \mathcal{A}_{\mathcal{J}}{}^{\mathcal{IVR}} \, \mathcal{A}_{\mathcal{V}}{}^{\mathcal{STU}} + \mathcal{A}^{\mathcal{IV}} \, \delta_{\mathcal{J}}^{\mathcal{R}} \, \mathcal{A}_{\mathcal{V}}{}^{\mathcal{STU}} - \mathcal{A}^{\mathcal{IR}} \, \mathcal{A}_{\mathcal{J}}{}^{\mathcal{STU}} \right) & = & 0 \ , \\[3mm] 
\hspace{-5mm} & - \frac{1}{8} \, \delta_{\mathcal{I}}^{\mathcal{N}} \left( \mathcal{A}_{\mathcal{R}}{}^{\mathcal{STM}} \mathcal{A}^{\mathcal{R}}{}_{\mathcal{STJ}} - \mathcal{A}_{\mathcal{J}}{}^{\mathcal{RST}} \mathcal{A}^{\mathcal{M}}{}_{\mathcal{RST}} \right) + \frac{1}{8} \, \delta_{\mathcal{J}}^{\mathcal{M}} \left( \mathcal{A}_{\mathcal{R}}{}^{\mathcal{STN}} \mathcal{A}^{\mathcal{R}}{}_{\mathcal{STI}} - \mathcal{A}_{\mathcal{I}}{}^{\mathcal{RST}} \mathcal{A}^{\mathcal{N}}{}_{\mathcal{RST}} \right) &  & \\[1mm]
\hspace{-5mm} &+\, \mathcal{A}_{\mathcal{I}}{}^{\mathcal{RSM}} \mathcal{A}^{\mathcal{N}}{}_{\mathcal{JRS}} - \mathcal{A}_{\mathcal{J}}{}^{\mathcal{RSN}} \mathcal{A}^{\mathcal{M}}{}_{\mathcal{IRS}} + 4 \, \mathcal{A}^{(\mathcal{M}}{}_{\mathcal{IJR}} \mathcal{A}^{\mathcal{N})\mathcal{R}} - 4 \, \mathcal{A}_{(\mathcal{I}}{}^{\mathcal{MNR}} \mathcal{A}_{\mathcal{J})\mathcal{R}}  & = & 0 \ , \\[3mm]
\hspace{-5mm} & - 9 \, \mathcal{A}_{[\mathcal{I}}{}^{\mathcal{R}[\mathcal{MN}} \mathcal{A}^{\mathcal{P}]}{}_{\mathcal{JK}]\mathcal{R}} - 9 \, \delta_{[\mathcal{I}}^{[\mathcal{M}} \, \mathcal{A}_{\mathcal{J}}{}^{\mathcal{RS}|\mathcal{N}} \mathcal{A}^{\mathcal{P}]}{}_{\mathcal{K}]\mathcal{RS}}  - 9 \, \delta_{[\mathcal{IJ}}^{[\mathcal{MN}} \, \mathcal{A}_{\mathcal{R}}{}^{\mathcal{P}]\mathcal{ST}} \mathcal{A}^{\mathcal{R}}{}_{\mathcal{K}]\mathcal{ST}} &  & \\[1mm]
\hspace{-5mm} &+ \, \delta_{\mathcal{IJK}}^{\mathcal{MNP}} \, |\mathcal{A}_{2}|^{2}  + \mathcal{A}_{\mathcal{R}}{}^{\mathcal{MNP}} \mathcal{A}^{\mathcal{R}}{}_{\mathcal{IJK}}  & = & 0 \ ,
\end{array}
\eeq
living in the $\,{\bf 63}\,$ (the first two), $\,{\bf 70} + {\bf 378} + {\bf 3584}\,$, $\,{\bf 945} + {\bf \overline{945}}\,$ and $\,{\bf 2352}\,$ irrep's of $\textrm{SU}(8)$, respectively. The above set of quadratic constraints in (\ref{QC_SU8}) automatically guarantees the consistency of the maximal gauged supergravity.
\\[-4mm]

The $\,\textrm{SO}(8)\,$ gauging \cite{SO8_old} as well as the $\,\textrm{CSO}(p,q,r)\,$ with $p+q+r=8$ contractions thereof \cite{Roest:2009tt,DallAgata:2011aa} are either simple gauge groups or straightforward contractions thereof. For this reason it has been relatively easy to explore these from a pure supergravity viewpoint, irrespective of their realisation in string theory. However, as the relation between gauged supergravities and flux compactifications of string theory became better understood \cite{Samtleben:2008pe}, more complicated non-semisimple gauge groups other than the CSO gaugings have gained interest both in maximal \cite{deWit:2003hq} and half-maximal \cite{Angelantonj_list,Dall'Agata:2009gv,Roest:2009dq,Dibitetto:2010rg,Dibitetto:2011gm} supergravity. The reason is that, as we will show later, the CSO gaugings turn out to correspond to non-geometric flux backgrounds for which an origin in string theory remains unknown\footnote{Nevertheless, some of them can still be obtained from M-theory reductions, as the $\textrm{SO}(8)$ gauging that appears after reducing eleven dimensional supergravity on a $S^{7}$ sphere.}, whereas gaugings corresponding to flux backgrounds with a higher-dimensional origin are in general not CSO. The latter are called geometric backgrounds and include fluxes associated to the Neveu-Schwarz-Neveu-Schwarz (NS-NS) and Ramond-Ramond (R-R) gauge fields present in the spectrum of the string \cite{deWit:2003hq} together with a metric flux associated to a spin connection in the internal space. However, the $\textrm{SU}(8)$ formulation of maximal supergravity is not the most intuitive when it comes to describe gauged supergravities arising from flux compactifications of string theory. Instead, an alternative formulation in terms of $\textrm{SL}(2) \times \textrm{SO}(6,6)$ tensors becomes more adequate as we discuss in the next section.

\subsection{Fluxes and the $\,\textrm{SL}(2) \times \textrm{SO}(6,6)\,$ formulation}
\label{sec:fluxes_A1D6}

Compactifications of string theory in the presence of background fluxes has become a very active research line when it comes to address the problem of moduli stabilisation. Non-geometric fluxes and their further extension to \textit{generalised fluxes} or \textit{dual fluxes} were originally introduced in order to recover invariance of four-dimensional supergravity under the action of duality transformations: more concretely, under non-perturbative S-duality and target space T-duality. The combined action of S-duality and T-duality relates ``apparently'' different four-dimensional backgrounds amongst themselves via an $\,\textrm{SL}(2) \,\times\, \textrm{SO}(6,6)\,$ transformation. This group of transformations corresponds with the global symmetry group of half-maximal $\,\mathcal{N}=4\,$ supergravity in four dimensions \cite{Schon:2006kz}. The relation between half-maximal supergravity and string compactifications with fluxes has been explored in refs~\cite{Aldazabal:2008zza,Dibitetto:2010rg,Dibitetto:2011gm}. As a speculative remark -- and up to quantum requirements such as the discrete nature of the gaugings when understood as fluxes --, by covering the different $\,\textrm{SL}(2) \times \textrm{SO}(6,6)\,$ orbits of half-maximal supergravities, one might have access to intrinsically \textit{stringy} effects involving winding modes and/or dyonic backgrounds, even though it is formulated as a supersymmetric field theory of point-like particles. 

Both S-duality and T-duality belong to a larger U-duality group, the $\,\textrm{E}_{7(7)}\,$ global symmetry group of maximal supergravity in four dimensions. Consequently, in order to go from half-maximal to maximal supergravity \cite{Dibitetto:2011eu}, one has to enlarge the field content of the theory, i.e. vectors, scalars and embedding tensor components, to complete irrep's of $\,\textrm{E}_{7(7)}$. It is at this point where an alternative formulation of maximal supergravity in terms of $\,\textrm{SL}(2) \,\times\, \textrm{SO}(6,6)\,$ tensors becomes mandatory in order to understand the relation between flux compactifications of string theory and maximal supergravity. 

\subsubsection*{Bosonic field content}

Complementary to the $\textrm{SU}(8)$ formulation of the previous section, maximal supergravity can also be unambiguously expressed in terms of $\,\textrm{SL}(2) \,\times\, \textrm{SO}(6,6)\,$ tensors since that is a maximal subgroup of $\,\textrm{E}_{7(7)}$ as well. Under $\,\textrm{SL}(2) \,\times\, \textrm{SO}(6,6)\,$, we now have the following branching 
\beq
\begin{array}{cclcl}
\textrm{E}_{7(7)} & \supset &  \textrm{SL}(2) \times \textrm{SO}(6,6) & \hspace{5mm} &   \hspace{20mm} \textrm{\textsc{fields}}\\[2mm]
\textbf{56} & \rightarrow & \hspace{-1mm}(\textbf{2},\textbf{12}) + (\textbf{1},\textbf{32}) &  & \hspace{-8mm} \textrm{vectors: } \,\,\,\,\,\,V_{\alpha M} \oplus V_{\mu}\\[2mm]
\textbf{133} & \rightarrow & \hspace{-1mm}(\textbf{1},\textbf{66}) + (\textbf{3},\textbf{1}) + (\textbf{2},\textbf{32'}) & & \hspace{-8mm} \textrm{scalars: } \,\,\,\,\,\,\phi_{MN} \oplus \phi_{\alpha \beta} \oplus \phi_{\alpha \dot{\mu}} \\[2mm]
\textbf{912} & \rightarrow & \hspace{-1mm} (\textbf{2},\textbf{220}) +(\textbf{2},\textbf{12}) +(\textbf{1},\textbf{352'}) +(\textbf{3},\textbf{32}) &  & \hspace{-8mm} \textrm{emb tens: } f_{\alpha MNP} \oplus \xi_{\alpha M} \oplus F_{M \dot{\mu}} \oplus \Xi_{\a \b \m}
\end{array}
\nonumber
\eeq
for the $\textrm{E}_{7(7)}$ representations associated to vectors, scalars and embedding tensor respectively. We follow the conventions in ref.~\cite{Dibitetto:2011eu} for the indices: $\,\alpha=+,-\,$ is a fundamental $\,\textrm{SL}(2)\,$ index, $\,M=1,...,12\,$ is a fundamental $\,\textrm{SO}(6,6)\,$ index and $\,\mu \,\,(\dot{\mu})=1,...,32\,$ denotes a left (right) Majorana-Weyl spinor transforming in the $\textbf{32} \,\, (\textbf{32'})$ of $\,\textrm{SO}(6,6)$. In order to fit the irrep's, the above set of fields come out with the following symmetry properties:

\begin{itemize}

\item[$i)$] The real vectors $\,V_{\alpha M}\,$ and $\,V_{\mu}\,$ are unrestricted.

\item[$ii)$] The scalars satisfy $\,\phi_{MN}=\phi_{[MN]}\,$ and $\,\phi_{\a \b}=\phi_{(\a \b)}\,$ whereas $\,\phi_{\a \dot{\mu}}\,$ remain unrestricted. However, in contrast to the $\,\textrm{SU}(8)\,$ formulation, the $\,70\,$ physical scalars no longer fit an irrep of $\,\textrm{SL}(2) \times \textrm{SO}(6,6)\,$. Instead, $38$ of them parameterise an element of the coset space $\,\frac{\textrm{SL}(2)}{\textrm{SO}(2)} \times \frac{\textrm{SO}(6,6)}{\textrm{SO}(6) \times \textrm{SO}(6)}\,$, whereas the remaining 32 extend it to an $\,\frac{\textrm{E}_{7(7)}}{\textrm{SU}(8)}\,$ coset element. At the origin of the moduli space, i.e. $\,{\phi_{MN}=\phi_{\a \b}=\phi_{\a \dot{\mu}}=0}\,$, we are left with a symmetric $\,\textrm{E}_{7(7)}\,$ scalar matrix of the form
\beq
\label{M_origin}
\mathcal{M}_{\mathbb{M} \mathbb{N}} \Big|_{\textrm{origin}} = 
\left( 
\begin{array}{c|c}
\d_{\a \b} \, \d_{MN} & 0 \\[1mm] 
\hline
\\[-4mm]
0 & B_{\m \n}
\end{array}
\right) 
 \ ,
\eeq
where the symmetric and unitary matrix $\,B_{\mu \nu}\,$ is the conjugation matrix introduced in appendix~\ref{App:spinors} to define a reality condition upon gamma matrices of $\,\textrm{SO}(6,6)\,$. 

It is worth mentioning here that the explicit form of $\,B_{\mu \nu}\,$ crucially depends on the choice of the gamma matrices representation. For instance, if taking the real representation presented in appendix~\ref{App:spinors}, then $\,B_{\m \n}=\mathds{1}_{32}\,$. On the other hand, if taking the complex representation that makes $\,\textrm{SU}(4) \times \textrm{SU}(4)\,$ covariance explicit (see also the appendix~\ref{App:spinors}), one finds that
$\,B_{\m \n}=
{\footnotesize{
\left( 
\begin{array}{cc}
0 & \mathds{1}_{16} \\
\mathds{1}_{16} & 0
\end{array}
\right) }}
\,$. Then, different choices of gamma matrices representation do change the notion of what is called the origin of the moduli space according to its definition in (\ref{M_origin}), even though they are related via a unitary $\,\textrm{U}(32)\,$ transformation. 

In order to avoid confusion, we will adopt the convention of $\,B_{\m \n}=\mathds{1}_{32}\,$ when referring to the origin of the moduli space, hence being compatible with the natural choice of
\beq
\label{M_origin_choice}
\mathcal{M}_{\mathbb{MN}}\Big|_{\textrm{origin}}=\mathds{1}_{56} \ ,
\eeq 
as the origin of field space.

\item[$iii)$] The different pieces of the embedding tensor satisfy $\,f_{\a MNP}=f_{\a [MNP]}\,$ together with  $\,\Xi_{\a \b \mu}=\Xi_{(\a \b) \mu}\,$ and $\,\slashed{F}^{\mu} \equiv F_{M \dot{\nu}} \, [\gamma^{M}]^{\mu \dot{\nu}}=0\,$.

\end{itemize}

In the $\,\textrm{SL}(2) \times \textrm{SO}(6,6)\,$ formulation, the $\,\textrm{Sp}(56,\mathbb{R})\,$ skew-symmetric invariant matrix $\,\Omega_{\mathbb{M} \mathbb{N}}\,$ becomes block-diagonal and reads
\beq
\label{Omega_matrix}
\Omega_{\mathbb{M} \mathbb{N}} = 
\left( 
\begin{array}{c|c}
\Omega_{\a M \b N} & 0 \\[1mm] 
\hline
\\[-4mm]
0 & \Omega_{\m \n}
\end{array}
\right) 
=
\left( 
\begin{array}{c|c}
\epsilon_{\a \b}\, \eta_{MN} & 0 \\[1mm] 
\hline
\\[-4mm]
0 & \mathcal{C}_{\m \n}
\end{array}
\right) \ ,
\eeq
where $\,\epsilon_{\a \b}\,$ is the Levi-Civita $\,\textrm{SL}(2)$-invariant tensor (normalised as $\,\epsilon_{+-}=1\,$) and where $\,\eta_{MN}\,$ and $\,\mathcal{C}_{\m \n}\,$ are the metric and the charge conjugation matrix of $\,\textrm{SO}(6,6)$, respectively. We have summarised our conventions for spinorial representations, gamma matrices, etc. of $\,\textrm{SO}(6,6)\,$ in the appendix~\ref{App:spinors}.

\subsubsection*{Fluxes and the embedding tensor}

The decomposition of the $\,\textbf{56}\,$ of $\,\textrm{E}_{7(7)}\,$ under $\,\textrm{SL}(2) \times \textrm{SO}(6,6)\,$ translates into the index splitting $\,\mathbb{M}=\alpha M \oplus \mu\,$. When expressed in terms of the different pieces of the embedding tensor, the tensor $\,X_{\mathbb{MNP}}\,$ entering the gauge brackets in (\ref{gauge_algebra}) can then be split into components involving an even number of fermionic indices
\beq
\label{Xbosonic}
\begin{array}{cclc}
X_{\a M \b N \g P} & = & - \, \epsilon_{\b \g} \, f_{\a MNP} - \, \epsilon_{\b \g} \, \eta_{M [N}\, \, \xi_{ \a P]} - \, \epsilon_{\a (\b} \, \xi_{\g) M} \, \eta_{NP}& , \\[4mm]
X_{\a M \m \n} & = & -\dfrac{1}{4}  \, f_{\a MNP} \, \left[ \g^{NP} \right]_{\m \n}  -  \dfrac{1}{4} \, \xi_{\a N} \, \left[ {\g_{M}}^{N} \right]_{\m \n}         & , \\[4mm]
X_{\m \a M \n} = X_{\m \n \a M} & = & \dfrac{1}{8} \, f_{\a MNP} \, \left[ \g^{NP} \right]_{\m \n} - \dfrac{1}{24} \, f_{\a NPQ} \, \left[ {\g_{M}}^{NPQ} \right]_{\m \n} \\[3mm]
& + & \dfrac{1}{8} \, \xi_{\a N} \, \left[ {\g_{M}}^{N} \right]_{\m \n} - \dfrac{1}{8} \, \xi_{\a M} \, \cC_{\m \n}        & , \\[2mm]
\end{array}
\eeq
which turn out to be sourced by $\,f_{\a MNP}\,$ and $\,\xi_{\a M}$, together with those involving an odd number of them
\beq
\label{Xfermionic}
\begin{array}{cclc}
X_{\mu \nu \rho}  & = & - \dfrac{1}{2} \, {F_{M}}_{\dot{\n}} \, {{[\gamma_{N}]}_{\mu}}^{\dot{\n}} \, \left[ \g^{MN} \right]_{\n \r} & , \\[4mm]
X_{\mu \a M \b N}  & = & -2 \, \epsilon_{\a \b} \, F_{[M \dot{\n}} \, {\left[ \g_{N]} \right]_{\m}}^{\dot{\n}} \, - \, 2 \,  \eta_{MN} \, \Xi_{\a \b \m}   & , \\[4mm]   
X_{\a M \m \b N} = X_{\a M \b N \m} & = & \epsilon_{\a \b} \, {[\gamma_{N}]_{\mu}}^{\dot{\n}} \, F_{M\dot{\n}} \, +  \, \Xi_{\a \b \n} \, {\left[ \g_{MN} \right]^{\n}}_{\m} \, +  \, \Xi_{\a \b \m} \, \eta_{MN}  & ,
\end{array}
\eeq
which are sourced by $\,F_{M \dot{\mu}}\,$ and $\,\Xi_{\a \b \mu}\,$. The set of components in (\ref{Xbosonic}) specifies how half-maximal supergravity is embedded inside maximal \cite{Dibitetto:2011eu}, whereas the remaining components in (\ref{Xfermionic}) represent the completion from half-maximal to maximal supergravity. A derivation of the expression in (\ref{Xbosonic}) and (\ref{Xfermionic}) can be found in the appendix~\ref{App:spinors}. 

The brackets of the gauge algebra in (\ref{gauge_algebra}) involving the $\,X_{\mathbb{M}}\,=\, X_{\a M}\,\oplus\,X_{\mu}\,$ generators in maximal supergravity then takes the form
\beq
\label{algebra_maximal}
\begin{array}{ccclclc}
\left[ X_{\a M} , X_{\b N} \right] & = &  -  & {X_{\a M \b N}}^{\g P} \, X_{\g P} &-& {X_{\a M \b N}}^{\r} \, X_{\r}  & , \\[2mm]
\left[ X_{\a M} , X_{\m} \right] & = &  - & {X_{\a M \m}}^{\g P} \, X_{\g P} &-& {X_{\a M \m}}^{\r} \, X_{\r}  & , \\[2mm]
\left[ X_{\m} , X_{\n} \right] & = &  - & {X_{\m \n}}^{\g P} \, X_{\g P} &-& {X_{\m \n}}^{\r} \, X_{\r}  & , 
\end{array}
\eeq
where, if looking at the part only involving the $\,X_{\a M}\,$ generators, namely
\beq
\label{algebra_half-maximal}
\begin{array}{ccc}
\left[ X_{\a M} , X_{\b N} \right]  &=&   - \,\, {X_{\a M \b N}}^{\g P} \, X_{\g P} \ ,
\end{array}
\eeq
we rediscover the gauge algebra of half-maximal supergravity \cite{Schon:2006kz} where the $\,X_{\m}\,$ generators have been projected out of the theory.

Moving to explicit string constructions, the $\,f_{\a MNP}\,$ and $\,\xi_{\a M}$ embedding tensor pieces have been related to different background fluxes, e.g. to gauge, geometric and non-geometric fluxes, in compactifications of type II and Heterotic strings producing half-maximal supergravities \cite{Aldazabal:2008zza,Dall'Agata:2009gv,Dibitetto:2010rg}. These fluxes restore the invariance of the four-dimensional supergravity under T- and S-duality, i.e. under $\,\textrm{SL}(2) \times \textrm{SO}(6,6)\,$ transformations. The $\,F_{M \dot{\mu}}\,$ and $\,\Xi_{\a \b \mu}\,$ embedding tensor pieces are related to additional background fluxes which restore the invariance of the theory under U-duality, i.e. under $\,\textrm{E}_{7(7)}\,$ transformations \cite{Aldazabal:2010ef}. Nevertheless, the identification between embedding tensor components and fluxes strongly depends on the string theory under consideration. For instance, a component of the embedding tensor corresponding to a metric flux $\,\omega\,$ in a type IIA construction might correspond to a non-geometric $Q$ flux in a type IIB one and vice versa (see tables \ref{table:unprimed_fluxes} and \ref{table:primed_fluxes} in appendix~\ref{App:fluxes}). We will take this fact into account in the last section when analysing specific type II flux models.

\subsubsection*{Quadratic constraints}

Plugging the expression for the components of the tensor $\,X_{\mathbb{MNP}}\,$ in (\ref{Xbosonic}) and (\ref{Xfermionic}) into the quadratic constraints in (\ref{quadratic_const}) one finds a set of quadratic relations for the embedding tensor pieces $\,f_{\a MNP}\,$, $\,\xi_{\a M}\,$, $\,F_{M \dot{\mu}}\,$ and $\,\Xi_{\a \b \mu}\,$. In doing so we use the $\,\textrm{Sp}(56,\mathbb{R})\,$  invariant matrix $\,\Omega_{\mathbb{M} \mathbb{N}}\,$ in (\ref{Omega_matrix}). As for the $\,\textrm{SU}(8)\,$ formulation in the previous section, let us use the $\,\textrm{E}_{7(7)}\supset \textrm{SL}(2) \times \textrm{SO}(6,6)\,$ branching relations
\beqa
\label{irrep_133}
\textbf{133} & \longrightarrow  & (\textbf{1},\textbf{66})\,+\,(\textbf{3},\textbf{1})\,+\,(\textbf{2},\textbf{32'}) \ , \\[2mm]
\label{irrep_8645}
\textbf{8645}&\longrightarrow & (\textbf{1},\textbf{66})\,+\,(\textbf{1},\textbf{2079})\,+\,(\textbf{3},\textbf{66})\,+\,(\textbf{3},\textbf{495})\,+\,(\textbf{3},\textbf{1})\,+\,(\textbf{1},\textbf{462'}) \,+ \nonumber \\[1mm]
  & &  + \,\,(\textbf{2},\textbf{32'})\,+\,(\textbf{2},\textbf{352})\,+\,(\textbf{2},\textbf{1728'}) \,+\,(\textbf{4},\textbf{32'}) \ ,
\eeqa
as an organising principle for the quadratic constraints. After a straightforward but tedious computation, one finds the following set of quadratic constraints:
\beqa
\label{QC1}
&\hspace{-15mm} i)  & \xi_{\a M}\,\xi_{\b}^{\phantom{a}M} \,+\,4\, \epsilon^{\g \d}\, \mathcal{C}^{\mu \nu} \,  \Xi_{\alpha \gamma \mu} \, \Xi_{\beta \delta \nu} = 0 \ , \\[2mm]
\label{QC2}
&\hspace{-15mm} ii) & \xi_{(\a}^{\phantom{a}P}\,f_{\b)PMN} \, - \, 4\,\Xi_{\a \b \m} \, F_{[M \dot{\n}} \, [\gamma_{N]}]^{\mu \dot{\n}} = 0 \ , \\[2mm]
\label{QC3}
&\hspace{-15mm} iii)& 3\,f_{\a R[MN}\,f_{\b PQ]}^{\phantom{abcde}R}\,+\,2\,\xi_{(\a [M}\,f_{\b)NPQ]}   \\ \nonumber
& & - \, 4\,\Xi_{\a \b \m} \, F_{[M \dot{\n}} \, [\gamma_{NPQ]}]^{\mu \dot{\n}} \,-\, \epsilon^{\g \d}\, \Xi_{\alpha \gamma \mu} \, \Xi_{\beta \delta \nu} \, [\gamma_{MNPQ}]^{\mu \nu} = 0 \ , \\[2mm]
\label{QC4}
&\hspace{-15mm} iv) & \epsilon^{\a \b}\left(\xi_{\a}^{\phantom{a}P}\,f_{\b PMN}\,+\,\xi_{\a M}\,\xi_{\b N}\right)  \\ \nonumber 
& & - \, 4\, F_{M\dot{\m}} \, {F_{N}}^{\dot{\m}}  \, - \, F_{P \dot{\mu}} \, {F^{P}}_{\dot{\nu}} \, [\gamma_{MN}]^{\dot{\mu} \dot{\nu}} \, + \, 2\, \eps^{\a \g} \, \eps^{\b \d} \, \Xi_{\a \b \m}  \, \Xi_{\g \d \n} \, [\g_{MN}]^{\m \n} = 0  \ , \\[2mm]
\label{QC5}
&\hspace{-15mm} v) & \epsilon^{\a \b}\left(f_{\a MNR}\,f_{\b PQ}^{\phantom{abcde}R}\,-\,\xi_{\a}^{\phantom{a}R}\,f_{\b R[M[P}\,\eta_{Q]N]}\,-\,\xi_{\a [M}\,f_{\b N]PQ}\,+\,\xi_{\a [P}\,f_{\b Q]MN}\right)  \\ \nonumber
& & + \, 4\, F_{[M\dot{\m}}\,[\g_{N][P}]^{\dot{\m}\dot{\n}}\,F_{Q] \dot{\n}} \, - \, F_{R \dot{\m}} \,  {F^{R}}_{\dot{\n}} \,[\bar{\g}_{[M}\,\eta_{N][P}\,\g_{Q]}]^{\dot{\m} \dot{\n}} \, + \, 2\,\Xi_{\a\b\m}\,{\Xi^{\a\b}}_{\n}\,[\g_{[M}\,\eta_{N][P}\,\bar{\g}_{Q]}]^{\m\n}  = 0 \ , \\[2mm]
\label{QC6}
&\hspace{-15mm} vi)  & f_{\a MNP} \, {f_{\b}}^{MNP} \,+\,30 \, \epsilon^{\g \d}\, \mathcal{C}^{\mu \nu} \,  \Xi_{\alpha \gamma \mu} \, \Xi_{\beta \delta \nu} = 0 \ , \\[2mm]
\label{QC7} &\hspace{-15mm} vii)  & \left. \epsilon^{\a \b}\,\,
f_{\a [MNP} \, f_{\b QRS]} \,\, \right|_{\textrm{SD}} \, - \,
\frac{1}{40}  \, F_{T \dot{\m}} \, {F^{T}}_{\dot{\n}} \,
[\gamma_{MNPQRS}]^{\dot{\m}\dot{\n}}  = 0  \ , \eeqa
associated to the irrep's
\beq
\begin{array}{cccc}
i)\,\,(\textbf{3},\textbf{1}) & \hspace{10mm}   ii)\,\,(\textbf{3},\textbf{66}) & \hspace{10mm} iii)\,\,(\textbf{3},\textbf{495}) & \hspace{10mm}  iv)\,\,(\textbf{1},\textbf{66})
\end{array}
\eeq
\vspace{-5mm}
\beq
\begin{array}{ccc}
v)\,\,(\textbf{1},\textbf{66}) \, + \, (\textbf{1},\textbf{2079}) &\hspace{20mm}  vi)\,\,(\textbf{3},\textbf{1})  & \hspace{20mm} vii)\,(\textbf{1},\textbf{462'}) \ ,
\end{array}
\eeq
together with three additional ones
\beqa
\label{QC8}
&\hspace{-15mm} viii)  & f_{(\a MNP}\,\Xi_{\b\g)\n}\,[\g^{MNP}]^{\n\dot{\m}}\,-\,15\,\xi_{(\a M}\,\Xi_{\b\g)\n}\,[\g^{M}]^{\n\dot{\m}} = 0 \ , \\[2mm]
\label{QC9}
&\hspace{-15mm} ix) & - 3\, \xi_{\a M} \,F^{M \dot{\m}}  \, + \, \epsilon^{\beta \gamma} \, \left( \frac{1}{2} \, \xi_{\b M} \, [\gamma^{M}]^{\nu \dot{\m}} \, + \, \frac{1}{6} \, f_{\b MNP} \, [\gamma^{MNP}]^{\nu \dot{\m}}  \right) \, \Xi_{\a \g \n} = 0 \ , \\[2mm]
\label{QC10} &\hspace{-15mm} x)& {f_{\a MN}}^{P} \,
{F_{P}}^{\dot{\m}} \, + \, \frac{1}{4} \, f_{\a PQ [M} \, F_{N]
\dot{\n}} \, [\gamma^{PQ}]^{\dot{\n} \dot{\m}} \,-\, \frac{1}{12} \,
f_{\a PQR} \, F_{[M \dot{\n}} \, [{\gamma_{N]}}^{PQR}]^{\dot{\n}
\dot{\m}} \\ \nonumber & & + \, \frac{1}{4} \, \xi_{\a P} \, F_{[M
\dot{\n}} \, [{\gamma_{N]}}^{P}]^{\dot{\n} \dot{\m}} \, + \, \epsilon^{\beta \gamma} \, \left( f_{\b MNP}
\, [\gamma^{P}]^{\nu \dot{\m}} \, - \, \xi_{\b [M} \,
[\gamma_{N]}]^{\nu \dot{\m}} \right) \, \Xi_{\a \g \n} \\ \nonumber
& & - \, \frac{5}{4} \, \xi_{\a [M} \, {F_{N]}}^{\dot{\m}} \, = 0
\eeqa
associated to
\beq
\begin{array}{ccc}
 viii)\,\,(\textbf{4},\textbf{32'}) & \hspace{10mm} ix)\,\,(\textbf{2},\textbf{32'}) & \hspace{10mm} x)\,\,(\textbf{2},\textbf{32'}) \, + \, (\textbf{2},\textbf{352}) \, + \, (\textbf{2},\textbf{1728'}) \ .
\end{array}
\eeq
Let us comment a bit more about the above set of quadratic constraints. If we refer to the embedding tensor pieces $\,f_{\a MNP}\,$ and $\,\xi_{\a M}\,$  as ``bosonic'' and to $\,F_{M\dot{\m}}\,$ and $\,\Xi_{\a \b \m}\,$ as ``fermionic'', then the first seven conditions can be understood as $\,\textrm{(bos $\times$ bos) $+$ (fermi $\times$ fermi}) = 0\,$ quadratic constraints whereas the last three are of the form $\,\textrm{(bos $\times$ fermi) } = 0\,$. As a check of consistency, the first seven conditions reduce to those of the form $\,\textrm{(bos $\times$ bos) } = 0\,$ in ref.~\cite{Dibitetto:2011eu} by setting $\,F_{M\dot{\m}} \,=\, \Xi_{\a \b \m} = 0\,$, namely, by switching off fluxes associated to $\,\textrm{SO}(6,6)$-fermi irrep's of the embedding tensor. In this case, the last three conditions are trivially satisfied.

As mentioned in the introduction, looking for a higher-dimensional origin of dual fluxes is becoming a very exciting line of research. As far as fluxes related to the $\,f_{\a MNP}\,$ components of the embedding tensor are concerned, only purely electric (or equivalently magnetic) $\,\textrm{SO}(6,6)\,$ gaugings have been so far addressed by Double Field Theory \cite{DFT_N=4}. However, the explicit twelve-dimensional twist matrices producing such gaugings have only been built in some particular cases \cite{Hull&Edwards,Dall'Agata:2007sr}. In order to firstly extend to $\,\textrm{SL}(2) \times \textrm{SO}(6,6)\,$ gaugings including fluxes related to $\,\xi_{\a M}\,$ and secondly to $\,\textrm{E}_{7(7)}\,$ gaugings involving also fluxes related to the $\,F_{M\dot{\m}}\,$ and $\,\Xi_{\a \b \m}\,$ components (such as R-R gauge fluxes amongst others), a generalisation to a $56$-dimensional ``twisted megatorus'' reduction has been proposed \cite{Dall'Agata:2007sr}. The restrictions upon the $56$-dimensional twist matrices on this megatorus have not been worked out yet, but, when expressed in terms of fluxes, they should at least imply those in (\ref{QC1})-(\ref{QC7}) and (\ref{QC8})-(\ref{QC10}) whenever the twist is compatible with maximal supersymmetry in four dimensions.

\subsection{Connecting different formulations}
\label{sec:connecting_A7&A1D6}

In order to relate the  $\,\textrm{SL}(2) \,\times \, \textrm{SO}(6,6)\,$ and the $\,\textrm{SU}(8)\,$ formulations of maximal supergravity, it is mandatory to derive the expression of the $\,X_{\mathbb{MNP}}\,$ tensor entering the brackets in (\ref{algebra_maximal}) as a function of the fermionic mass terms in (\ref{Fermi_Lagrangian}).\, This can be done in a two-step procedure as follows:

\begin{itemize}

\item[$1)$] By using the tensors $\mathcal{A}^{\mathcal{IJ}}$ and ${\mathcal{A}_{\mathcal{I}}}^{\mathcal{JKL}}$, we can build the so-called  $T$-tensor \cite{deWit:2007mt,Diffon:2011wt}. The components of this $T$-tensor take the form
\beq
\hspace{-12mm}
\begin{array}{ccll}
\label{T(A)}
T_{\mathcal{IJ KL MN}} & = & \frac{1}{24} \, \epsilon_{\mathcal{KLMNRSTU}}\, \delta_{[\mathcal{I}}^{\mathcal{R}} \, {\mathcal{A}_{\mathcal{J}]}}^{\mathcal{STU}} & , \\[4mm]  
{T_{\mathcal{IJ KL}}}^{\mathcal{MN}} & = & \phantom{-} \frac{1}{2} \,\delta_{[\mathcal{K}}^{[\mathcal{M}} \, {\mathcal{A}^{\mathcal{N}]}}_{\mathcal{L}]\mathcal{IJ}}  + \delta_{[\mathcal{I}[\mathcal{K}}^{\mathcal{MN}} \, \mathcal{A}_{\mathcal{L}]\mathcal{J}]} & , \\[4mm]
T_{\mathcal{IJ \phantom{KL} MN}}^{\mathcal{\phantom{IJ} KL}} & = & - \frac{1}{2} \,\delta_{[\mathcal{M}}^{[\mathcal{K}} \, {\mathcal{A}^{\mathcal{L}]}}_{\mathcal{N}]\mathcal{IJ}}  - \delta_{[\mathcal{I}[\mathcal{M}}^{\mathcal{KL}} \, \mathcal{A}_{\mathcal{N}]\mathcal{J}]} & ,\\[4mm]
{T_{\mathcal{IJ}}}^{\mathcal{KL MN}} & = & \delta_{[\mathcal{I}}^{[\mathcal{K}} \, {\mathcal{A}_{\mathcal{J}]}}^{\mathcal{LMN}]} & , \\[2mm]  
\end{array}
\eeq
together with their complex conjugates. We can arrange them into a $\,T_{\underline{\mathbb{M}} \underline{\mathbb{N}}}{}^{\underline{\mathbb{P}}}\,$ tensor by using the decomposition $\,\underline{\mathbb{M}}=[\mathcal{IJ}] \oplus [\mathcal{\bar{I}\bar{J}}] \equiv \left\lbrace _{\mathcal{IJ}} \, , \, ^{\mathcal{IJ}} \right\rbrace\,$ of the $\,\bold{56}\,$ of $\,\textrm{E}_{7(7)}\,$ under $\,\textrm{SU}(8)\,$.

\item[$2)$] The constant $\,X_{\mathbb{MNP}}\,$ tensor in the $\,\textrm{SU}(8)\,$ formulation, let us denote it $\,\widetilde{X}_{\mathbb{MNP}}\,$ to avoid confusion with that in the $\,\textrm{SL}(2) \times \textrm{SO}(6,6)\,$ formulation, can then be obtained by removing the dependence of the $\,T_{\underline{\mathbb{M}} \underline{\mathbb{N}} \underline{\mathbb{P}}}\,$ tensor on the scalar fields (see footnote~1)
\beq
\label{tilde_X(T)}
\widetilde{X}_{\mathbb{MNP}} = 2 \, \widetilde{\mathcal{V}}_{\mathbb{M}}^{\phantom{\mathbb{M}}\underline{\mathbb{Q}}} \, \widetilde{\mathcal{V}}_{\mathbb{N}}^{\phantom{\mathbb{N}}\underline{\mathbb{R}}} \, \widetilde{\mathcal{V}}_{\mathbb{P}}^{\phantom{\mathbb{P}}\underline{\mathbb{S}}}\,\,\, T_{\underline{\mathbb{Q}\mathbb{R}\mathbb{S}}} \ ,
\eeq
where $\,\widetilde{\mathcal{V}}_{\mathbb{M}}^{\phantom{\mathbb{M}}\underline{\mathbb{Q}}}\,$ is the $\,\textrm{E}_{7(7)}/\textrm{SU}(8)\,$ vielbein in the $\,\textrm{SU}(8)\,$ formulation \cite{deWit:2007mt}. After removing the scalar dependence, the $\,\widetilde{X}_{\mathbb{MNP}}\,$ and $\,X_{\mathbb{MNP}}\,$ constant tensors in the $\textrm{SU}(8)$ and $\,{\textrm{SL}(2) \times \textrm{SO}(6,6)}\,$ formulations are related via a constant change of basis
\beq
\label{X(tilde_X)}
X_{\mathbb{MNP}} = \mathring{\mathcal{V}}_{\mathbb{M}}^{\phantom{\mathbb{M}}\mathbb{Q}} \, \mathring{\mathcal{V}}_{\mathbb{N}}^{\phantom{\mathbb{N}}\mathbb{R}} \, \mathring{\mathcal{V}}_{\mathbb{P}}^{\phantom{\mathbb{P}}\mathbb{S}}\,\,\, \widetilde{X}_{\mathbb{QRS}} \ .
\eeq 
Composing (\ref{tilde_X(T)}) and (\ref{X(tilde_X)}), the resulting vielbein\footnote{For more details on the vielbein $\,\mathcal{V}^{\phantom{\mathbb{M}}\underline{\mathbb{N}}}_{\mathbb{M}}\,$, see appendix~\ref{App:vielbein}.} $\,\mathcal{V}_{\mathbb{M}}^{\phantom{\mathbb{M}}\underline{\mathbb{N}}} = \mathring{\mathcal{V}}_{\mathbb{M}}^{\phantom{\mathbb{M}}\mathbb{P}} \, \widetilde{\mathcal{V}}_{\mathbb{P}}^{\phantom{\mathbb{P}}\underline{\mathbb{N}}}\,$ directly connects the tensors $\,X_{\mathbb{MNP}}\,$ and $\,T_{\underline{\mathbb{MNP}}}\,$ 
\beq
\label{X(T)}
X_{\mathbb{MNP}} = 2 \, \mathcal{V}_{\mathbb{M}}^{\phantom{\mathbb{M}}\underline{\mathbb{Q}}} \, \mathcal{V}_{\mathbb{N}}^{\phantom{\mathbb{N}}\underline{\mathbb{R}}} \, \mathcal{V}_{\mathbb{P}}^{\phantom{\mathbb{P}}\underline{\mathbb{S}}}\,\,\, T_{\underline{\mathbb{QRS}}} \ .
\eeq 
\end{itemize}

Schematically, the connection between the two formulations of maximal supergravity works in the following way
\beq
\label{mapping_summary}
\begin{array}{ccccc}
 \underbrace{\left( \mathcal{A}^{\mathcal{IJ}} \,\,,\,\, {\mathcal{A}_{\mathcal{I}}}^{\mathcal{JKL}} \right)}_{\textrm{fermi. masses}} \hspace{5mm} & \vspace{-5mm} \Longrightarrow  &  T_{\underline{\mathbb{MNP}}}  & \Longrightarrow  & \hspace{5mm} \underbrace{\hspace{5mm}X_{\mathbb{MNP}}\hspace{5mm}}_{\textrm{flux background}} \ .\\[2mm]
  & (\ref{T(A)}) &  & (\ref{X(T)})  
\end{array}
\eeq
By inverting the above chain\footnote{The inversion of the relations in (\ref{T(A)}) gives $\,\mathcal{A}^{\mathcal{IJ}}=\frac{4}{21} \, {T^{\mathcal{IKJL}}}_{\mathcal{KL}}\,$ and $\,{\mathcal{A}_{\mathcal{I}}}^{\mathcal{JKL}}=2 \, {T_{\mathcal{MI}}}^{\mathcal{MJKL}}$, showing that there is some redundancy in the $T$-tensor components.} we are able to relate a flux background given in terms of $\,f_{\a MNP}\,$, $\,\xi_{\a M}\,$, $\,F_{M \dot{\mu}}\,$ and $\,\Xi_{\a \b \mu}\,$ to certain fermionic mass terms $\mathcal{A}^{\mathcal{IJ}}$ and ${\mathcal{A}_{\mathcal{I}}}^{\mathcal{JKL}}\,$. This amounts to know the relations
\beq
\label{A's_fluxes}
\begin{array}{ccr}
\mathcal{A}^{\mathcal{IJ}} &=& \mathcal{A}^{\mathcal{IJ}} \, \left( \, f_{\a MNP} \,,\, \xi_{\a M} \,,\, F_{M \dot{\mu}} \,,\, \Xi_{\a \b \mu} \,\, ; \,\, \mathcal{V}_{\phantom{\mathbb{M}}\underline{\mathbb{N}}}^{\mathbb{M}} \, \right) \phantom{\ .} \\[2mm]
{\mathcal{A}_{\mathcal{I}}}^{\mathcal{JKL}} &=& {\mathcal{A}_{\mathcal{I}}}^{\mathcal{JKL}} \, \left( \, f_{\a MNP} \,,\, \xi_{\a M} \,,\, F_{M \dot{\mu}} \,,\, \Xi_{\a \b \mu} \,\, ; \,\, \mathcal{V}_{\phantom{\mathbb{M}}\underline{\mathbb{N}}}^{\mathbb{M}} \, \right) \ ,
\end{array}
\eeq
where $\,\mathcal{V}_{\phantom{\mathbb{M}}\underline{\mathbb{N}}}^{\mathbb{M}}=(\mathcal{V}^{-1})_{\underline{\mathbb{N}}}^{\phantom{\mathbb{N}}\mathbb{M}}\,$ is the inverse vielbein. Having the relations (\ref{A's_fluxes}), we can make use of (\ref{V_SU8}), (\ref{scalars_eom}) and (\ref{Mass-matrix}) in order to compute the scalar potential, the E.O.M's and the masses of the $70$ physical scalars for a specific flux background. 

\subsection*{The $\,F_{M\dot{\m}} \,=\, \Xi_{\a \b \m} = 0\,$ case}

Let us derive the relations (\ref{A's_fluxes}) between fermionic mass terms and embedding tensor components when $\,F_{M\dot{\m}} \,=\, \Xi_{\a \b \m} \,=\, 0\,$. In the string theory side, this means that fluxes related to fermionic components of the embedding tensor are set to zero, so that
\beq
\label{A's_fluxes_Bosonic}
\begin{array}{ccr}
\mathcal{A}^{\mathcal{IJ}} &=& \mathcal{A}^{\mathcal{IJ}} \, \left( \, f_{\a MNP} \,,\, \xi_{\a M}  \,\, ; \,\, \mathcal{V}_{\phantom{\mathbb{M}}\underline{\mathbb{N}}}^{\mathbb{M}} \, \right) \phantom{\ .} \\[2mm]
{\mathcal{A}_{\mathcal{I}}}^{\mathcal{JKL}} &=& {\mathcal{A}_{\mathcal{I}}}^{\mathcal{JKL}} \, \left( \, f_{\a MNP} \,,\, \xi_{\a M}  \,\, ; \,\, \mathcal{V}_{\phantom{\mathbb{M}}\underline{\mathbb{N}}}^{\mathbb{M}} \, \right) \ .
\end{array}
\eeq

Before presenting the explicit form of the relations in (\ref{A's_fluxes_Bosonic}), we want to point out an issue that appears during the computation, the way to overcome it and the corresponding price to pay: 
\begin{itemize}

\item[$i)$] In the $\,\textrm{SL}(2) \times \textrm{SO}(6,6)\,$ formulation of maximal supergravity, the scalar fields split into ``bosonic'' $\,\{\phi_{\a\b} \,,\, \phi_{MN}\}\,$ and ``fermionic'' $\,\phi_{\a\dot{\m}}\,$ ones. While the former enter the vielbein $\,{\mathcal{V}_{\mathbb{M}}}^{\underline{\mathbb{N}}}\,$ in a simple way, the latter do it in a very complicated way. In the derivation of the relations (\ref{A's_fluxes_Bosonic}), we will set $\,\phi_{\a\dot{\m}}=0\,$ which means that all the ``fermionic'' scalars are fixed to their values at the origin of the moduli space. Therefore, the relation between fluxes and fermionic mass terms that we present here is only valid in the submanifold of the moduli space where $\,\phi_{\a\dot{\m}}=0\,$. 

\item[$ii)$] Being tight to the submanifold with $\,\phi_{\a\dot{\m}}=0\,$ is perfectly consistent with embedding $\,\cN=4\,$ flux compactifications (and truncations thereof) inside $\,\cN=8\,$ supergravity, since ``fermionic'' scalars are projected out (set to zero) when truncating from maximal to half-maximal supergravity in four dimension \cite{Dibitetto:2011eu}. A special point in this submanifold is the origin of the moduli space defined in (\ref{M_origin}), where both ``bosonic'' and ``fermionic'' scalars are set to zero.

\item[$iii)$] One of the main consequences of taking $\,F_{M\dot{\m}} \,=\, \Xi_{\a \b \m} \,=\, 0\,$ as well as $\,\phi_{\a\dot{\m}}=0\,$ is that the method introduced in ref.~\cite{Dibitetto:2011gm} (and further exploited in ref.~\cite{DallAgata:2011aa}) for charting critical points of the scalar potential becomes more subtle. This method relies on the fact that the manifold spanned by the scalars, i.e. $\,\textrm{E}_{7(7)}/\textrm{SU}(8)\,$ in the case of maximal supergravity, is homogeneous so any critical point can be brought back to the origin of the moduli space by applying an $\,\textrm{E}_{7(7)}\,$ transformation. However, neither $\,{F_{M\dot{\m}} \,=\, \Xi_{\a \b \m} \,=\, 0}\,$ nor $\,\phi_{\a\dot{\m}}=0\,$ are U-duality covariant conditions: $\,\textrm{E}_{7(7)}\,$ transformations will mix ``fermionic'' and ``bosonic'' embedding tensor components and scalars, hence rendering the relations in (\ref{A's_fluxes_Bosonic}) no longer valid. We will be back to this point in the last section when discussing specific flux backgrounds yielding maximal supergravities.

\end{itemize}

Taking $\,F_{M\dot{\m}} \,=\, \Xi_{\a \b \m} \,=\, 0\,$ together with $\,\phi_{\a\dot{\m}}=0\,$ has strong implications for the mapping between fluxes and fermionic mass terms. By virtue of the decompositions (\ref{A1D6_decomp}) and (\ref{A7_decomp}), only those components (as well as their c.c.) of the form $\,\{ {\mathcal{V}_{\a M}}^{ij} \,,\,  {\mathcal{V}_{\a M}}^{\hi \hj} \,,\,  {\mathcal{V}_{i \hj}}^{k \hl}  \}\,$ inside the vielbein $\,\mathcal{V}^{\phantom{\mathbb{M}}\underline{\mathbb{N}}}_{\mathbb{M}}\,$ are non-vanishing. They are expressed in terms of an $\,\textrm{SL}(2)\,$ complexified vielbein $\,\mathcal{V}_{\a}\,$ and an $\,\textrm{SO}(6,6)\,$ vielbein $\,\mathcal{V}_{M}=\{ {\mathcal{V}_{M}}^{ij} , {\mathcal{V}_{M}}^{\hi \hj} \}\,$ where $\,i=1,...,4\,$ and $\,\hi=1,...,4\,$ respectively denote $\,\textrm{SU}(4)_{\textrm{time-like}}\,$ and $\,\textrm{SU}(4)_{\textrm{space-like}}\,$ fundamental indices (see appendix~\ref{App:vielbein}). 

Considering this reduced set of vielbein components, we can build the explicit mapping between fermionic mass terms and fluxes by following the prescription in (\ref{mapping_summary}). It will be useful to define the tensors	
\beq
\label{fermi_mass_N=4}
\begin{array}{ccl}
A_{1}^{ij} & = & \epsilon^{\alpha \beta} \, (\mathcal{V}_{\alpha})^{*} \, {\mathcal{V}^{M}}_{kl} \, \mathcal{V}^{N ik} \, \mathcal{V}^{P jl} \, f_{\beta MNP} \\[2mm]
A_{2}^{ij} & = & \epsilon^{\alpha \beta} \,\,\, \mathcal{V}_{\alpha} \,\,\,\,\,\, {\mathcal{V}^{M}}_{kl} \, \mathcal{V}^{N ik} \, \mathcal{V}^{P jl} \, f_{\beta MNP} \,+ \frac{3}{2} \, \epsilon^{\alpha \beta} \, \mathcal{V}_{\alpha} \, {\mathcal{V}}^{M ij} \, {\xi_{\beta M}} \\[2mm]
{A_{2\,\hi \hj i}}^{j} & = & \epsilon^{\alpha \beta} \,\,\, \mathcal{V}_{\alpha} \,\,\,\,\,\, {\mathcal{V}^{M}}_{\hi \hj} \, {\mathcal{V}^{N}}_{ik} \, \mathcal{V}^{P jk} \, f_{\beta MNP} - \frac{1}{4} \,\delta_{i}^{j} \, \epsilon^{\alpha \beta} \, \mathcal{V}_{\alpha} \, {\mathcal{V}^{M}}_{\hi \hj} \, {\xi_{\beta M}}
\end{array}
\eeq
which reproduce the fermionic mass terms in $\,\cN=4\,$ supergravity \cite{Schon:2006kz}, together with their counterparts
\beq
\label{fermi_mass_N=4_extension}
\begin{array}{ccl}
A_{1}^{\hi \hj} & = & \epsilon^{\alpha \beta} \,\,\, \mathcal{V}_{\alpha} \,\,\,\,\,\, {\mathcal{V}^{M}}_{\hk \hl} \, \mathcal{V}^{N \hi \hk} \, \mathcal{V}^{P \hj \hl} \, f_{\beta MNP} \\[2mm]
A_{2}^{\hi \hj} & = & \epsilon^{\alpha \beta} \, (\mathcal{V}_{\alpha})^{*} \, {\mathcal{V}^{M}}_{\hk \hl} \, \mathcal{V}^{N \hi \hk} \, \mathcal{V}^{P \hj \hl} \, f_{\beta MNP} \,- \frac{3}{2} \, \epsilon^{\alpha \beta} \, (\mathcal{V}_{\alpha})^{*} \, {\mathcal{V}}^{M \hi \hj} \, {\xi_{\beta M}} \\[2mm]
{A_{2\,i j \hi}}^{\hj} & = & \epsilon^{\alpha \beta} \, (\mathcal{V}_{\alpha})^{*} \, {\mathcal{V}^{M}}_{i j} \, {\mathcal{V}^{N}}_{\hi \hk} \, \mathcal{V}^{P \hj \hk} \, f_{\beta MNP} + \frac{1}{4} \,\delta_{\hi}^{\hj} \, \epsilon^{\alpha \beta} \, (\mathcal{V}_{\alpha})^{*} \, {\mathcal{V}^{M}}_{i j} \, {\xi_{\beta M}}
\end{array}
\eeq
which complete the $\,\mathcal{N}=8\,$ theory. In terms of these, the relation between fluxes and fermionic mass terms is given by
\beq
g \, \mathcal{A}^{\mathcal{IJ}} = \frac{1}{3\,\sqrt{2}} \left( 
\begin{array}{c|c}
A_{1}^{ij} & 0 \\[1mm] 
\hline
\\[-4mm]
0 & i \, A_{1}^{\hi \hj} 
\end{array}
\right)
\eeq
for the components inside $\,{\mathcal{A}}^{\mathcal{IJ}}\,$ and
\beq
\begin{array}{cclcccl}
g \,  {\mathcal{A}_{i}}^{jkl} & = &  - \dfrac{1}{3\,\sqrt{2}} \, \epsilon^{jklm} \, A_{2 \, m i} & \hspace{5mm} , \hspace{5mm} & g \,  {\mathcal{A}_{\hi}}^{\hj \hk \hl} & = &  \dfrac{i}{3\,\sqrt{2}} \, \epsilon^{\hj \hk \hl \hm} \, A_{2 \, \hm \hi} \\[4mm]
g \,  {\mathcal{A}_{i}}^{j \hk \hl} & = &  \dfrac{i}{2\,\sqrt{2}} \, \epsilon^{\hk \hl \hi \hj} \, {A_{2 \, \hi \hj i}}^{j} & \hspace{5mm} , \hspace{5mm} & g \,  {\mathcal{A}_{\hi}}^{\hj k l} & = &  -\dfrac{1}{2\,\sqrt{2}} \, \epsilon^{k l i j} \, {A_{2 \, i j \hi}}^{\hj}
\end{array}
\eeq
for those inside $\,{\mathcal{A}_{\mathcal{I}}}^{\mathcal{JKL}}$. Further components involving an odd number of $\,\hat{i}\,$ indices, e.g. $\,\mathcal{A}^{i\hat{j}}\,$ or $\,{\mathcal{A}_{ijk}}^{\hat{l}}\,$, are sourced by fermionic fields and fluxes, thus vanishing in our set-up.

In the next section we present a series of consistent truncations of maximal supergravity yielding simpler theories with a smaller set of fields and embedding tensor components. Later, in the last section, we will investigate the lifting of (solutions of) these truncations to maximal supergravity  making use of the explicit correspondence between flux backgrounds and fermionic mass terms derived here.

\section{A web of group-theoretical truncations}
\label{sec:truncations}

Starting from gauged maximal supergravity in four dimensions, we present a net of group-theoretical truncations (see figure~\ref{fig:truncations}) which connects various supergravity theories preserving different amounts of supersymmetry with different field contents and sets of deformation parameters (embedding tensor components). By group-theoretical truncation, we mean the following procedure: a certain subgroup $H$ of the global symmetry group $G$ of the theory under consideration is chosen, the branching of $G$ irrep's in which fields and deformations live are computed and only the fields and deformations which are singlets with respect to $H$ are kept.

\begin{figure}[h]
\hspace{-7mm}
\renewcommand{\arraystretch}{1.5}
\scalebox{0.70}[0.70]{\xymatrix{*+[F-,]{\begin{array}{cc} \cN=8 : &
\dfrac{\textrm{E}_{7(7)}}{\textrm{SU}(8)}\\\phantom{S} &
\phantom{(\textbf{1},\textbf{1},\textbf{3})+
(\textbf{1},\textbf{3},\textbf{1})+(\textbf{3},\textbf{1},\textbf{1})}
\\[-8mm] \hline \textrm{vectors} & \textbf{56} \vspace{2mm}\\ \hdashline
 \textrm{scalars} & \textbf{133} \vspace{2mm}\\
\hdashline \textrm{emb tens} & \textbf{912}\\
 & \hspace{-16mm}\text{--1--}\end{array}} &
*+[F-,]{\begin{array}{cc}\cN=2 : &
\left(\dfrac{\textrm{SL}(2)}{\textrm{SO}(2)}\right)_{T}\times
\dfrac{\textrm{G}_{2(2)}}{\textrm{SO}(4)}\\ \phantom{S} &
\phantom{(\textbf{1},\textbf{1},\textbf{3})+
(\textbf{1},\textbf{3},\textbf{1})+(\textbf{3},\textbf{1},\textbf{1})}\\[-8mm]
\hline
\textrm{vectors} & (\textbf{4},\textbf{1}) \vspace{2mm}\\\hdashline \textrm{scalars} &
(\textbf{3},\textbf{1})+
(\textbf{1},\textbf{14})\vspace{2mm}\\\hdashline \textrm{emb tens} &
(\textbf{2},\textbf{1})+(\textbf{4},\textbf{14})+(\textbf{2},\textbf{7})\\
 & \hspace{-16mm}\text{--2--}\end{array}}
&*+[F-,]{\begin{array}{cc}\cN=2 : &
\left(\dfrac{\textrm{SL}(2)}{\textrm{SO}(2)}\right)_{T}\times
\dfrac{\textrm{SU}(2,1)}{\textrm{SU}(2)\times\textrm{U}(1)_{U}}\\\phantom{S}
& \phantom{(\textbf{1},\textbf{1},\textbf{3})+
(\textbf{1},\textbf{3},\textbf{1})+(\textbf{3},\textbf{1},\textbf{1})}
\\[-8mm]
\hline \textrm{vectors} & (\textbf{4},\textbf{1}) \vspace{2mm}\\\hdashline \textrm{scalars} &
(\textbf{3},\textbf{1})+
(\textbf{1},\textbf{8})\vspace{2mm}\\\hdashline \textrm{emb tens} &
(\textbf{2},\textbf{1})+(\textbf{4},\textbf{8})+(\textbf{2},\textbf{1})\\
 & \hspace{-16mm}\text{--3--}\end{array}}\\
*+[F-,]{\begin{array}{cc}\cN=4 : &
\left(\dfrac{\textrm{SL}(2)}{\textrm{SO}(2)}\right)_{S}\times\dfrac{\textrm{SO}(6,6)}{\textrm{SO}(6)\times\textrm{SO}(6)}\\\phantom{S}
& \phantom{(\textbf{1},\textbf{1},\textbf{3})+
(\textbf{1},\textbf{3},\textbf{1})+(\textbf{3},\textbf{1},\textbf{1})}
\\[-8mm]
\hline \textrm{vectors} & (\textbf{2},\textbf{12}) \vspace{2mm}\\\hdashline \textrm{scalars} &
(\textbf{3},\textbf{1})+
(\textbf{1},\textbf{66})\vspace{2mm}\\\hdashline \textrm{emb tens} &
(\textbf{2},\textbf{12})+(\textbf{2},\textbf{220})\\
 & \hspace{-16mm}\text{--4--}\end{array}}
& *+[F-,]{\begin{array}{cc}\cN=1 : &
\displaystyle\prod_{\Phi=S,T,U} \left(\dfrac{\textrm{SL}(2)}{\textrm{SO}(2)}\right)_{\Phi} \\[3mm]
\hline \textrm{vectors} &
\text{---} \vspace{2mm} \\\hdashline \textrm{scalars} &
(\textbf{3},\textbf{1},\textbf{1})+(\textbf{1},\textbf{3},\textbf{1})+(\textbf{1},\textbf{1},\textbf{3})\vspace{2mm}\\\hdashline
\textrm{emb tens} &
(\textbf{2},\textbf{2},\textbf{2})+(\textbf{2},\textbf{4},\textbf{4})\\
 & \hspace{-16mm}\text{--5--}\end{array}}
&*+[F-,]{\begin{array}{cc}\cN=1 : &
\displaystyle\prod_{\Phi=S,T}\left(\dfrac{\textrm{SL}(2)}{\textrm{SO}(2)}\right)_{\Phi} \, \times \, \mathbb{R}_{U}\\ \phantom{S}
& \phantom{(\textbf{3},\textbf{1},\textbf{1})+(\textbf{1},\textbf{3},\textbf{1})+(\textbf{1},\textbf{1},\textbf{3})}\\[-8mm]
\hline
\textrm{vectors} & \text{---} \vspace{2mm}\\\hdashline \textrm{scalars} &
(\textbf{1},\textbf{1}) + (\textbf{3},\textbf{1}) +(\textbf{1},\textbf{3}) \vspace{2mm}\\\hdashline \textrm{emb tens}
& (\textbf{2},\textbf{4}) +(\textbf{2},\textbf{4}) \\
 & \hspace{-16mm}\text{--6--}\end{array}}
}}\caption{Starting from gauged maximal supergravity (box --1-- in
the above diagram), one can move step by step downwards or towards
the right by performing group-theoretical truncations which are
described below in detail. The labels $S$, $T$ and $U$ are introduced in order to keep track of the different group factors along the truncations.} 
\label{fig:truncations}
\end{figure}

\begin{itemize}

\item \textbf{Step from --1-- to --2--:} a truncation with respect to an $\,H=\textrm{SO}(3)\,$ subgroup of the $\,G=\textrm{E}_{7(7)}\,$ global symmetry of maximal supergravity is performed by making use of the following chain of maximal subgroups
\beq 
\textrm{E}_{7(7)} \,\,\supset \,\,\textrm{SL}(2)_{T} \,\times \,\textrm{F}_{4(4)} \,\,\supset \,\, \textrm{SL}(2)_{T} \,\times \,\textrm{G}_{2(2)} \,\times \, \textrm{SO}(3) \ .
\eeq
By looking at the decomposition of the fundamental representation of the $\textrm{SU}(8)$ \text{R-symmetry} group in --1-- under the $\,\textrm{SO}(3)\,$ diagonal subgroup\footnote{The diagonal subgroup $\,\textrm{SO}(3)_{\textrm{diag}}\,$ in the chain (\ref{chain:SO3_trunc}) is obtained by identifying the two $\,\textrm{SO}(3)\,$ factors, namely, the fundamental representation of the first with the fundamental of the second.}
\beq 
\label{chain:SO3_trunc}
\textrm{SU}(8) \,\,\supset \,\,\textrm{SU}(4) \,\times \,\textrm{SU}(4) \,\,\supset \,\, \textrm{SU}(3) \,\times \,\textrm{SU}(3) \,\,\supset \,\, \textrm{SO}(3) \,\times \,\textrm{SO}(3) \,\,\supset \,\, \textrm{SO}(3)_{\textrm{diag}} \ ,
\eeq
one finds $\,\textbf{8} = \textbf{3} + \textbf{3} + \textbf{1} + \textbf{1}\,$, hence containing two singlets. This implies that the theory preserves $\cN=2$ supersymmetry with
\beq 
\cM_{\textrm{SK}}= \left(\frac{\textrm{SL}(2)}{\textrm{SO}(2)}\right)_{T} \qquad\textrm{and}\qquad \cM_{\textrm{QK}}=\frac{\textrm{G}_{2(2)}}{\textrm{SO}(4)} \ ,
\eeq
being the special K\"ahler (SK) and the quaternionic K\"ahler (QK) manifolds associated to one vector multiplet and two hyper multiplets respectively. From the diagram in figure~\ref{fig:truncations} one reads that the vector fields in maximal supergravity, transforming in the $\,\textbf{56}\,$ of $\,\textrm{E}_{7(7)}$, are branched into the sum of several irrep's of $\,{\textrm{SL}(2)_{T}\times\textrm{G}_{2(2)} \times \textrm{SO}(3)}$, of which, though, only the ones transforming in the $\,(\textbf{4},\textbf{1})\,$ of $\,\textrm{SL}(2)_{T}\times\textrm{G}_{2(2)}\,$ are $\,\textrm{SO}(3)$ singlets hence surviving the truncation. The resulting theory then comprises four vectors out of which only two (the graviphoton plus an extra vector coming from the vector multiplet) are linearly independent due to the $\,\textrm{Sp}(4,\mathbb{R})\,$ electric-magnetic duality. In addition, the theory contains $2 + 8$ physical scalars spanning $\cM_{\textrm{SK}}$ and $\cM_{\textrm{QK}}$ respectively, together with $72$ deformation parameters associated to the embedding tensor components surviving the truncation.  
\\[-6mm]

\noindent Due to the presence of vectors, this theory might have interesting applications in holographic superconductivity as well as in Cosmology as far as the existence of de Sitter solutions via D-terms uplifting is concerned. We hope to come back to these two issues in the near future.  

\item \textbf{Step from --2-- to --3--:} the truncation is now with respect to an $\,H=\mathbb{Z}_3\,$ discrete subgroup of the $\,G=\textrm{SL}(2)_{T}\times\textrm{G}_{2(2)}\,$ global symmetry of the previous $\,\cN=2\,$ theory via the chain
\beq
\label{chainN2Z3} 
\textrm{SL}(2)_{T}\times \textrm{G}_{2(2)} \,\,\supset \,\,\textrm{SL}(2)_{T}\times\textrm{SU}(2,1) \,\,\supset \,\, \textrm{SL}(2)_{T} \times \textrm{SU}(2) \times \textrm{U}(1)_{U} \ .
\eeq
More concretely we mod-out the different fields in the theory by a $\,\mathbb{Z}_{3}\,$ element of the form $e^{i \frac{2 \pi}{3} \, q}$, where $\,q\,$ $\textrm{mod}(3)$ denotes the charge of the fields with respect to the $\,\textrm{U}(1)_{U}\,$ factor in (\ref{chainN2Z3}). The field content inside the box --3-- follows from the $\,\textrm{G}_{2(2)}\,$ irrep decompositions
\beq
\begin{array}{ccccl}
\hspace{-1mm}\textrm{G}_{2(2)} & \supset & \textrm{SU}(2,1) & \supset & \textrm{SU}(2) \times \textrm{U}(1)_{U} \\[2mm]  
\textbf{1} & \rightarrow & \textbf{1} & \rightarrow & \textbf{1}_{(0)} \\[2mm]
\textbf{7} & \rightarrow & \textbf{1} + \textbf{3} + \overline{\textbf{3}} & \rightarrow & \textbf{1}_{(0)} + (\, \textbf{1}_{(-2)} + \textbf{2}_{(1)} \,) + (\, \textbf{1}_{(2)} + \textbf{2}_{(-1)}  \, ) \\[2mm]
\textbf{14} & \rightarrow & \textbf{8} + \textbf{3} + \overline{\textbf{3}} & \rightarrow &\hspace{-1mm} ( \,\textbf{1}_{(0)} + \textbf{2}_{(0)} + \textbf{2}_{(0)} + \textbf{3}_{(0)} \, ) + ( \, \textbf{1}_{(-2)} + \textbf{2}_{(1)} \, ) + (\, \textbf{1}_{(2)} + \textbf{2}_{(-1)} \,)
\end{array}
\nonumber
\eeq
where the subindex in $\,\textbf{\textit{n}}_{(q)}\,$ refers to the $\,\textrm{U}(1)_{U}\,$ charge $\,q\,$ of the $\,\textrm{SU}(2)\,$ irrep $\,\textbf{\textit{n}}\,$. The truncated theory has an $\,\textrm{SL}(2)_{T}\times\textrm{SU}(2,1)\,$ global symmetry and still keeps $\,\cN=2\,$ supersymmetry. This fact can be seen by obtaining the theory directly from an $\,\textrm{SU}(3)\,$ truncation of maximal supergravity without any intermediate step, as we see next. 

\item \textbf{Step from --1-- to --3--:} truncating maximal supergravity with respect to a compact $\,H=\textrm{SU}(3)\,$ subgroup of its $\,G=\textrm{E}_{7(7)}\,$ global symmetry via the chain
\beq 
\textrm{E}_{7(7)} \,\,\supset \,\,\textrm{SL}(2)_{T} \,\times \,\textrm{F}_{4(4)} \,\,\supset \,\, \textrm{SL}(2)_{T} \,\times \,\textrm{SU}(2,1) \,\times \, \textrm{SU}(3) \ ,
\eeq
gives rise to the theory inside the box --3-- of figure~\ref{fig:truncations}. The truncation preserves $\,{\cN=2}\,$ supersymmetry as results from the decomposition $\,\textbf{8} = \textbf{3} \,+\, \textbf{3} \,+\, \textbf{1} \,+\, \textbf{1}\,$ of the fundamental representation of the $\textrm{SU}(8)$ \text{R-symmetry} group of the maximal theory under the $\,\textrm{SU}(3)\,$ diagonal subgroup\footnote{This time the diagonal subgroup $\,\textrm{SU}(3)_{\textrm{diag}}\,$ in the chain (\ref{chain:SU3_trunc}) is obtained by anti-identifying the two $\,\textrm{SU}(3)\,$ factors, namely, the fundamental representation of the first with the anti-fundamental of the second.}
\beq 
\label{chain:SU3_trunc}
\textrm{SU}(8) \,\,\supset \,\,\textrm{SU}(4) \,\times \,\textrm{SU}(4) \,\,\supset \,\, \textrm{SU}(3) \,\times \,\textrm{SU}(3) \,\,\supset  \,\, \textrm{SU}(3)_{\textrm{diag}} \ .
\eeq
The $\,\cN=2\,$ truncated theory involves the special K\"ahler and the quaternionic K\"ahler manifolds
\beq 
\cM_{\textrm{SK}}= \left(\frac{\textrm{SL}(2)}{\textrm{SO}(2)}\right)_{T} \qquad\textrm{and}\qquad \cM_{\textrm{QK}}=\frac{\textrm{SU}(2,1)}{\textrm{SU}(2) \times \textrm{U}(1)_{U}} \ ,
\eeq
associated to one vector multiplet and one hyper multiplet respectively. This theory can therefore be seen as a truncation of that in box  --2-- where one of the hyper multiplets is projected out after modding out by the $\,\mathbb{Z}_{3}\,$ discrete subgroup previously introduced. The same truncation was explored in ref.~\cite{Warner:1006} and further investigated in refs~\cite{non-relativistic} as gravity dual of non-relativistic field theories.

\item \textbf{Step from --1-- to --4--:} this is the truncation connecting maximal supergravity and half-maximal supergravity coupled to six vector multiplets. It can be seen as a truncation with respect to an $\,H=\mathbb{Z}_2\,$ discrete subgroup of the $\,G=\textrm{E}_{7(7)}\,$ global symmetry of the maximal theory. The resulting theory keeps $\,\cN=4\,$ supersymmetry due to the branching of the R-symmetry group of the maximal theory
\beq
\textrm{SU}(8) \,\, \supset \,\, \textrm{SU}(4) \times \textrm{SU}(4) \, \sim \, \textrm{SO}(6) \times \textrm{SO}(6) \ ,
\eeq
where one of the $\,\textrm{SU}(4)\,$ factors, let us say the first, is parity even under the $\,\mathbb{Z}_2\,$ and the other is parity odd. The fundamental representation of the R-symmetry group of maximal supergravity then decomposes as $\,\textbf{8} = (\textbf{4} , \textbf{1}) + (\textbf{1} , \textbf{4}) = \textbf{4}_{\textrm{even}} + \textbf{4}_{\textrm{odd}}\,$ hence keeping only half of the supersymmetries, namely, those related to $\,\textbf{4}_{\textrm{even}}\,$. The action of the $\,\mathbb{Z}_2\,$ on the fundamental representation of $\textrm{E}_{7(7)}$ becomes more transparent by looking at the branching
\beq
\begin{array}{ccl}
\textrm{E}_{7(7)} & \supset & \textrm{SL}(2)_{S} \times \textrm{SO}(6,6) \\[2mm]  
\textbf{56} & \rightarrow & (\textbf{2},\textbf{12}) + (\textbf{1},\textbf{32})
\end{array}
\nonumber
\eeq
Under the $\,\mathbb{Z}_2\,$, the different $\textrm{E}_{7(7)}$ irrep's are modded-out according to the $\textrm{SO}(6,6)$ irrep's appearing in their branchings. In particular, $\textrm{SO}(6,6)$ bosonic irrep's, e.g. the $\bf{1}$, $\bf{12}$, $\bf{66}$, etc., are parity even and survive the truncation. In contrast, $\textrm{SO}(6,6)$ fermionic irrep's involving an odd number of Majorana-Weyl indices, e.g. the $\bf{32}$ and $\textbf{32'}$, are parity odd and are projected out whereas those involving an even number of them are parity even hence surviving the truncation. As a result, the vectors $\,V_{\mu}\,$, the scalars $\,\phi_{\alpha \dot{\mu}}\,$ and the embedding tensor pieces $\,F_{M \dot{\mu}}\,$ and $\,\Xi_{\alpha \beta \mu}\,$ in the bosonic field content of maximal supergravity are truncated away when going to half-maximal. The remaining fields then describe an $\,\cN=4\,$ supergravity coupled to six vector multiplets with an associated
\beq
\cM_{\textrm{scalar}} = \left(\frac{\textrm{SL}(2)}{\textrm{SO}(2)}\right)_{S} \times \frac{\textrm{SO}(6,6)}{\textrm{SO}(6) \times \textrm{SO}(6)} \ ,
\eeq
coset space spanned by the $\,2 + 36\,$ physical scalars in the theory belonging to the gravity multiplet and the six vector multiplets respectively. Further details about this truncation can be found in ref.~\cite{Dibitetto:2011eu}.

\item \textbf{Step from --4-- to --5--:} this step corresponds to a truncation with respect to an $\,H=\textrm{SO}(3)\,$ subgroup of the $\,G=\textrm{SL}(2)_{S} \times \textrm{SO}(6,6)\,$ global symmetry of half-maximal supergravity coupled to six vector multiplets following the chain
\beq 
\textrm{SL}(2)_{S} \times \textrm{SO}(6,6) \,\,\supset \,\,\textrm{SL}(2)_{S} \times \textrm{SO}(2,2) \times \textrm{SO}(3) \,\sim\,  \prod_{\Phi=S,T,U} \textrm{SL}(2)_{\Phi} \,\times\, \textrm{SO}(3) \ .
\eeq
The truncation breaks half-maximal to minimal $\,\cN=1\,$ supergravity due to the decomposition $\,\textbf{4} = \textbf{1} + \textbf{3}\,$ of the fundamental representation of the $\,\textrm{SU}(4)\,$ R-symmetry group in $\cN=4$ supergravity under the $\,\textrm{SO}(3)\,$ subgroup
\beq 
\textrm{SU}(4) \,\,\supset \,\, \textrm{SU}(3) \,\,\supset \,\, \textrm{SO}(3) \ .
\eeq
The resulting theory does not contain vectors since there are no $\,\textrm{SO}(3)$-singlets in the decomposition $\,\textbf{12}=(\textbf{4},\textbf{3})\,$ of the fundamental of $\,\textrm{SO}(6,6)\,$ under $\,\textrm{SO}(2,2) \times \textrm{SO}(3)$. The physical scalar fields span the coset space
\beq 
\cM_{\textrm{scalar}}= \prod_{\Phi=S,T,U}\left( \frac{\textrm{SL}(2)}{\textrm{SO}(2)} \right)_{\Phi} \ ,
\eeq
involving three $\,\textrm{SL}(2)/\textrm{SO}(2)\,$ factors each of which can be parameterised by a complex scalar $\,\Phi=(S , T , U)$. In addition, the embedding tensor of the theory contains $\,40\,$ independent components fitting two irrep's of the $\,\textrm{SL}(2)_{S} \times \textrm{SL}(2)_{T} \times \textrm{SL}(2)_{U}\,$ global symmetry group, as shown inside box --5-- in figure~\ref{fig:truncations}. We will come back to this truncation in the next section when studying type II string models.
\\[-6mm]

\noindent This $\,\cN=1\,$ supergravity theory has been extensively investigated because of its direct connection to string theory via type II orientifold compactifications with fluxes \cite{STU_models(1),STU_models(2),STU_models(3)}. The resulting supergravity models are referred to as $STU$-models and different background fluxes in the string theory side correspond with different embedding tensor configurations in the supergravity side. However, not all the embedding tensor configurations in the supergravity side have a higher-dimensional interpretation since most of them corresponds to non-geometric flux backgrounds for which an origin in string theory, if possible, remains to be found. 

\noindent It is worth noticing here that this theory can be lifted to that in box --2-- by completing it with the fermionic irrep's removed by the $\,\mathbb{Z}_{2}\,$ truncation taking from the box --1-- to the box --4-- in figure~\ref{fig:truncations}.

\item \textbf{Step from --5-- to --6--:} this truncation is with respect to an $\,H=\mathbb{Z}_3\,$ discrete subgroup of the $\,G=\textrm{SL}(2)_{S} \times \textrm{SL}(2)_{T} \times \textrm{SL}(2)_{U}\,$ global symmetry in --5-- via the chain
\beq
\label{chainN2} 
\textrm{SL}(2)_{S} \times \textrm{SL}(2)_{T} \times \textrm{SL}(2)_{U} \,\,\supset \,\, \textrm{SL}(2)_{S} \times \textrm{SL}(2)_{T} \times \textrm{U}(1)_{U} \ .
\eeq
As happened when truncating from --2-- to --3-- before, we mod-out again the different fields in the theory by a $\,\mathbb{Z}_{3}\,$ element of the form $e^{i \frac{2 \pi}{3} \, q}$, with $\,q\,$ $\textrm{mod}(3)$ being this time the charge of the fields with respect to the $\,\textrm{U}(1)_{U}\,$ factor in (\ref{chainN2}). Now, the relevant branchings in order to derive the field content inside the box --6-- are
\beq
\begin{array}{ccl}
\textrm{SL}(2)_{U} & \supset & \textrm{U}(1)_{U} \\[2mm]  
\textbf{1} & \rightarrow & \textbf{1}_{(0)} \\[2mm]
\textbf{2} & \rightarrow & \textbf{1}_{(-1)} + \textbf{1}_{(1)} \\[2mm]
\textbf{3} & \rightarrow & \textbf{1}_{(-2)} + \textbf{1}_{(0)} + \textbf{1}_{(2)}\\[2mm]
\textbf{4} & \rightarrow & \textbf{1}_{(0)} + \textbf{1}_{(-1)} + \textbf{1}_{(1)} + \textbf{1}_{(0)}
\end{array}
\nonumber
\eeq
where, as before, the subindex in $\,\textbf{1}_{(q)}\,$ refers to the $\,\textrm{U}(1)_{U}\,$ charge $\,q\,$ of the state. The truncated theory still has $\,\cN=1\,$ supersymmetry since the gravitino in the parent theory was already a singlet with respect to both $\,\textrm{U}(1)_{T}\,$ and $\,\textrm{U}(1)_{U}\,$.

\noindent The scalars in the truncated theory span the scalar manifold
\beq
\label{scalar_manifold_6}
\cM_{\textrm{scalar}}= \prod_{\Phi=S,T}\left( \frac{\textrm{SL}(2)}{\textrm{SO}(2)} \right)_{\Phi}  \,\, \times \,\, \mathbb{R}_{U} \ .
\eeq
It can be parameterised by two complex scalars $\,S\,$ and $\,T\,$ associated to the $\,\textrm{SL}(2)/\textrm{SO}(2)\,$ factors plus an extra real scalar associated to the Cartan generator (rescalings) inside the $\,\textrm{SL}(2)_{U}\,$ factor in the global symmetry group of the parent theory. This scalar relates to the $\,\mathbb{R}_{U}\,$ factor in the scalar manifold (\ref{scalar_manifold_6}). As summarised inside the  box --6-- in figure~\ref{fig:truncations}, the embedding tensor consists of two pieces sitting in the same irrep of the global symmetry group of the theory.

\end{itemize}

In the next section we concentrate on the $\,\cN=1\,$ theory inside box --5-- which can be seen as a truncation of the $\,\cN=4\,$ theory inside box --4-- . We will investigate the lifting of some vacuum solutions to $\,\cN=8\,$ supergravity (box --1--) making use of the relations (\ref{A's_fluxes_Bosonic}) between fermionic masses and flux backgrounds when $\,F_{M\dot{\m}}=\Xi_{\a \b \m}=0\,$, i.e. when spinorial fluxes do vanish. It would be interesting to explore the phenomenology of fluxes related to embedding tensor components fitting these fermionic irrep's and still having a higher-dimensional origin in string theory as gauge fluxes or metric fluxes.

\section{Exceptional flux backgrounds}
\label{sec:typeII_examples}

When compactifying type II ten-dimensional supergravities down to four dimensions, background fluxes threading the internal space can be switched on during the compactification procedure giving rise to gauged maximal supergravity models. As introduced in section~\ref{sec:fluxes_A1D6}, flux backgrounds on the string side correspond to deformation parameters related to the $\,f_{\a MNP}\,$, $\,\xi_{\a M}\,$, $\,F_{M \dot{\mu}}\,$ and $\,\Xi_{\a \b \mu}\,$ pieces of the embedding tensor on the supergravity side. For the sake of simplicity, we will restrict our study to the case 
\beq
\label{F&Xi=0}
F_{M\dot{\m}} \,\,=\,\, \Xi_{\a \b \m} \,\,=\,\, 0 \ ,
\eeq
this is, to string backgrounds not including fluxes associated to $\,\textrm{SO}(6,6)\,$ fermionic irrep's of the embedding tensor. However, even though the remaining $\,f_{\a MNP}\,$ and $\,\xi_{\a M}\,$ pieces reproduce those of half-maximal supergravity, the set of quadratic constraints they are restricted by will be that of maximal supergravity derived in section~\ref{sec:fluxes_A1D6}. Setting to zero spinorial fluxes as in (\ref{F&Xi=0}) does not amount to modding out maximal supergravity by a $\,\mathbb{Z}_2\,$ symmetry. While the former does not affect other fields in the theory (as scalars and vectors), the latter projects out some of them in order to truncate from maximal to half-maximal supergravity. On the string theory side, modding out by this $\,\mathbb{Z}_2\,$ symmetry is commonly referred to as applying an orientifold projection.

\subsection{String theory embedding \,vs\, moduli stabilisation}
\label{sec:typeII_models}

Thus far, we have discussed in detail the correspondence between maximal gauged supergravities and type II flux compactifications. However, one might also be interested in the interplay between gaugings, fluxes and moduli stabilisation: in short, fluxes were introduced in order to achieve moduli stabilisation. Sketchily, the picture in this respect seems to be the following

\begin{center}
\scalebox{0.85}[0.9]{\xymatrix{
*+[F-,]{\begin{array}{c}
\textbf{\textrm{semisimple gaugings}}\\[1mm]
\hline \\[-3mm]
\textrm{moduli stabilisation } $\ding{51}$ \\[2mm] 
\textrm{string embedding } $\ding{55}$ \\[2mm] 
\end{array} } 
& \leftrightarrow &
*+[F-,]{\begin{array}{c}
\textbf{\textrm{intermediate gaugings}}\\[1mm]
\hline \\[-3mm]
\textrm{moduli stabilisation } ? \\[2mm] 
\textrm{string embedding } ? \\[2mm] 
\end{array} } 
& \leftrightarrow &
*+[F-,]{\begin{array}{c}
\textbf{\textrm{nilpotent gaugings}}\\[1mm]
\hline \\[-3mm]
\textrm{moduli stabilisation } $\ding{55}$ \\[2mm] 
\textrm{string embedding } $\ding{51}$ \\[2mm] 
\end{array} }
}}
\end{center}
\vspace{2mm}

\noindent Semisimple gaugings are likely to produce critical points and moduli stabilisation \cite{Hull:1988jw,DallAgata:2011aa,Fischbacher_numerical}, but we will show that their embedding as type II flux compactifications involves highly non-geometric backgrounds. On the other hand, nilpotent gaugings can be obtained from type II compactifications including gauge fluxes \cite{deWit:2003hq}, but they seem not to be enough to get moduli stabilisation. Intermediate gaugings containing a semisimple part and an Abelian part have recently been found in ref.~\cite{DallAgata:2011aa}, although their embedding into string theory/M-theory has not been explored yet. 

Here we will present a novel intermediate gauging consisting of a semisimple and a nilpotent part which allows for moduli stabilisation and can be embedded into string theory as a type IIA flux compactification including gauge and metric fluxes.

\subsection*{Setting up the flux models}

Our starting point is the $\,\mathcal{N}=1\,$ supergravity theory inside box --5-- in figure~\ref{fig:truncations}. As explained in section~\ref{sec:truncations}, this theory can be obtained by truncating the $\,\mathcal{N}=4\,$ supergravity in box --4-- with respect to an $\,\textrm{SO}(3)\,$ subgroup which, in turn, can be obtained by a $\,\mathbb{Z}_{2}\,$ truncation of $\,\mathcal{N}=8\,$ supergravity in box --1--. As summarised in appendix~{\ref{App:fluxes}}, all the deformation parameters of this theory belong to the $\,f_{\alpha MNP}\,$ piece of the embedding tensor which comes out with forty independent components \cite{Dibitetto:2011gm}. These can be arranged into a tensor $\,{\Lambda_{\a ABC}=\Lambda_{\a (ABC)}}\,$, where $\,\a=+,-\,$ denotes an $\,\textrm{SL}(2)_{S}$ electric-magnetic index and $\,{A=1,...,4}\,$ denotes an $\,\textrm{SO}(2,2) \sim \textrm{SL}(2)_{T} \times \textrm{SL}(2)_{U}\,$ fundamental index of the global symmetry group. The theory comprises three complex scalars $\,S\,$, $\,T\,$ and $\,U\,$ parameterising the complex K\"ahler manifold
\beq
\cM_{\textrm{scalar}}= \left( \frac{\textrm{SL}(2)}{\textrm{SO}(2)} \right)_{S}  \times  \left( \frac{\textrm{SL}(2)}{\textrm{SO}(2)} \right)_{T}   \times  \left( \frac{\textrm{SL}(2)}{\textrm{SO}(2)} \right)_{U}  \ ,
\eeq
and no vector fields since they are projected out in the truncation. This supergravity theory can be obtained from type II orientifolds of $\,\mathbb{Z}_{2} \times \mathbb{Z}_{2}\,$ orbifold compactifications in the presence of generalised flux backgrounds, and the scalar potential can be derived from the $\,\mathcal{N}=1$ flux-induced superpotential in (\ref{W_fluxes}). 

Now we want to lift this $\,\cN=1\,$ theory firstly to $\,\cN=4\,$ by removing the $\,\textrm{SO}(3)$ truncation and secondly to $\,\cN=8\,$ by also removing the $\,\mathbb{Z}_2\,$ orientifold projection. This amounts to re-introduce the $\,28\,$ physical vectors in maximal supergravity and to complete the number of scalars from $\,6\,$ to $\,70\,$ without changing the set of embedding tensor components, in other words, without modifying the flux backgrounds. Nevertheless, in order for a flux background to be liftable to maximal supergravity, the set of quadratic constraints found in section~\ref{sec:fluxes_A1D6} must be imposed.

\begin{figure}[t!]
\vspace{-15mm}
\begin{center}
\scalebox{0.45}[0.45]{
\includegraphics[keepaspectratio=true]{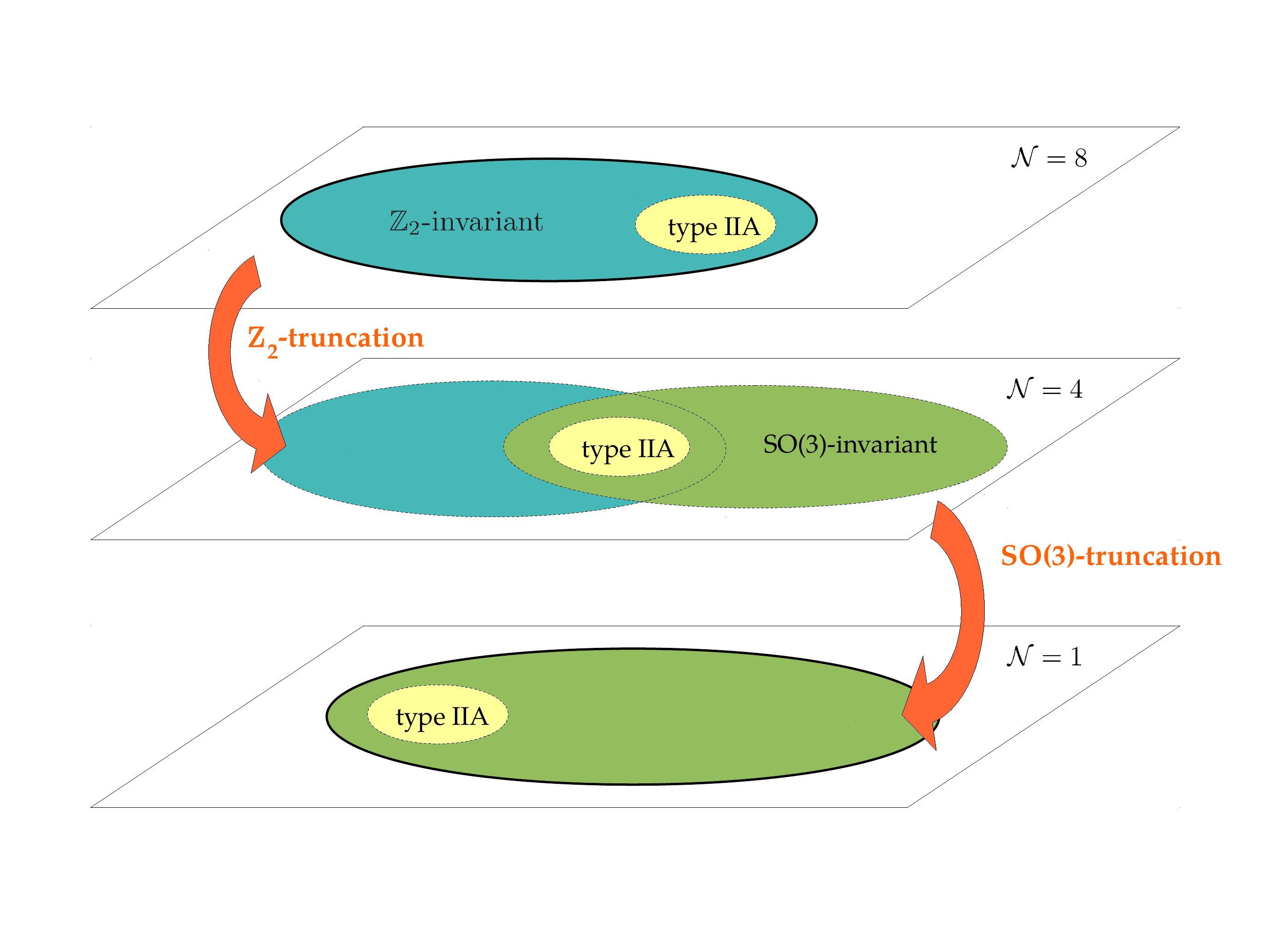}} 
\end{center}
\vspace{-15mm}
\caption{Diagram of the two-step lifting of $\,\cN=1\,$ flux backgrounds firstly to $\,\cN=4\,$ by removing the $\,\textrm{SO}(3)\,$ truncation and secondly to $\,\cN=8\,$ by removing the $\,\mathbb{Z}_2\,$ orientifold projection. As depicted in the figure, only a subset of $\,\cN=4\,$ theories can be truncated to $\,\cN=1\,$ theories via an $\,\textrm{SO}(3)\,$ truncation. On the other hand, only a subset of $\,{\cN=4}$ theories can be obtained from $\,\cN=8\,$ supergravity via a $\,\mathbb{Z}_{2}\,$ orientifold projection. The relevant fact is that the intersection between these two subsets of $\,\cN=4\,$ theories happens not to be empty and, furthermore, contains some theories for which a realisation in terms of type IIA string theory is known.}
\label{fig:field/flux_pic}
\end{figure}

In the rest of the section we will concentrate on two specific type II fluxes models which are relevant from a string theory point of view:
\begin{itemize}
\item Type IIB non-geometric flux backgrounds with an $\,\textrm{SO}(3,3) \,\times\, \textrm{SO}(3,3)\,$ splitting in $\,{\cN=4}\,$ supergravity and lifting to $\,\textrm{SO}(8)\,$, $\,\textrm{SO}(4,4)\,$, $\,\textrm{SO}(3,5)\,$ and $\,\textrm{CSO}(2,0,6)\,$ gaugings in $\,\cN=8\,$ supergravity.

\item Type IIA geometric flux backgrounds lifting to $\,\textrm{ISO}(3) \,\ltimes\, \textrm{U}(1)^{6}\,$ gaugings in $\,{\cN=4}\,$ supergravity and further lifting to $\,\textrm{SO}(4) \,\ltimes\, \textrm{Nil}_{22}\,$ gaugings in $\,\cN=8\,$ supergravity.

\end{itemize}

We will reduce our search of critical points to the origin of the moduli space and discuss the issues of stability and supersymmetry at those solutions. However, due to the restriction (\ref{F&Xi=0}), our classification of critical points will no longer be exhaustive since spinorial fluxes might produce new solutions we do not have access to by only looking at the origin of the moduli space \cite{Dibitetto:2011gm,DallAgata:2011aa}.

\subsection{$\textrm{CSO}(p,q,r)\,$ gaugings from type IIB with non-geometric fluxes}
\label{sec:typeIIB_models}

Let us concentrate on a set of type IIB flux backgrounds for which one has the direct product splitting $\,\textrm{SL}(2)_{S} \times \textrm{SO}(6,6) \supset \textrm{SO}(3,3)_{+} \times \textrm{SO}(3,3)_{-}\,$, where the labels $\,+\,$ and $\,-\,$ stand for $\,\textrm{SL}(2)_{S}\,$ electric and magnetic pieces, respectively. These backgrounds can be obtained from type IIB flux compactifications including the following set of generalised fluxes: R-R and NS-NS gauge fluxes $\,(F_{3},H_{3})\,$, non-geometric fluxes $\,(Q,P)\,$ and their primed counterparts which have been less studied in the literature. 

The above set of fluxes gives rise to maximal gauged supergravities based on $\,\textrm{CSO}(p,q,r)\,$ gauge groups with $\,p+q+r=8\,$. By applying the $\,\mathbb{Z}_{2}\,$ orientifold projection truncating from maximal to half-maximal supergravity, the $\,\textrm{CSO}(p,q,r)\,$ gauge groups get broken to the direct product of two smaller $\,\textrm{CSO}_{\pm}\,$ groups as
\beq
\label{CSO_chain}
\begin{array} {ccc}
\mathcal{N}=8 &   & \mathcal{N}=4 \\[2mm]
\textrm{CSO}(p,q,r) & \longrightarrow & \textrm{CSO}_{+}(p_{+},q_{+},r_{+}) \, \times \, \textrm{CSO}_{-}(p_{-},q_{-},r_{-}) \ , 
\end{array}
\eeq
with $\,p_{\pm} + q_{\pm} + r_{\pm} =4\,$. As explained in ref.~\cite{Roest:2009tt}, each of the $\,\textrm{CSO}_{\pm}\,$ factors in the r.h.s of (\ref{CSO_chain}) can be parameterised in terms of two real symmetric $\,4 \times 4\,$ matrices $M_{\pm}$ and $\tilde{M}_{\pm}\,$ which determine their embedding into an $\,\textrm{SO}(3,3)_{\pm}\,$ group, respectively. In terms of generalised flux components, these matrices read
\beq
M_{+} = 
\left( 
\begin{array}{cc}
-a'_{0} & 0 \\
0 & \tilde{c}_{1} \times \mathds{1}_{3} 
\end{array}
\right)_{(\,F'_{3}\,,\,Q\,)}
\hspace{6mm} \textrm{ and } \hspace{6mm}
\tilde{M}_{+} = 
\left( 
\begin{array}{cc}
-a_{0} & 0 \\
0 & \tilde{c}'_{1} \times \mathds{1}_{3} 
\end{array}
\right)_{(\,F_{3}\,,\,Q'\,)}
\eeq
together with
\beq
M_{-} = 
\left( 
\begin{array}{cc}
b'_{3} & 0 \\
0 & \tilde{d}_{2} \times \mathds{1}_{3} 
\end{array}
\right)_{(\,H'_{3}\,,\,P\,)}
\hspace{10mm} \textrm{ and } \hspace{10mm}
\tilde{M}_{-} = 
\left( 
\begin{array}{cc}
b_{3} & 0 \\
0 & \tilde{d}'_{2} \times \mathds{1}_{3} 
\end{array}
\right)_{(\,H_{3}\,,\,P'\,)}
\eeq
where the concrete identification between flux entries in $M_{\pm}$ and $\tilde{M}_{\pm}\,$ and embedding tensor components $\,f_{\a MNP}\,$ can be read off from tables~\ref{table:unprimed_fluxes} and \ref{table:primed_fluxes} in appendix~\ref{App:fluxes}. By substituting into the set of quadratic constraints derived in section~\ref{sec:fluxes_A1D6}, one finds three families of solutions:
\begin{itemize}
\item[$i)$] Flux matrices corresponding to a $\,(Q,F_{3})\,$-flux background 
\beq
M_{+} = 
\left( 
\begin{array}{cc}
0 & 0 \\
0 & \lambda_{1} \times \mathds{1}_{3} 
\end{array}
\right)
\hspace{2mm} \textrm{ , } \hspace{2mm}
\tilde{M}_{+} = 
\left( 
\begin{array}{cc}
\lambda_{2} & 0 \\
0 & 0  \times \mathds{1}_{3} 
\end{array}
\right)
\hspace{5mm} \textrm{ and } \hspace{5mm}
M_{-} = \tilde{M}_{-} =0
\eeq

\item[$ii)$] Flux matrices corresponding to a $\,(P,H_{3})\,$-flux background
\beq
M_{-} = 
\left( 
\begin{array}{cc}
0 & 0 \\
0 & \lambda_{1} \times \mathds{1}_{3} 
\end{array}
\right)
\hspace{2mm} \textrm{ , } \hspace{2mm}
\tilde{M}_{-} = 
\left( 
\begin{array}{cc}
\lambda_{2} & 0 \\
0 & 0 \times \mathds{1}_{3}  
\end{array}
\right)
\hspace{5mm} \textrm{ and } \hspace{5mm}
M_{+} = \tilde{M}_{+} =0
\eeq

\item[$iii)$] Flux matrices corresponding to a $\,(F'_{3}\,,\,Q\,,\,H_{3}\,,\,P')\,$-flux background
\beq
M_{+} = \textrm{unrestricted}
\hspace{5mm} \textrm{ , } \hspace{5mm}
\tilde{M}_{-} = \textrm{unrestricted}
\hspace{5mm} \textrm{ and } \hspace{5mm}
\tilde{M}_{+} = M_{-} = 0
\eeq

\end{itemize}
together with three additional ones obtained by swapping $\,M_{\pm} \leftrightarrow \tilde{M}_{\pm}\,$. This amounts to interchange primed and unprimed fluxes, i.e. to apply six T-dualities along the internal space directions, so the resulting theories are physically equivalent.

\begin{table}[h!]
\renewcommand{\arraystretch}{1.50}
\begin{center}
\scalebox{0.67}[0.7]{
\begin{tabular}{ | c | c | c | c | c  | c |}
\hline
\textrm{ID} & $\frac{1}{\lambda} \, M_{+}\,\,\,$ and $\,\,\,\frac{1}{\lambda} \, \tilde{M}_{-}$ & $\,\cN=8\,$ gauging & $\,\cN=4\,$ gauging &  $\frac{1}{\lambda^{2}} \, V_0$  & Mass spectrum  \\[1mm]
\hline \hline
$1$ & $\begin{array}{l} M_{+}=(1,1,1,1) \\[-3mm] \tilde{M}_{-}=(1,1,1,1) \end{array} $ & \multirow{3}{*}{$\textrm{SO}(8)$} & \multirow{3}{*}{$\textrm{SO}(4)^2$}  & $-\dfrac{3}{2}$ & $(70 \, \times)\, -\frac{2}{3}$  \\[1mm]
\cline{1-2}\cline{5-6} $2$ & $\begin{array}{l} M_{+}=(5 ,1,1,1) \\[-3mm] \tilde{M}_{-}=(1,1,1,1) \end{array} $ &  &   & $-\dfrac{5}{2}$ & $2$ , $(27 \, \times)\, -\frac{4}{5}$ , $(35 \, \times)\, -\frac{2}{5}$ , $(7 \, \times) \, 0$ \\[1mm]
\hline
\hline
$3$ & $\begin{array}{l} M_{+}=(1,1,1,1) \\[-3mm] \tilde{M}_{-}=(1,-3,-3,-3) \end{array} $ & $\textrm{SO}(5,3)$ & $\textrm{SO}(4) \times \textrm{SO}(1,3)$  & $\dfrac{3}{2}$ & $-2$ , $(5 \, \times)\, 4$ , $(30 \, \times)\, 2$ , $(14 \, \times) \, \frac{4}{3}$ , $(5 \, \times) \, -\frac{2}{3}$ , $(15 \, \times) \, 0$ \\[1mm]
\hline
\hline
$4$ & $\begin{array}{l} M_{+}=(1,-1,-1,-1) \\[-3mm] \tilde{M}_{-}=(-1,1,1,1) \end{array} $ & \multirow{3}{*}{$\textrm{SO}(4,4)$} & $\textrm{SO}(1,3) \times \textrm{SO}(3,1)$  & \multirow{3}{*}{$\dfrac{1}{2}$} & \multirow{3}{*}{$(2 \, \times)\, -2$ , $(36 \, \times)\, 2$ , $(16 \, \times) \, 1$ , $(16\, \times) \, 0$} \\[1mm]
\cline{1-2}\cline{4-4} $5$ & $\begin{array}{l} M_{+}=(1,1,1,1) \\[-3mm] \tilde{M}_{-}=(-1,-1,-1,-1) \end{array} $ &  & $\textrm{SO}(4,0)\times \textrm{SO}(0,4)$  &  &  \\[1mm]
\hline
\hline
$6$ & $\begin{array}{l} M_{+}=(1,0,0,0) \\[-3mm] \tilde{M}_{-}=(1,0,0,0) \end{array} $ & $\textrm{CSO}(2,0,6)$ & $\textrm{CSO}(1,0,3)^2$  & $0$ & $(20 \, \times)\, \frac{\lambda^{2}}{8}$ , $(2 \, \times)\, \frac{\lambda^{2}}{2}$ , $(48\, \times) \, 0$ \\[1mm]
\hline
\end{tabular}
}
\end{center}
\caption{Set of critical points of the scalar potential for generalised type IIB flux backgrounds compatible with an $\,\textrm{SL}(2)_{S} \times \textrm{SO}(6,6) \supset \textrm{SO}(3,3)_{+} \times \textrm{SO}(3,3)_{-}\,$ splitting. The first two correspond to Anti-de Sitter solutions, the next three to de Sitter solutions and the last one is a Minkowski solution. By looking for solutions of the Killing equations (\ref{Killing_equations}), we find that all the solutions break all the supersymmetries except the first one which preserves $\,\cN=8\,$ supersymmetry.} 
\label{table:typeIIB_vacua}
\end{table}

The next step is to build the fermionic mass terms $\,\mathcal{A}^{\mathcal{IJ}}\,$ and $\,{\mathcal{A}_{\mathcal{I}}}^{\mathcal{JKL}}\,$ in maximal supergravity as a function of the matrices $M_{\pm}$ and $\tilde{M}_{\pm}\,$ by using the relations (\ref{A's_fluxes_Bosonic}). Plugging them into the set of E.O.M's for the scalars (\ref{scalars_eom}), it turns out that gaugings belonging to the first and the second family of solutions to the quadratic constraints do not generate critical points at the origin of the moduli space. This is related to the fact that they are purely electric and magnetic gaugings in half-maximal supergravity, so moduli stabilisation is not possible \cite{deRoo:1985jh}. In contrast, six inequivalent patterns\footnote{Additional solutions apart from those shown in table~\ref{table:typeIIB_vacua} can be obtained by exchanging the flux matrices $\,M_{+} \leftrightarrow \tilde{M}_{-}\,$. However, they go back to those in the table via the composition of an S-duality and three T-duality transformations, hence being physically equivalent.} of flux matrices belonging to the third family of solutions, i.e. $\,M_{+} , \tilde{M}_{-} \neq 0\,$ and $\,\tilde{M}_{+} = M_{-} = 0\,$, are compatible with moduli stabilisation at the origin of the moduli space (\ref{M_origin_choice}). They correspond to dyonic gaugings in half-maximal supergravity even though their parent CSO gaugings in the maximal theory turn out to be purely electric. These are determined by the signature of the block-diagonal flux matrix
\beq
M_{\textrm{electric}} = 
\left( 
\begin{array}{c|c}
M_{+} & 0 \\[1mm] 
\hline
\\[-4mm]
0 & \tilde{M}_{-} 
\end{array}
\right) \ .
\eeq
We have computed the value of the energy\footnote{We are setting $\,g=\frac{1}{2}\,$ in analogy to ref.~\cite{Dibitetto:2011gm}.} $\, V_{0}\,$, the normalised mass spectrum (for those solutions with $\,V_{0} \neq 0\,$) and the amount of residual supersymmetry at the solutions together with the corresponding gauge group in maximal and half-maximal supergravity according to the chain (\ref{CSO_chain}). The results are summarised in table~\ref{table:typeIIB_vacua}, matching perfectly those in ref.~\cite{DallAgata:2011aa}.

\subsection{Type IIA with gauge and metric fluxes}
\label{sec:typeIIA_models}  

Now we investigate specific flux backgrounds having a higher-dimensional interpretation in terms of type IIA string compactifications including geometric fluxes: these are R-R $\,F_{0,2,4,6}\,$ and NS-NS $\,H_{3}\,$ gauge fluxes together with a metric flux $\,\omega\,$ associated to the spin connection of the internal space.

By using again the fluxes/embedding tensor correspondence of table~\ref{table:unprimed_fluxes} and the relations~(\ref{A's_fluxes_Bosonic}), we can build the fermionic mass terms $\,\mathcal{A}^{\mathcal{IJ}}\,$ and $\,{\mathcal{A}_{\mathcal{I}}}^{\mathcal{JKL}}\,$ in maximal supergravity associated to these type IIA backgrounds. Imposing the
set of $\,\cN=8\,$ quadratic constraints (\ref{QC_SU8}) and the E.O.M's for the scalar fields (\ref{scalars_eom}), we obtain sixteen AdS critical points at the origin of the moduli space (\ref{M_origin_choice}) which are collected in table~\ref{table:typeIIA_vacua}. As anticipated in ref.~\cite{Dibitetto:2011qs}, they can be seen as the uplifting to
maximal supergravity of half-maximal supergravity solutions compatible with the total absence of sources\footnote{The presence of sources as O$6$-planes and D$6$-branes in these type IIA scenarios modifies the set of $\,\cN=8\,$ quadratic constraints.}.

\begin{table}[h!]
\renewcommand{\arraystretch}{1.80}
\begin{center}
\scalebox{0.77}[0.77]{
\begin{tabular}{ | c || c | c |c | c | c | c |c | c |}
\hline
\textrm{\textsc{id}} & $a_{0}$ & $a_{1}$ & $a_{2}$ & $a_{3}$ & $b_{0}$ & $b_{1}$ & $c_{0}$ & $c_{1}=\tilde{c}_{1}$ \\[1mm]
\hline \hline
$1_{(s_1,s_2)}$ & $s_{2} \,  \dfrac{3 \,\sqrt{10}}{2}\, \lambda $ & $s_{1} \,\dfrac{\sqrt{6}}{2} \, \lambda$ & $ - s_{2} \,\dfrac{\sqrt{10}}{6} \, \lambda$ & $s_{1} \, \dfrac{5\,\sqrt{6}}{6} \, \lambda$ & $-s_{1} \,s_{2} \, \dfrac{\sqrt{6}}{3} \, \lambda$ & $\dfrac{\sqrt{10}}{3}\,\lambda$ & $s_{1} \,s_{2} \, \dfrac{\sqrt{6}}{3}\,\lambda$ & $\sqrt{10} \, \lambda$   \\[1mm]
\hline \hline
$2_{(s_1,s_2)}$ & $s_{2} \,\dfrac{16 \, \sqrt{10}}{9} \,\lambda$ & $0$ & $0$ & $s_{1} \, \dfrac{16 \, \sqrt{2}}{9} \, \lambda$ & $0$ & $\dfrac{16 \, \sqrt{10}}{45} \, \lambda$ & $0$ & $\dfrac{16 \, \sqrt{10}}{15} \, \lambda$   \\[1mm]
\hline
$3_{(s_1,s_2)}$ & $s_{2} \,\dfrac{4\,\sqrt{10}}{5}\,\lambda$ & $-s_{1} \, \dfrac{4\,\sqrt{30}}{15}\,\lambda$ & $s_{2} \, \dfrac{4\,\sqrt{10}}{15}\,\lambda$ & $s_{1} \,\dfrac{4\,\sqrt{30}}{15}\,\lambda$ & $s_{1} \, s_{2} \,\dfrac{4\,\sqrt{30}}{15}\,\lambda$ & $\dfrac{4\,\sqrt{10}}{15}\,\lambda$ & $-s_{1} \, s_{2} \,\dfrac{4\,\sqrt{30}}{15}\,\lambda$ & $\dfrac{4\,\sqrt{10}}{5}\,\lambda$   \\[1mm]
\hline
$4_{(s_1,s_2)}$ & $s_{2} \,\dfrac{16 \, \sqrt{10}}{9} \,\lambda$ & $0$ & $0$ & $s_{1} \,\dfrac{16 \, \sqrt{2}}{9} \,\lambda$ & $0$ & $\dfrac{16 \, \sqrt{2}}{9} \,\lambda$ & $0$ & $\dfrac{16 \, \sqrt{2}}{9} \,\lambda$  \\
\hline
\end{tabular}
}
\end{center}
\caption{List of the critical points at the origin of the moduli space generated by R-R $\,(a_{0,1,2,3})\,$, NS-NS $\,(b_{0},c_{0})\,$ and metric $\,(b_{1},c_{1},\tilde{c}_{1})\,$ flux backgrounds in type IIA scenarios. They can be organised into four groups each of which consists of four equivalent solutions labelled by a pair $\,(s_{1},s_{2})\equiv\left\lbrace (+,+) , (+,-) , (-,+) , (-,-) \right\rbrace$. The quantity $\,\lambda\,$ is a free parameter setting the AdS energy scale $\,V_{0} \propto -\lambda^{2}\,$ at the solutions.} \label{table:typeIIA_vacua}
\end{table}

With the fermionic mass terms $\,\mathcal{A}^{\mathcal{IJ}}\,$ and $\,{\mathcal{A}_{\mathcal{I}}}^{\mathcal{JKL}}\,$ at our disposal, we can now compute the different values of the cosmological constant (\ref{V_SU8}) at the above set type IIA solutions. These are given by
\beq
\label{V0_IIA}
V_{0}\left[ 1_{(s_1,s_2)} \right] = -\lambda^{2}
\hspace{3mm} , \hspace{3mm}
V_{0}\left[ 2_{(s_1,s_2)} \right] = V_{0}\left[ 4_{(s_1,s_2)} \right]= -\dfrac{32 \, \lambda^{2}}{27}
\hspace{3mm} , \hspace{3mm}
V_{0}\left[ 3_{(s_1,s_2)} \right] = -\dfrac{8 \, \lambda^{2}}{15}  \ .
\eeq
In addition, we can also obtain the complete mass spectrum for the $70$ physical scalars by using the mass formula (\ref{Mass-matrix}) and check stability as well as the amount of supersymmetry preserved. The mass spectrum at the critical points in table~\ref{table:typeIIA_vacua} turns out to be the following:
\begin{itemize}
\item At the solution $\,1_{(s_{1},s_{2})}\,$, the normalised scalar field masses and their multiplicities are given by
\beq
\begin{array}{lcrr}
\dfrac{1}{9} \left(47 \pm \sqrt{159}\right)\,\,(\times 1)
\hspace{8mm} , \hspace{8mm} \dfrac{1}{3} \left(4 \pm
\sqrt{6}\right)\,\,(\times 1) & , &
\dfrac{29}{9}\,\,(\times 3) & ,\\[4mm]
\dfrac{1}{18} \left( \, 89 + 5 \, \sqrt{145} \pm \sqrt{606 + 30 \, \sqrt{145}} \, \right)\,\,(\times 5)
& \hspace{5mm} , \hspace{5mm} &
0\,\,(\times 10) & , \\[4mm]
\dfrac{1}{18} \left(\, 89 - 5 \, \sqrt{145} \pm \sqrt{606 - 30 \, \sqrt{145}} \, \right)\,\,(\times 5)
& \hspace{5mm} ,\hspace{5mm} &
-\dfrac{2}{3}\,\,(\times 1) & ,
\end{array}
\nonumber
\eeq
for the $\,38\,$ scalars surviving the truncation from maximal to half-maximal supergravity, together with
\beq
\dfrac{1}{3} \left(4 \pm \sqrt{6}\right)\,\,(\times 3)
\hspace{5mm} , \hspace{5mm}
6\,\,(\times 3)
\hspace{5mm} , \hspace{5mm}
\dfrac{13}{3} \,\,(\times 5)
\hspace{5mm} , \hspace{5mm}
-\dfrac{2}{3}\,\,(\times 1)
\hspace{5mm} , \hspace{5mm}
0 \,\,(\times 17) \ , \nonumber
\eeq
for the additional $\,32\,$ scalars in the maximal theory. There are two tachyons in the spectrum both with the same normalised mass $\,m^2= - \frac{2}{3}\,$, so this AdS solution is completely stable since it satisfies the B.F. bound in (\ref{BF_bound}).

\item At the solution $\,2_{(s_{1},s_{2})}\,$, the values of the normalised scalar masses and their multiplicities read
\beq
\begin{array}{c}
\dfrac{1}{15} \left(77 \pm 5 \, \sqrt{145}\right)\,\,(\times
5) \hspace{4mm} , \hspace{4mm} \dfrac{2}{15} \left(31 \pm
\sqrt{145}\right)\,\,(\times 5) \hspace{4mm} , \hspace{4mm}
\dfrac{64}{15}\,\,(\times 1) \hspace{4mm} , \hspace{4mm}
\dfrac{20}{3}\,\,(\times 1) \ , \\[4mm]
\dfrac{46}{15}\,\,(\times 3) \hspace{5mm} , \hspace{5mm}
2\,\,(\times 1) \hspace{5mm} , \hspace{5mm}
0\,\,(\times 10) \hspace{5mm} , \hspace{5mm}
-\dfrac{2}{5}\,\,(\times 1) \hspace{5mm} , \hspace{5mm}
-\dfrac{4}{5}\,\,(\times 1) \ ,
\end{array}
\nonumber
\eeq
for the scalars surviving the truncation to half-maximal supergravity, and also
\beq
6\,\,(\times 3)
\hspace{4mm} , \hspace{4mm}
4\,\,(\times 5)
\hspace{4mm} , \hspace{4mm}
2\,\,(\times 3)
\hspace{4mm} , \hspace{4mm}
-\frac{4}{5}\,\,(\times 1)
\hspace{4mm} , \hspace{4mm}
0\,\,(\times 20) \ ,
\nonumber
\eeq
for the scalars been projected out by the $\mathbb{Z}_{2}$ orientifold projection. There are three tachyons in the spectrum, two of them with $\,m^2=-4/5\,$. This value is below the B.F. bound in (\ref{BF_bound}), rendering this AdS solution unstable.

\item At the solution $\,3_{(s_{1},s_{2})}\,$, the normalised scalar field masses and their multiplicities take the values of
\beq
\frac{1}{3} \left(19\pm\sqrt{145}\right)\,\,(\times
10) \hspace{4mm} , \hspace{4mm} \frac{20}{3}\,\,(\times 2)
\hspace{4mm} , \hspace{4mm} \frac{14}{3}\,\,(\times 3)
\hspace{4mm} , \hspace{4mm} 2\,\,(\times 2)
\hspace{4mm} , \hspace{4mm} 0\,\,(\times 11) \ , \nonumber
\eeq
for those scalars still present in half-maximal supergravity, and
\beq
2\,\,(\times 6)
\hspace{4mm} , \hspace{4mm}
6\,\,(\times 5)
\hspace{4mm} , \hspace{4mm}
8\,\,(\times 3)
\hspace{4mm} , \hspace{4mm}
0\,\,(\times 18) \ ,
\nonumber
\eeq
for those ones completing to maximal supergravity. It is worth noticing that all the masses are non-negative at this critical point, hence corresponding to an AdS stable extremum of the theory.

\item At the solution $\,4_{(s_{1},s_{2})}\,$, the set of normalised scalar masses and their multiplicities are
\beq
\dfrac{20}{3}\,\,(\times 1) \hspace{4mm} , \hspace{4mm}
6\,\,(\times 6) \hspace{4mm} , \hspace{4mm}
\dfrac{8}{3}\,\,(\times 5) \hspace{4mm} , \hspace{4mm}
2\,\,(\times 4) \hspace{4mm} , \hspace{4mm}
\dfrac{4}{3}\,\,(\times 6) \hspace{4mm} , \hspace{4mm}
0\,\,(\times 16) \,\ .
\nonumber
\eeq
again for the $38$ scalars present in half-maximal supergravity, as well as
\beq
\dfrac{8}{3}\,\,(\times 5) \hspace{4mm} , \hspace{4mm}
6\,\,(\times 3) \hspace{4mm} , \hspace{4mm}
-\dfrac{4}{3}\,\,(\times 1) \hspace{4mm} , \hspace{4mm}
2\,\,(\times 3) \hspace{4mm} , \hspace{4mm}
\dfrac{4}{3}\,\,(\times 3) \hspace{4mm} , \hspace{4mm}
0\,\,(\times 17) \,\ .
\nonumber
\eeq
for the $32$ extra ones in the maximal theory. Amongst the latter, there is a tachyon with a normalised mass $\,m^2=-4/3\,$ lying below the B.F. bound (\ref{BF_bound}). Therefore, this solution, while stable with respect to the scalars in half-maximal supergravity, becomes unstable when lifted to the maximal theory.
\end{itemize}

In order to determine the amount of residual supersymmetry preserved by the above set of critical points, we have to look for solutions to the Killing equations (\ref{Killing_equations}). The final outcome is that supersymmetry becomes completely broken in all the solutions except in $\,1_{(s_{1},s_{2})}\,$ which preserves $\,\cN=1\,$ supersymmetry. Let us go deeper into the way in which minimal supersymmetry is preserved by the $\,1_{(s_{1},s_{2})}\,$ solution. Recalling the decomposition of the $\,\textrm{SU}(8)\,$ R-symmetry group of maximal supergravity first under the $\,\mathbb{Z}_{2}\,$ orientifold projection truncating to half-maximal supergravity and then under the $\,\textrm{SO}(3)\,$ truncation yielding minimal supergravity
\beq
\begin{array}{ccccc}
 \textrm{SU}(8) & \overset{\mathbb{Z}_{2}}{\supset} & \textrm{SU}(4)_{\textrm{even}} \times \textrm{SU}(4)_{\textrm{odd}}  & \supset  & \textrm{SO}(3)_{\textrm{even}}  \times \textrm{SO}(3)_{\textrm{odd}} \\[2mm]
\textbf{8} & \rightarrow & (\textbf{4},\textbf{1})_{\textrm{even}} + (\textbf{1},\textbf{4})_{\textrm{odd}}  & \rightarrow & (\textbf{1},\textbf{1})_{\textrm{even}} + (\textbf{3},\textbf{1})_{\textrm{even}} + (\textbf{1},\textbf{1})_{\textrm{odd}} + (\textbf{1},\textbf{3})_{\textrm{odd}}
\end{array}
\nonumber
\eeq
one observes that there are two invariant (covariantly constant) spinors associated to the $\,(\textbf{1},\textbf{1})_{\textrm{even}}\,$ and $\,(\textbf{1},\textbf{1})_{\textrm{odd}}\,$ irrep's respectively. This implies that there are two possible $\,\cN=1\,$ residual supersymmetry that can be preserved by the $\,1_{(s_{1},s_{2})}\,$ configurations. However, since the $\,\textrm{SU}(4)\,$ R-symmetry group of half-maximal supergravity is identified with $\,\textrm{SU}(4)_{\textrm{even}}\,$ and \textit{not} with $\,\textrm{SU}(4)_{\textrm{odd}}\,$, only those configurations preserving the $\,\cN=1\,$ supersymmetry associated to the $\,(\textbf{1},\textbf{1})_{\textrm{even}}\,$ irrep ($\,1_{(+,+)}\,$ and $\,1_{(-,+)}\,$) can still be truncated to half-maximal supergravity as ${\,\cN=1\,}$ supersymmetric solutions. In contrast, solutions preserving the ${\,\cN=1\,}$ supersymmetry associated to the $\,(\textbf{1},\textbf{1})_{\textrm{odd}}\,$ irrep ($\,1_{(-,-)}\,$ and $\,1_{(+,-)}\,$) appear as non-supersymmetric solutions when truncated to half-maximal supergravity. Nevertheless, they are fake supersymmetric in the sense that they are supersymmetric with respect to the ``wrong'' R-symmetry group, hence inheriting all the stability properties associated to supersymmetric solutions. 

At this point, the nature of the two $\mathbb{Z}_{2}$ factors labelled by $(s_{1},s_{2})$ becomes clear. The first one, as already pointed out in ref.~\cite{Dibitetto:2011gm}, is a symmetry of the $\,\cN=4\,$ theory and hence it does not really label different solutions, whereas, at this level, the second $\mathbb{Z}_{2}$ seems to appear as an accidental symmetry forcing
the value of the energy and the mass spectra of inequivalent critical points to be identical. When lifting these solutions to maximal supergravity, the second $\mathbb{Z}_{2}$ becomes a symmetry as well: it corresponds precisely to the $\,\textrm{SU}(8)\,$ element interchanging $\,\textrm{SU}(4)_{\textrm{time-like}}\,$ with $\,\textrm{SU}(4)_{\textrm{space-like}}\,$, thus relating equivalent solutions. As a consequence, the number of inequivalent critical points reduces to four and they can be seen as different solutions of the same maximal gauged supergravity.

Let us now identify the gauge group underlying these type IIA geometric backgrounds in a maximal gauged supergravity context. Since we set $\,F_{M \dot{\mu}}\,=\,\Xi_{\a \b \mu}=0\,$, the $X_{\,\mathbb{MNP}}$ components in (\ref{Xfermionic}) do vanish. Then, the brackets (\ref{gauge_algebra}) of the gauge group $\,G_{0}\,$ take the simpler form
\beq
\label{algebra_bos_general}
\begin{array}{ccclcc}
\left[ X_{\a M} , X_{\b N} \right] & = &  -  & {X_{\a M \b N}}^{\g P} \, X_{\g P} & , \\[2mm]
\left[ X_{\a M} , X_{\m} \right] & = &  - & {X_{\a M \m}}^{\r} \, X_{\r}  & , \\[2mm]
\left[ X_{\m} , X_{\n} \right] & = &  - & {X_{\m \n}}^{\g P} \, X_{\g P}  & .
\end{array}
\eeq
The $12$-dimensional subgroup $\,G_{\textrm{bos}} \subset G_{0}\,$ spanned by the linearly independent\footnote{Only $12$ out the $24$ bosonic generators $X_{\a M}$ are linearly independent hence entering the gauging. Adopting the same choice that in ref.~\cite{Dibitetto:2011gm}, we decide to write the magnetic generators as a function of the electric ones, i.e.,  $\,X_{-M}(X_{+M})\,$.} $\,X_{\a M}\,$ bosonic generators in (\ref{algebra_IIA}) turns out to be
\beq
\label{subalgebra_bos}
G_{\textrm{bos}}=\textrm{ISO}(3) \ltimes \textrm{U}(1)^6 \ .
\eeq
This is the gauge group of the half-maximal theory in which fermionic generators $\,X_{\m}\,$ are projected out by the $\,\mathbb{Z}_{2}\,$ orientifold projection \cite{Dibitetto:2011gm}. The $16$ linearly independent fermionic generators extend $\,G_{\textrm{bos}}\,$ in (\ref{subalgebra_bos}) to the complete $28$-dimensional gauge group $\,G_{0}\,$ of maximal supergravity which is identified with
\beq
\label{algebra_IIA}
G_{0}=\textrm{SO}(4) \ltimes \textrm{Nil}_{(22)} \ ,
\eeq
for these type IIA geometric flux backgrounds. To be more concrete about the structure of the gauge group, let us split the $28$ linearly independent generators into $6$ generators $\,\{   T_{i}^{(0)} \, , \, T_{a}^{(0)} \} \,$ spanning the semisimple $\,\textrm{SO}(4) \sim \textrm{SU}(2)_{i}  \times \textrm{SU}(2)_{a}\,$ part in (\ref{algebra_IIA}) and $22$ generators $\,\{  T_{i}^{(1)} \, , \, T_{a}^{(1)} \, , \, T_{i}^{(2)} \, , \, T_{a}^{(2)} \, , \, T_{ia} \, , \,T \} \,$ associated to the nilpotent ideal $\,\textrm{Nil}_{(22)}\,$. The index structure of the generators, where $\,i,a=1,2,3\,$, reflects their transformation properties with respect to the semisimple part of the gauge group. In the appropriate basis, we can write the non-vanishing gauge brackets as
\beq
[ \, T_{i}^{(0)} , T_{j}^{(p)} ] = \epsilon_{ijk} \, T_{k}^{(p)} \hspace{5mm} (p=0,1,2)
\hspace{15mm} \textrm{ and } \hspace{15mm}
[ \, T_{i}^{(0)} , T_{ja} ] = \epsilon_{ijk} \, T_{ka} \ ,
\eeq
which involve the semisimple generators $\,T_{i}^{(0)}$, together with
\beq
[ \, T_{i}^{(1)} , T_{j}^{(2)} ] = \delta_{ij} \, T
\hspace{5mm} \textrm{ , } \hspace{5mm}
[ \, T_{i}^{(1)} , T^{(1)}_{a} ] = - T_{ia} \ ,
\hspace{5mm} \textrm{ and } \hspace{5mm}
[ \, T_{i}^{(1)} , T_{ja} ] = \delta_{ij} \, T^{(2)}_{a} \ ,
\eeq
involving generators in the nilpotent part. When non-equivalent, the above set of brackets must be supplemented with additional ones obtained by exchanging $\,T_{i}^{(p)} \leftrightarrow T_{a}^{(p)}\,$, $\,\epsilon_{ijk} \leftrightarrow \epsilon_{abc}\,$ and $\,\delta_{ij} \leftrightarrow \delta_{ab}\,$. By inspection of the above brackets, one finds that the $\,\textrm{Nil}_{(22)}\,$ piece is a nilpotent $22$-dimensional ideal of order three (four steps) with lower central series
\beq
\left\lbrace   T_{i}^{(1)} \, , \, T_{a}^{(1)} \, , \, T_{i}^{(2)} \, , \, T_{a}^{(2)} \, , \, T_{ia} \, , \,T \right\rbrace 
\supset
\left\lbrace T_{i}^{(2)} \, , \, T_{a}^{(2)} \, , \, T_{ia} \, , \,T \right\rbrace  
\supset 
\left\lbrace T_{i}^{(2)} \, , \, T_{a}^{(2)} \, , \,T  \right\rbrace 
\supset 
\left\lbrace T \right\rbrace
\supset
\left\lbrace 0 \right\rbrace \ . \nonumber
\eeq

As an aside remark, we have taken the real realisation of gamma matrices (see appendix~\ref{App:spinors}) when building the structure constants of the gauge algebra in (\ref{algebra_bos_general}). Otherwise, if taking the $\,{\textrm{SU}(4) \times \textrm{SU}(4)}\,$ covariant realisation, the structure constants turn out to be complex an so the gauge generators in the adjoint representation. Thus, one still would have to impose a reality condition upon vectors when it comes to identify the gaugings.

Because of all the aforementioned, we conclude that the gauge group in (\ref{algebra_IIA}) gives rise to $\,\cN=1\,$ supersymmetric and non-supersymmetric AdS stable solutions of maximal supergravity at the origin of the moduli space which can be embedded in string theory as type IIA flux compactifications in the presence of geometric fluxes.

\section{Conclusions}
\label{sec:conclusions}

We have presented the SL($2)\,\times\,$SO($6,6$) formulation of gauged maximal supergravity in four dimensions, in which we have written down the embedding tensor components by consistently solving the linear constraint and deriving the quadratic constraints. These turn out to be divided into bosonic ($\mathbb{Z}_{2}$ even) and fermionic ($\mathbb{Z}_{2}$ odd) irrep's of SO($6,6$). The bosonic embedding tensor components contain the usual generalised fluxes obtained in orientifolds of type II flux compactifications, whereas the fermionic ones happen to involve additional fluxes which are removed by the orientifold projection. Furthermore, we have presented the link with the standard formulation of $\,\cN=8\,$ supergravity given in terms of SU($8$) irrep's, by means of which the equations of motion in the origin of moduli space and the mass matrix for all the 70 scalars are given. This allows one to study exceptional flux backgrounds, for which all tadpole conditions are satisfied without the inclusion of branes.

Building on these results, we have constructed the uplift to maximal supergravity of a set of four AdS vacua of type IIA geometric flux compactifications. These solutions can be seen as
different critical points of the same theory with gauge group $G_{0}\,=\,\textrm{SO}(4) \ltimes \textrm{Nil}_{(22)}$. One of those is supersymmetric, whereas the other three are non-supersymmetric. However, amongst the non-supersymmetric solutions, we find that one is not only perturbatively stable, but it even turns out to be a minimum.

Additionally, we have studied the uplift of a particular class of type IIB non-geometric flux backgrounds by requiring any brane sources to be absent. We have compared the results with the list of critical points of maximal supergravity which are already present in the literature and we have analysed their mass spectra. Our results are in perfect agreement with ref.~\cite{DallAgata:2011aa}.

As a future line of research, it would be of great interest to carry out the analysis of critical points in a set-up with non-vanishing spinorial flux components. As we have seen, such fluxes are necessary for U-duality covariance, and hence they might shed a new light on the role of U-duality as an organising principle in flux compactifications. Furthermore, configurations with spinorial flux components would correspond to theories which cannot be truncated to $\,\cN=4\,$ and therefore to flux compactifications with no orientifold planes and/or branes. This implies that all the \emph{no-go} results along the lines of e.g.~\cite{no-go_theorems}, related to the minimal set of ingredients for finding dS solutions, would not apply there. In particular, it cannot be excluded that there might be dS extrema by only including metric and gauge fluxes both in the bosonic and fermionic sectors. For these reasons, exceptional flux compactifications with $\textrm{SO}(6,6)$ fermionic fluxes certainly deserve further investigations.

%
%

\section*{Acknowledgments}

We are grateful to A.~Borghese, D.~Geissb\"uhler, D.~Marqu\'es and H.~Samtleben for very stimulating discussions. The work of the authors is supported by a VIDI grant from the Netherlands Organisation for Scientific Research (NWO).

%
%

\appendix

\section{Summary of indices}
\label{App:indices}

All through the text we extensively make use of indices of different
groups. Here we give a list of the notations retained in this work

\be\label{Indices} \begin{array}{cccl}
\mathpzc{A}\,,\,\,\mathpzc{B}\,,\,\,\cdots & & & \textrm{adjoint of
E}_{7(7)}\\
\mathbb{M}\,,\,\,\mathbb{N}\,,\,\,\cdots & & & \textrm{fundamental
of
E}_{7(7)}\textrm{ (global)}\\
\underline{\mathbb{M}}\,,\,\,\underline{\mathbb{N}}\,,\,\,\cdots & &
& \textrm{fundamental of
E}_{7(7)}\textrm{ (local)}\\
\mathcal{I}\,,\,\,\mathcal{J}\,,\,\,\cdots & & & \textrm{fundamental
of SU}(8)\\
\a\,,\,\,\b\,,\,\,\cdots & & & \textrm{fundamental of SL}(2)\\
M\,,\,\,N\,,\,\,\cdots & & & \textrm{fundamental of SO}(6,6)\\
\m\,,\,\,\n\,,\,\,\cdots & & & \textrm{M-W spinor of
SO}(6,6)\textrm{
(L)}\\
\dot{\m}\,,\,\,\dot{\n}\,,\,\,\cdots & & & \textrm{M-W spinor of
SO}(6,6)\textrm{
(R)}\\
A\,,\,\,B\,,\,\,\cdots & & & \textrm{fundamental of SO}(2,2)\\
m\,,\,\,n\,,\,\,\cdots & & & \textrm{fundamental of SO}(6)_{\textrm{time-like}}\\
a\,,\,\,b\,,\,\,\cdots & & & \textrm{fundamental of
SO}(6)_{\textrm{space-like}}\\
i\,,\,\,j\,,\,\,\cdots & & & \textrm{fundamental of SU}(4)_{\textrm{time-like}}\\
\hat{i}\,,\,\,\hat{j}\,,\,\,\cdots & & & \textrm{fundamental of
SU}(4)_{\textrm{space-like}}\,\,\,.
\end{array} \ee

\section{Majorana-Weyl spinors of $\textrm{SO}(6,6)$}
\label{App:spinors}

Our starting point is a Majorana spinor in $6+6$ dimensions carrying $2^{6}=64$ real degrees of freedom. For Majorana spinors there exists a real
representation of the $\Gamma$-matrices $\{\Gamma_M\}_{M=1,\cdots,12}$
such that they satisfy
\bea\label{Clifford}
\left\{\Gamma_M,\Gamma_N\right\}=2\,\eta_{MN}\,\mathds{1}_{64}\,,
\eea
where $\,\eta_{MN}=\eta^{MN}\,$ is the $\,\textrm{SO}(6,6)\,$ invariant metric. We adopt a set of conventions in which Majorana spinors are naturally objects
of the form $\chi^{a}$ and hence $\Gamma$-matrices carry indices $\left[\Gamma_M\right]^{a}_{\phantom{a}b}\,$. In addition to the $\Gamma$-matrices, we
introduce two antisymmetric matrices, $\,\cC_{a b}\,$ and $\,\cC^{a b}\,$, which turn out to represent the components of the transposed charge conjugation
matrix $\,\cC\,$ and its inverse respectively. We will use these objects in order to raise and lower spinorial indices according to
the so-called SouthWest-NorthEast (SW-NE) conventions \cite{VanProeyen:1999ni}. This translates into the following rules
\bea 
\chi^{a}=\chi_b \, \cC^{b a} \hspace{8mm},\hspace{8mm}
\chi_a=\cC_{a b} \, \chi^b \ , 
\eea
and the consistency of the two rules implies 
\bea 
\cC^{a b}\,\cC_{c
b}=\delta^{a}_{\phantom{a}c}\hspace{8mm}\textrm{and}\hspace{8mm}\cC_{b
a}\,\cC^{b c}=\delta_{a}^{\phantom{a}c} \ . 
\eea
The charge conjugation matrix mentioned above relates $\Gamma$-matrices to their transpose in the following way
\bea
\label{GammaT} 
\left(\Gamma_M\right)^T =-\, \cC\,\Gamma_M\,\cC^{-1} \ ,
\eea
whereas one can also define a conjugation matrix $\,B=-A^{-T} \, \mathcal{C}\,$, such that
\beq
\label{B64}
B^{*} B = \mathds{1}_{64}   \hspace{10mm} \textrm{ and } \hspace{10mm} \Gamma_{M}^{*}= -B \, \Gamma_{M} \, B^{-1} \hspace{15mm} \textrm{with} \hspace{10mm} A = \Gamma_{1} ... \Gamma_{6} \ .
\eeq

Majorana spinors live in the $\,\textbf{64}\,$ of $\,\textrm{SO}(6,6)\,$ which is not an irrep and can be decomposed in terms of left- and right-handed Majorana-Weyl (M-W) spinors. These are  related to the $\,\textbf{32}\,$ and $\,\textbf{32'}\,$ irrep's, respectively. In a basis in which $\,\Gamma_{13}=\Gamma_1\cdots\Gamma_{12}\,$ takes the form $\Gamma_{13}=\textrm{diag}(+\mathds{1}_{32},-\mathds{1}_{32})\,$, one can introduce the so-called 2-component formalism such that
\bea \chi^a=\left(\begin{array}{c} \chi^{\m}\\
\chi_{\dm}\end{array}\right)\,, \eea
where the indices $\,\m\,$ and $\,\dm\,$ respectively denote left- and right-handed M-W spinors. Accordingly to this decomposition, the $\Gamma$-matrices split into $32\times 32$ blocks as follows
\bea
\label{Gamma_splitting}
\left[\Gamma_M\right]^{a}_{\phantom{a}b}=\left(\begin{array}{cc} 0 &
\left[\g_M\right]^{\m\dn}\\
\left[\bar{\g}_M\right]_{\dm\n} & 0
\end{array}\right) \ , 
\eea
and the charge conjugation and conjugation matrices become
\bea
\label{B&C_splitting}
\cC_{ab}=\left(
\begin{array}{cc} 
\cC_{\m\n} & 0\\
0 & \cC^{\dm\dn}
\end{array}
\right) 
\hspace{10mm} \textrm{and}  \hspace{10mm}
B_{ab}=\left(
\begin{array}{cc} 
B_{\m\n} & 0\\
0 & B^{\dm\dn}
\end{array}
\right) \ . 
\eea
In terms of these $\,32\times 32\,$ gamma matrices, the relations \eqref{Clifford} and \eqref{GammaT} can be respectively written as
\bea
\left[{\g}_{(M}\right]^{\m\dot{\rho}}\,\left[\bar{\g}_{N)}\right]_{\dot{\rho}\n}=\eta_{MN}\,\delta^{\m}_{\n}
\hspace{10mm}\textrm{and}\hspace{10mm}
\left[\bar{\g}_M\right]_{\dm\n} = \left[{\g}_M\right]_{\n\dm}= \cC_{\n \sigma}\left[{\g}_M\right]^{\sigma\dot{\rho}}\cC_{\dot{\rho} \dm}\ .
\eea

Antisymmetrised products of two gamma matrices can be defined both for left- and right-handed M-W representations as
\bea
[\g_{MN}]^{\m}_{\phantom{a}\n}\equiv[\g_{[M}]^{\m\dot{\rho}}\,[\bar{\g}_{N]}]_{\dot{\rho}\n}
\hspace{15mm} \textrm{and} \hspace{15mm}
[\g_{MN}]_{\dm}^{\phantom{a}\dn}\equiv[\bar{\g}_{[M}]_{\dm\rho}\,[\g_{N]}]^{\rho\dn} \ ,
\eea
and further extended to antisymmetrised products of an even number of gamma matrices. However, only those up to degree six are linearly independent since higher-degree ones (from $7$ to $12$) are related to them by Hodge duality\footnote{The limit case of the antisymmetrised product of six gamma matrices turns out to be anti-selfdual (ASD) when involving undotted indices and self-dual (SD) when involving dotted indices.}. After defining all the products of gamma matrices, one can make use of $\,\cC_{\m\n}$\,, $\,\cC^{\dm\dn}\,$ and their inverse transpose in order to rise and lower indices. As a result, antisymmetrised products of \emph{two} and \emph{six} gamma matrices are symmetric, whereas the ones with \emph{four} are antisymmetric.

\subsection*{$\textrm{SU}(4) \times \textrm{SU}(4)$ covariant formulation of M-W spinors}

Decomposing the vector and the left- and right-handed M-W spinor irrep's of $\,\textrm{SO}(6,6)\,$ under its maximal compact subgroup
\beq
\label{A3A3_in_D6}
\textrm{SU}(4)_{\textrm{time-like}} \times \textrm{SU}(4)_{\textrm{space-like}} \sim \textrm{SO}(6)_{\textrm{time-like}} \times \textrm{SO}(6)_{\textrm{space-like}} \subset \textrm{SO}(6,6) \ ,
\eeq
yields the following branching relations
\beq
\begin{array}{ccccc}
\textrm{index} \,\,\,\,\,\,& \textrm{SO}(6,6) & \supset & \textrm{SU}(4) \times \textrm{SU}(4) & \\[2mm]
M \,\,\,\,\,\, & \textbf{12} & \rightarrow &  (\textbf{6},\textbf{1}) + (\textbf{1},\textbf{6})  & \\[2mm]
\m \,\,\,\,\,\, & \textbf{32} & \rightarrow &  (\textbf{4},\textbf{4}) + (\bar{\textbf{4}},\bar{\textbf{4}})  & \\[2mm]
\dot{\m} \,\,\,\,\,\, & \textbf{32'} & \rightarrow &  (\textbf{4},\bar{\textbf{4}}) + (\bar{\textbf{4}},\textbf{4})  &  .
\end{array}
\eeq
At the level of indices, this translates into the splittings $\,M=m \oplus a\,$, with $\,m,a=1,...,6\,$, together with $\,\mu = \left \lbrace _{i \hj} \,\oplus\, ^{i \hj} \right \rbrace\,$ and $\,\dot{\nu} = \left \lbrace {_{i}}^{\hj} \oplus {^{i}}_{\hj} \right \rbrace\,$ with $\,i,\hi=1,...,4\,$. The fundamental $\,\textrm{SO}(6)\,$ indices $\,m\,$ and $\,a\,$ respectively correspond to the time-like and space-like parts of the block-diagonal $\,\eta_{MN}=\eta^{MN}\,$ metric of $\,\textrm{SO}(6,6)\,$ in Lorentzian coordinates 
\beq
\label{eta_Lorentzian}
\eta_{MN} = 
\left( 
\begin{array}{c|c}
-\delta_{mn} & 0 \\[1mm] 
\hline
\\[-4mm]
0 & \delta_{ab} 
\end{array}
\right) \ .
\eeq
When written in terms of $\,\textrm{SU}(4)\sim\textrm{SO}(6)\,$ invariant tensors, the antisymmetric charge conjugation matrices in (\ref{B&C_splitting}) take the block off-diagonal form
\beq
\mathcal{C}_{\mu \nu} = 
\left( 
\begin{array}{c|c}
0 & {{\mathcal{C}}_{i \hj}}^{k \hl} = -i \, \delta_{i}^{k} \delta_{\hj}^{\hl}  \\[1mm] 
\hline
\\[-4mm]
{\mathcal{C}^{i \hj}}_{k \hl} =  i \,  \delta^{i}_{k} \delta^{\hj}_{\hl} & 0 
\end{array}
\right)
\hspace{1mm},\hspace{1mm}
{\mathcal{C}}^{\dot{\mu} \dot{\nu}} = 
\left( 
\begin{array}{c|c}
0 & \mathcal{C}^{i \phantom{\hj} \phantom{k} \hl}_{\phantom{i} \hj k \phantom{\hl}} = -i \,  \delta^{i}_{k} \delta^{\hl}_{\hj}  \\[1mm] 
\hline
\\[-4mm]
\mathcal{C}^{\phantom{i} \hj k \phantom{\hl}}_{i \phantom{\hj} \phantom{k} \hl}  =  i \, \delta_{i}^{k} \delta_{\hl}^{\hj} & 0 
\end{array}
\right) \ ,
\eeq
whereas the gamma matrices in (\ref{Gamma_splitting}) split into a set of time-like matrices with a block-diagonal structure
\beq
\label{gamma_1}
[\gamma_{m}]^{\mu \dot{\nu}} = 
\left( 
\begin{array}{c|c}
[\gamma_{m}]^{i \hj k \phantom{\hl}}_{\phantom{i} \phantom{\hj} \phantom{k} \hl} = [G_{m}]^{ik}  \delta_{\hl}^{\hj} & 0 \\[1mm] 
\hline
\\[-4mm]
0 & [\gamma_{m}]^{\phantom{i} \phantom{\hj} \phantom{k} \hl}_{i \hj k \phantom{\hl}}  = [G_{m}]_{ik}  \delta_{\hj}^{\hl} 
\end{array}
\right) \ ,
\eeq
\beq
\label{gamma_2}
[\bar{\gamma}_{m}]_{\dot{\mu} \nu} = 
\left( 
\begin{array}{c|c}
[\bar{\gamma}_{m}]^{\phantom{i}  \hj \phantom{k} \phantom{\hl}}_{i \phantom{\hj} k \hl} = [G_{m}]_{ik}  \delta_{\hl}^{\hj} & 0 \\[1mm] 
\hline
\\[-4mm]
0 & [\bar{\gamma}_{m}]^{i \phantom{\hj} k \hl}_{\phantom{i} \hj \phantom{k} \phantom{\hl}} = [G_{m}]^{ik}  \delta_{\hj}^{\hl} 
\end{array}
\right) \ ,
\eeq
and a set of space-like ones
\beq
\label{gamma_3}
[\gamma_{a}]^{\mu \dot{\nu}} = 
\left( 
\begin{array}{c|c}
0 & [\gamma_{a}]^{i \hj \phantom{k} \hl}_{\phantom{i} \phantom{\hj} k \phantom{\hl}}  =  [G_{a}]^{\hj \hl}  \delta_{k}^{i}  \\[1mm] 
\hline
\\[-4mm]
[\gamma_{a}]^{\phantom{i} \phantom{\hj} k \phantom{\hl}}_{i \hj \phantom{k} \hl} =  [G_{a}]_{\hj \hl}  \delta_{i}^{k}  & 0 
\end{array}
\right) \ ,
\eeq
\beq
\label{gamma_4}
[\bar{\gamma}_{a}]_{\dot{\mu} \nu}  = 
\left( 
\begin{array}{c|c}
0 & [\bar{\gamma}_{a}]^{\phantom{i} \hj k \hl}_{i \phantom{\hj} \phantom{k} \phantom{\hl}} = - [G_{a}]^{\hj \hl}  \delta^{k}_{i} \\[1mm] 
\hline
\\[-4mm]
[\bar{\gamma}_{a}]^{i \phantom{\hj} \phantom{k} \phantom{\hl}}_{\phantom{i} \hj k \hl}  =  - [G_{a}]_{\hj \hl}  \delta^{i}_{k}  & 0 
\end{array}
\right) \ ,
\eeq
with a block off-diagonal structure. The invariant tensors $\,G_{m}=[G_{m}]^{ij}\,$ and $\,G_{a}=[G_{a}]^{\hi \hj}\,$ are defined with upper indices and correspond to the gamma matrices for each of the $\,\textrm{SO}(6) \sim \textrm{SU}(4)\,$ factors in (\ref{A3A3_in_D6}). Often they are also called 't Hooft symbols and we take them to satisfy the (anti-)self-duality conditions\footnote{The non-vanishing parts of the scalar vielbeins $\,{\mathcal{V}_{M}}^{ij}\,$ and $\,{\mathcal{V}_{M}}^{\hi \hj}\,$ in (\ref{fermi_mass_N=4}) and (\ref{fermi_mass_N=4_extension}) reduce to $\,\frac{1}{2}[G_{m}]^{ij}\,$ and $\,\frac{1}{2}[G_{a}]^{\hi \hj}\,$ when evaluated at the origin of the moduli space, so them both must square to the identity in order to satisfy $\,\mathcal{V} \, \mathcal{V}^{T}=\mathds{1}\,$ at the origin.}
\beq
[G_{m}]_{ij} = - \frac{1}{2} \,\, \epsilon_{ijkl} \,\, [G_{m}]^{kl}
\hspace{10mm} \textrm{and} \hspace{10mm}
[G_{a}]_{\hi \hj} = \frac{1}{2} \,\, \epsilon_{\hi \hj \hk \hl} \,\, [G_{a}]^{\hk \hl} \ ,
\eeq
where $\,[G_{m}]_{ij} = ([G_{m}]^{ij})^{*}\,$ and $\,[G_{a}]_{\hi \hj} = ([G_{a}]^{\hi \hj})^{*}\,$. All along the present work, we have used the following explicit realisation of anti-self-dual $\,G_{m}\,$ 't Hooft symbols 
\begin{equation}
\begin{array}{c}
[G_{1}]=
{\scriptsize{\left[
\begin{array}{cccc}
 0 & 1 & 0 & 0 \\
 -1 & 0 & 0 & 0 \\
 0 & 0 & 0 & -1 \\
 0 & 0 & 1 & 0
\end{array}
\right]}}
\hspace{2mm} , \hspace{2mm}
[G_{3}]=
{\scriptsize{\left[
\begin{array}{cccc}
 0 & 0 & 1 & 0 \\
 0 & 0 & 0 & 1 \\
 -1 & 0 & 0 & 0 \\
 0 & -1 & 0 & 0
\end{array}
\right]}}
\hspace{2mm} , \hspace{2mm}
[G_{5}]=
{\scriptsize{\left[
\begin{array}{cccc}
 0 & 0 & 0 & 1 \\
 0 & 0 & -1 & 0 \\
 0 & 1 & 0 & 0 \\
 -1 & 0 & 0 & 0
\end{array}
\right]}} \ ,

\\[8mm]

[G_{2}]=
{\scriptsize{\left[
\begin{array}{cccc}
 0 & i & 0 & 0 \\
 -i & 0 & 0 & 0 \\
 0 & 0 & 0 & i \\
 0 & 0 & -i & 0
\end{array}
\right]}}
\hspace{2mm} , \hspace{2mm}
[G_{4}]=
{\scriptsize{\left[
\begin{array}{cccc}
 0 & 0 & i & 0 \\
 0 & 0 & 0 & -i \\
 -i & 0 & 0 & 0 \\
 0 & i & 0 & 0
\end{array}
\right]}}
\hspace{2mm} , \hspace{2mm}
[G_{6}]=
{\scriptsize{\left[
\begin{array}{cccc}
 0 & 0 & 0 & i \\
 0 & 0 & i & 0 \\
 0 & -i & 0 & 0 \\
 -i & 0 & 0 & 0
\end{array}
\right]}} \ ,
\end{array}
\end{equation}
together with the self-dual $\,G_{a}\,$ ones 
\begin{equation}
\begin{array}{c}
[G_{1}]=
{\scriptsize{\left[
\begin{array}{cccc}
 0 & 1 & 0 & 0 \\
 -1 & 0 & 0 & 0 \\
 0 & 0 & 0 & 1 \\
 0 & 0 & -1 & 0
\end{array}
\right]}}
\hspace{2mm} , \hspace{2mm}
[G_{3}]=
{\scriptsize{\left[
\begin{array}{cccc}
 0 & 0 & 1 & 0 \\
 0 & 0 & 0 & -1 \\
 -1 & 0 & 0 & 0 \\
 0 & 1 & 0 & 0
\end{array}
\right]}}
\hspace{2mm} , \hspace{2mm}
[G_{5}]=
{\scriptsize{\left[
\begin{array}{cccc}
 0 & 0 & 0 & 1 \\
 0 & 0 & 1 & 0 \\
 0 & -1 & 0 & 0 \\
 -1 & 0 & 0 & 0
\end{array}
\right]}} \ ,

\\[8mm]

[G_{2}]=
{\scriptsize{\left[
\begin{array}{cccc}
 0 & i & 0 & 0 \\
 -i & 0 & 0 & 0 \\
 0 & 0 & 0 & -i \\
 0 & 0 & i & 0
\end{array}
\right]}}
\hspace{2mm} , \hspace{2mm}
[G_{4}]=
{\scriptsize{\left[
\begin{array}{cccc}
 0 & 0 & i & 0 \\
 0 & 0 & 0 & i \\
 -i & 0 & 0 & 0 \\
 0 & -i & 0 & 0
\end{array}
\right]}}
\hspace{2mm} , \hspace{2mm}
[G_{6}]=
{\scriptsize{\left[
\begin{array}{cccc}
 0 & 0 & 0 & i \\
 0 & 0 & -i & 0 \\
 0 & i & 0 & 0 \\
 -i & 0 & 0 & 0
\end{array}
\right]}} \ .
\end{array}
\end{equation}
Notice that they are complex matrices and then will lead to a complex representation of gamma matrices in (\ref{gamma_1})-(\ref{gamma_4}). This is related to the fact that $\,\textrm{SO}(6)\,$ does not admit \mbox{M-W} spinors (it is $\,\textrm{SO}(3,3)\,$ which does), so a real representation of $\,\textrm{SO}(6,6)\,$ gamma matrices is no longer possible when moving to an $\,\textrm{SU}(4) \times \textrm{SU}(4)\,$ covariant formulation.

When arranged into $\,64 \times 64\,$ matrices according to the splittings (\ref{Gamma_splitting}) and (\ref{B&C_splitting}), one verifies that the defining relations (\ref{Clifford}) and (\ref{GammaT}) hold and also that $\,{\Gamma_{13}=\textrm{diag}(+\mathds{1}_{32},-\mathds{1}_{32})}\,$. Finally, the conjugation matrices (\ref{B&C_splitting}) entering the definition of the scalar matrix in (\ref{M_origin}) take the form
\beq
\label{B_A3A3}
B_{\mu \nu} = 
\left( 
\begin{array}{c|c}
0 & \mathds{1}_{16}  \\[1mm] 
\hline
\\[-4mm]
\mathds{1}_{16} & 0 
\end{array}
\right)
\hspace{5mm} \textrm{ , } \hspace{5mm}
B^{\dot{\mu} \dot{\nu}} = 
\left( 
\begin{array}{c|c}
0 & - \mathds{1}_{16}  \\[1mm] 
\hline
\\[-4mm]
- \mathds{1}_{16} & 0 
\end{array}
\right) \ ,
\eeq
producing a non-standard definition of the origin of the moduli space, as discussed in detail in the main text.

\subsection*{Real formulation of M-W spinors}

In addition to the $\,\textrm{SU}(4) \times \textrm{SU}(4)\,$ covariant formulation of M-W spinors described above, we can adopt another realisation such that: $i)$ it is a real realisation of M-W spinors $ii)$ it is compatible with the standard choice of (\ref{M_origin_choice}) as the origin of the moduli space. 

We build our real $\,64 \times 64\,$ $\Gamma$-matrices in a Majorana representation out of the $\,2 \times 2\,$ Pauli matrices $\,\sigma_{1,2,3}\,$ in the following way 
\beq
\label{Gamma_Real}
\begin{array}{lcc}
\Gamma_{1} & = & i \, \s_{2} \otimes \mathds{1}_{2} \otimes \mathds{1}_{2}\otimes \mathds{1}_{2}\otimes \mathds{1}_{2} \otimes \mathds{1}_{2} \\
\Gamma_{2} & = & i \, \s_{3} \otimes \s_{2}         \otimes \mathds{1}_{2}\otimes \mathds{1}_{2}\otimes \mathds{1}_{2} \otimes \mathds{1}_{2} \\
\Gamma_{3} & = & i \, \s_{3} \otimes \s_{3}         \otimes \s_{2}        \otimes \mathds{1}_{2}\otimes \mathds{1}_{2} \otimes \mathds{1}_{2} \\
\Gamma_{4} & = & i \, \s_{3} \otimes \s_{3}         \otimes \s_{3}        \otimes\s_{2}         \otimes \mathds{1}_{2} \otimes \mathds{1}_{2} \\
\Gamma_{5} & = & i \, \s_{3} \otimes \s_{3}         \otimes \s_{3}        \otimes \s_{3}        \otimes \s_{2}         \otimes \mathds{1}_{2} \\
\Gamma_{6} & = & i \, \s_{3} \otimes \s_{3}         \otimes \s_{3}        \otimes \s_{3}        \otimes \s_{3}         \otimes \s_{2} 
\end{array}
\hspace{3mm} , \hspace{3mm}
\begin{array}{lcc}
\Gamma_{7} & = & \s_{1} \otimes \mathds{1}_{2} \otimes \mathds{1}_{2}\otimes \mathds{1}_{2}\otimes \mathds{1}_{2} \otimes \mathds{1}_{2} \\
\Gamma_{8} & = & \s_{3} \otimes \s_{1}         \otimes \mathds{1}_{2}\otimes \mathds{1}_{2}\otimes \mathds{1}_{2} \otimes \mathds{1}_{2} \\
\Gamma_{9} & = & \s_{3} \otimes \s_{3}         \otimes \s_{1}        \otimes \mathds{1}_{2}\otimes \mathds{1}_{2} \otimes \mathds{1}_{2} \\
\Gamma_{10} & = & \s_{3} \otimes \s_{3}         \otimes \s_{3}        \otimes\s_{1}         \otimes \mathds{1}_{2} \otimes \mathds{1}_{2} \\
\Gamma_{11} & = & \s_{3} \otimes \s_{3}         \otimes \s_{3}        \otimes \s_{3}        \otimes \s_{1}         \otimes \mathds{1}_{2} \\
\Gamma_{12} & = & \s_{3} \otimes \s_{3}         \otimes \s_{3}        \otimes \s_{3}        \otimes \s_{3}         \otimes \s_{1} 
\end{array}
\eeq
where we decide to use a set of Pauli matrices satisfying $\,[\s_{i} , \s_{j}] = 2 \,  i \,\epsilon_{ijk} \, \s_{k}\,$. This corresponds to the choice
\beq
\s_{1} = 
\left( 
\begin{array}{cc}
0 & 1  \\ 
1 & 0 
\end{array}
\right)
\hspace{5mm} \textrm{ , } \hspace{5mm}
\s_{2} = 
\left( 
\begin{array}{cc}
0 & -i  \\ 
i & 0 
\end{array}
\right)
\hspace{5mm} \textrm{ , } \hspace{5mm}
\s_{3} = 
\left( 
\begin{array}{cc}
1 & 0  \\ 
0 & -1 
\end{array}
\right) \ .
\eeq
Building the $\,64 \times 64\,$ charge conjugation matrix as
\beq
\label{C_Real}
\begin{array}{lcc}
\mathcal{C} & = & -i\, \s_{2} \otimes \s_{1} \otimes \s_{2} \otimes \s_{1} \otimes \s_{2} \otimes \s_{1} \ ,
\end{array}
\eeq
one can easily check that (\ref{Gamma_Real}) and (\ref{C_Real}) automatically satisfy the conditions in (\ref{Clifford}) and (\ref{GammaT}) with the $\,\eta_{MN}\,$ metric given in (\ref{eta_Lorentzian}). By applying an $\,\textrm{SO}(64)\,$ rotation taking $\,{\Gamma_{13}=\textrm{diag}(+\mathds{1}_{32},-\mathds{1}_{32})}\,$, we go to a real M-W basis according to the splittings (\ref{Gamma_splitting}) and (\ref{B&C_splitting}). In this basis, the conjugation matrix in (\ref{M_origin}) happens to be $\,B_{\m \n}=-B^{\dm \dn}=\mathds{1}_{32}\,$, hence being compatible with the standard choice for the origin of the moduli space in (\ref{M_origin_choice}). We will use this real representation of M-W spinors when it comes to identify gaugings associated to critical points at the origin of the moduli space.

\section{$X_{\mathbb{M} \mathbb{N} \mathbb{P}}\,$ in the $\,\textrm{SL}(2) \times \textrm{SO}(6,6)\,$ formulation}
\label{App:X-tensor_components}

In this appendix we derive the explicit form of the components of the $\,X_{\mathbb{M} \mathbb{N} \mathbb{P}}\,$ tensor given in (\ref{Xbosonic}) and (\ref{Xfermionic}). Let us first start by giving the explicit form of the $\,\textrm{E}_{7(7)}\,$ symmetric generators $\,[t_{\mathpzc{A}}]_{\mathbb{MN}}\,$ in the fundamental representation following the conventions in ref.~\cite{Dibitetto:2011eu}. By virtue of the index splittings $\,{\mathbb{M}=\a M \oplus \mu}\,$ and $\,\mathpzc{A}=\a M \b N \oplus \g \dot{\mu}\,$ associated to the branching of the $\textbf{56}$ and $\textbf{133}$ irrep's of $\,\textrm{E}_{7(7)}\,$ under $\,\textrm{SL}(2) \times \textrm{SO}(6,6)$, they are given by
\beq
\label{E7_gen}
\begin{array}{cclc}
\left[t_{\a M \b N}\right]_{\g P \d Q} & = & \epsilon_{\a \b} \, \epsilon_{\g \d} \, \left[t_{MN}\right]_{PQ} + \eta_{MN}\, \, \eta_{PQ} \, \left[t_{\a \b}\right]_{\g \d} & , \\[3mm]
\left[t_{\a M \b N}\right]_{\m \n} & = &\dfrac{1}{4} \, \epsilon_{\a \b} \, \left[\g_{MN}\right]_{\m \n} & , \\[3mm]
\left[t_{\a \dm}\right]_{\b N \n} = \left[t_{\a \dm}\right]_{\n \b N} & = & \epsilon_{\a \b} \, \left[\bg_{N}\right]_{\dm \n} = \epsilon_{\a \b} \, \left[\g_{N}\right]_{\n \dm}& ,
\end{array}
\eeq
where $\,[t_{\a \b}]^{\g \d} = \delta_{\a}^{(\g}\delta_{\b}^{\d)}\,$ and $\,[t_{M N}]^{P Q} = \delta_{MN}^{PQ}\,$ are the generators of $\,\textrm{SL}(2)\,$ and $\,\textrm{SO}(6,6)\,$, respectively. 

On the other hand, we need the $\Theta$-components of the embedding tensor $\,{\Theta_{\mathbb{M}}}^{\mathpzc{A}}\,$ in order to compute $\,X_{\mathbb{MNP}}\,$. These can be split into those components involving an even number of fermionic indices 
\beq
\label{emb_tens_comp_old}
\begin{array}{cclc}
{\Theta_{\a M}}^{\b N \g P} & = & - \dfrac{1}{2} \, \epsilon^{\b \g} \, {f_{\a M}}^{NP} \, -  \, \dfrac{1}{2} \, \epsilon^{\b \g}  \, \delta_{M}^{[N} \, \xi_{\a}^{\,\,\,\,P]} \, + \,  \dfrac{1}{12} \,  \delta_{\a}^{(\b} \, \xi^{\g)}_{\,\,\,\,M} \, \eta^{N P} & , \\[4mm]
{\Theta_{\m}}^{\a \dn} & = & \dfrac{1}{24} \, \epsilon^{\a \b} \, f_{\b M N P} \, \left[  \g^{MNP} \right]_{\mu}^{\,\,\,\,\,\,\dn} \, - \, \dfrac{1}{8} \, \epsilon^{\a \b} \, \xi_{\b M} \left[ \gamma^{M}\right]_{\mu}^{\,\,\,\,\,\,\dn} & ,
\end{array}
\eeq
which were already derived in ref.~\cite{Dibitetto:2011eu}, together with a set of additional ones involving an odd number of fermionic indices. The most general ansatz for the latter according to the symmetry is given by
\beq
\label{emb_tens_comp_new}
\begin{array}{cclc}
{\Theta_{\a M}}^{\b \dot{\m}} & = & h_{1} \, \d_{\a}^{\b} \, {F_{M}}^{\dot{\m}} \, +  \, h_{2} \, \epsilon^{\b \g}  \, \Xi_{\a\g\n} \, [\g_{M}]^{\n \dot{\m}} & , \\[4mm]
{\Theta_{\m}}^{\a M \b N} & = & h_{3} \, \epsilon^{\a \b} \, {F^{[M}}_{\dot{\n}} \, \left[  \g^{N]} \right]_{\mu}^{\,\,\,\,\,\dn} \, + \, h_{4} \,\, {\Xi^{\a \b}}_{\m} \,\, \eta^{MN} & ,
\end{array}
\eeq
where $\,h_{i=1,2,3,4}\,$ are constant coefficients to be fixed as follows: $i)$ Following the definition in (\ref{gauge_algebra}) and using the form of the $\,\Omega_{\mathbb{MN}}\,$ matrix in (\ref{Omega_matrix}), we can build the $\,X_{\mathbb{MNP}}\,$ tensor as $\,X_{\mathbb{M} \mathbb{N} \mathbb{P}} = {\Theta_{\mathbb{M}}}^{\mathpzc{A}} \, {[t_{\mathpzc{A}}]_{\mathbb{N}}}^{\mathbb{R}} \, \Omega_{\mathbb{RP}}\,$.  $ii)$ By requiring the $\,X_{\mathbb{MNP}}\,$ tensor to live in the $\,\textbf{912}\,$ irrep of $\,\textrm{E}_{7(7)}\,$ we still have to impose the linear constraints in (\ref{linear_const}). This imposes the relations $\,h_{3}=h_{1}\,$ and $\,h_{4}=-\frac{1}{6} \, h_{2}\,$ on the coefficients. $iii)$ Finally we set the remaining free parameters to the values $\,h_{1}=-1\,$ and $\,h_{2}=1\,$, what fixes the relative normalisation between $\,F_{M \dot{\m}}\,$ and $\,\Xi_{\a \b \m}\,$. The final expression is then given by
\beq
\label{emb_tens_comp_new}
\begin{array}{cclc}
{\Theta_{\a M}}^{\b \dot{\m}} & = & - \d_{\a}^{\b} \, {F_{M}}^{\dot{\m}} \, +  \,  \epsilon^{\b \g}  \, \Xi_{\a\g\n} \, [\g_{M}]^{\n \dot{\m}} & , \\[4mm]
{\Theta_{\m}}^{\a M \b N} & = & - \epsilon^{\a \b} \, {F^{[M}}_{\dot{\n}} \, \left[  \g^{N]} \right]_{\mu}^{\,\,\,\,\,\dn} \, -\, \dfrac{1}{6} \,\, {\Xi^{\a \b}}_{\m} \,\, \eta^{MN} & ,
\end{array}
\eeq
and after some algebra, the set of independent components of the $\,X_{\mathbb{MNP}}\,$ tensor are those given in (\ref{Xbosonic}) and (\ref{Xfermionic}).

\section{The vielbein in the origin of the moduli space}
\label{App:vielbein}

As stated in section~\ref{sec:connecting_A7&A1D6}, the vielbein $\,\mathcal{V}_{\mathbb{M}}^{\phantom{\mathbb{M}}\underline{\mathbb{N}}}\,$ is the fundamental object connecting the $\,\,\textrm{SU}(8)\,$ and $\,\textrm{SL}(2) \times \textrm{SO}(6,6)\,$ formulations of maximal supergravity. Recalling the relation (\ref{X(T)})
\beq
\label{X(T)_2}
X_{\mathbb{MNP}} = 2 \, \mathcal{V}_{\mathbb{M}}^{\phantom{\mathbb{M}}\underline{\mathbb{Q}}} \, \mathcal{V}_{\mathbb{N}}^{\phantom{\mathbb{N}}\underline{\mathbb{R}}} \, \mathcal{V}_{\mathbb{P}}^{\phantom{\mathbb{P}}\underline{\mathbb{S}}}\,\,\, T_{\underline{\mathbb{QRS}}} \ ,
\eeq 
one concludes that the vielbein $\,\mathcal{V}_{\mathbb{M}}^{\phantom{\mathbb{M}}\underline{\mathbb{Q}}}\,$ has the row index $\,\mathbb{M}\,$ in the $\,\textrm{SL}(2)\times \textrm{SO}(6,6)\,$ basis and the column index $\,\underline{\mathbb{Q}}\,$ in the $\,\textrm{SU}(8)\,$ one. In order to determine the form of $\,\mathcal{V}_{\mathbb{M}}^{\phantom{\mathbb{M}}\underline{\mathbb{Q}}}\,$ we must go to a common basis where to simultaneously describe $\,\textrm{SL}(2)\times \textrm{SO}(6,6)\,$ and $\,\textrm{SU}(8)\,$ indices. This common basis turns out to be
\beq
\begin{array}{ccc}
\textrm{SO}(2) \times \textrm{SO}(6)_{\textrm{time}} \times \textrm{SO}(6)_{\textrm{space}} &\sim& \textrm{U}(1) \times \textrm{SU}(4)_{\textrm{time}} \times \textrm{SU}(4)_{\textrm{space}} \ ,
\end{array}
\eeq
since it is the only maximal subgroup being shared by them both. Furthermore, it coincides with their maximal compact subgroup. All our conventions related to the $\textrm{SU}(4) \times \textrm{SU}(4)\,$ covariant formulation of $\,\textrm{SO}(6,6)\,$ spinors, gamma matrices, etc. are summarised in appendix~\ref{App:spinors}.

In what follows we will make an extensive use of two different decompositions of an $\,\textrm{E}_{7(7)}\,$ fundamental index:
\begin{itemize}

\item[$i)$] the decomposition with respect to $\,\textrm{SL}(2)\times \textrm{SO}(6,6)\,$
\beq 
\label{A1D6_decomp}
\begin{array}{ccccc}
\textrm{E}_{7(7)} & \supset &  \textrm{SL}(2)\times \textrm{SO}(6,6) & \supset  & \textrm{SL}(2) \,\times\, \textrm{SO}(6)_{\textrm{time}} \,\times\, \textrm{SO}(6)_{\textrm{space}} \\[2mm]
\textbf{56} & \rightarrow &  (\textbf{2},\textbf{12}) + (\textbf{1},\textbf{32}) &  \rightarrow  & (\textbf{2},\textbf{6},\textbf{1}) + (\textbf{2},\textbf{1},\textbf{6}) + (\textbf{1},\textbf{4},\textbf{4}) + (\textbf{1},\bar{\textbf{4}},\bar{\textbf{4}})\\[2mm]
\mathbb{M} & = &  \a M \oplus \mu & =  & \a \, m \,\,\, \oplus \,\,\, \a \,a \,\,\, \oplus \,\,\, i \, \hat{j} \,\,\, \oplus \,\,\, \overline{i} \, \overline{\hat{j}}\\[2mm]
\end{array}
\eeq

\item[$ii)$] the decomposition with respect to $\,\textrm{SU}(8)\,$
\beq 
\label{A7_decomp}
\begin{array}{ccccc}
\textrm{E}_{7(7)} & \supset &  \textrm{SU}(8) & \supset  & \textrm{U}(1) \,\times\, \textrm{SU}(4)_{\textrm{time}} \,\times\, \textrm{SU}(4)_{\textrm{space}} \\[2mm]
\textbf{56} & \rightarrow &  \textbf{28} + \overline{\textbf{28}} &  \rightarrow  & (\textbf{6},\textbf{1})_{(-2)} + (\textbf{1},\textbf{6})_{(2)} + (\textbf{4},\textbf{4})_{(0)} \,\,\, + \,\,\, \textrm{c.c.}  \\[2mm]
\underline{\mathbb{M}} & = &  [\mathcal{I} \, \mathcal{J}] \, \oplus \, \textrm{c.c.} & =  & [i\, j] \,\,\, \oplus \,\,\, [\hat{i} \,\hat{j}] \,\,\, \oplus \,\,\,  i \, \hat{j} \,\,\, \oplus \,\,\, \textrm{c.c.} 
\end{array}
\eeq
\end{itemize}
where $\,_{(q)}\,$ in (\ref{A7_decomp}) denotes the $\,\textrm{U}(1)\,$ charge of the $\,\textrm{SU}(4) \,\times\, \textrm{SU}(4)\,$ irrep's. By comparing the decompositions in (\ref{A1D6_decomp}) and (\ref{A7_decomp}), the non-vanishing components of the vielbein\footnote{As discussed in section~\ref{sec:connecting_A7&A1D6}, we are setting all the ``fermionic'' scalars to the origin of the moduli space.} will correspond to
\beq
\label{vielbein_components}
\begin{array}{ccl}
\mathcal{V}_{\mathbb{M}}^{\phantom{\mathbb{M}}\underline{\mathbb{Q}}} & = & \left\lbrace \, {\mathcal{V}_{\a M}}^{[\mathcal{I}\mathcal{J}]} \,\, , \,\, \mathcal{V}_{\a M [\mathcal{I} \mathcal{J}]}  \,\, , \,\, {\mathcal{V}_{\mu}}^{[\mathcal{I} \mathcal{J}]} \,\, , \,\, \mathcal{V}_{\mu [\mathcal{I}\mathcal{J}]} \, \right\rbrace \\[2mm]
&  =  & \left\lbrace \, {\mathcal{V}_{\a m}}^{[ij]} \,\, , \,\, \mathcal{V}_{\a m [ij]} \,\, , \,\,  {\mathcal{V}_{\a a}}^{[\hat{i} \hat{j}]}  \,\, , \,\,  \mathcal{V}_{\a a [\hat{i} \hat{j}]} \,\, , \,\, {\mathcal{V}_{i \hat{j}}}^{k \hat{l}} \,\, , \,\, {\mathcal{V}^{i \hat{j}}}_{k \hat{l}} \, \right\rbrace \ , 
\end{array}
\eeq 
where $\,\a=+,-\,$ is an $\,\textrm{SL}(2)\,$ index raised and lowered by $\,\epsilon_{\a \b}\,$ and where $\,[ij]\,$, $\,[\hat{i} \hat{j}]\,$ and $\,i\hat{j}\,$ are pairs of fundamental $\,\textrm{SU}(4)\,$ indices with $\,i,\hat{i}=1,...,4\,$. The fundamental $\,\textrm{SO}(6)\,$ indices $\,m\,$ and $\,a\,$ correspond to the time-like and space-like parts of the diagonal metric $\,\eta_{MN}=(\underbrace{-1,...,-1}_{\textrm{$6$ times}},\underbrace{1,...,1}_{\textrm{$6$ times}})\,$ of $\,\textrm{SO}(6,6)\,$ in Lorentzian coordinates.
\\[-1mm]

If setting all the scalar fields to zero, i.e. moving to the origin of the moduli space, the set of vielbein components in (\ref{vielbein_components}) must reduce to the product of a constant $\,\textrm{SL}(2)\,$ complexified vielbein $\,L_{\a}\equiv\mathcal{V}_{\a}|_{\textrm{origin}} =\left( i \,,\, 1\right)\,$ satisfying
\beq
\label{LLstar}
L_{\a} \, L^{*}_{\b} = \d_{\a \b} + i \, \epsilon_{\a \b} \ ,
\eeq
with a set of $\,\textrm{SO}(6) \sim \textrm{SU}(4)\,$ invariant tensors. These are the \text{'t Hooft symbols} $\,{\mathcal{V}_{m}}^{ij}|_{\textrm{origin}}=\frac{1}{2} \, [G_{m}]^{ij}\,$, $\,{\mathcal{V}_{a}}^{\hi \hj}|_{\textrm{origin}}=\frac{1}{2} \, [G_{a}]^{\hat{i}\hat{j}}\,$ (see appendix~\ref{App:spinors}) and the Kronecker deltas $\,\delta_{i}^{j}\,$ and $\,\delta_{\hat{i}}^{\hat{j}}\,$. The non-vanishing components of the vielbein $\,{\mathcal{V}_{\mathbb{M}}}^{\underline{\mathbb{N}}}\,$ are given by
\beq
\begin{array}{lclclc}
{\mathcal{V}_{\a m}}^{ij} = \dfrac{-i}{2\sqrt{2}} \, (L_{\a})^{*} \, [G_{m}]^{ij} & \hspace{0mm} , \hspace{0mm} & \mathcal{V}_{\a m \, ij} = \dfrac{i}{2\sqrt{2}} \, L_{\a} \, [G_{m}]_{ij}
& \hspace{0mm} , \hspace{0mm} &   {\mathcal{V}_{i \hj}}^{k \hl} = \dfrac{(1+i)}{2} \, \d^{k}_{i} \, \d^{\hl}_{\hj} & , \\[5mm]
{\mathcal{V}_{\a a}}^{\hi \hj} = \dfrac{-1}{2\sqrt{2}} \, L_{\a} \, [G_{a}]^{\hi \hj} & \hspace{0mm} , \hspace{0mm} & \mathcal{V}_{\a a \, \hi \hj} = \dfrac{-1}{2\sqrt{2}} \, (L_{\a})^{*} \, [G_{a}]_{\hi \hj} & \hspace{0mm} , \hspace{0mm} & {\mathcal{V}^{i \hj}}_{k \hl} = \dfrac{(1-i)}{2} \, \d^{i}_{k} \, \d^{\hj}_{\hl} & ,
\end{array}
\eeq
whereas those of the inverse vielbein $\,{\mathcal{V}^{\mathbb{M}}}_{\underline{\mathbb{N}}}\,$ read
\beq
\begin{array}{lclclc}
{\mathcal{V}^{\a m}}_{ij} = \dfrac{-1}{2\sqrt{2}} \, L^{\a} \, [G^{m}]_{ij} & \hspace{0mm} , \hspace{0mm} & \mathcal{V}^{\a m \, ij} = \dfrac{-1}{2\sqrt{2}} \, (L^{\a})^{*} \, [G^{m}]^{ij}
& \hspace{0mm} , \hspace{0mm} &   {\mathcal{V}^{i \hj}}_{k \hl} = \dfrac{(1-i)}{2} \, \d_{k}^{i} \, \d_{\hl}^{\hj} & ,\\[5mm]
{\mathcal{V}^{\a a}}_{\hi \hj} = \dfrac{-i}{2\sqrt{2}} \, (L^{\a})^{*} \, [G^{a}]_{\hi \hj} & \hspace{0mm} , \hspace{0mm} & \mathcal{V}^{\a a \, \hi \hj} = \dfrac{i}{2\sqrt{2}} \, L^{\a} \, [G^{a}]^{\hi \hj} & \hspace{0mm} , \hspace{0mm} & {\mathcal{V}_{i \hj}}^{k \hl} = \dfrac{(1+i)}{2} \, \d_{i}^{k} \, \d_{\hj}^{\hl} & ,
\end{array}
\eeq
and completely specify the relations (\ref{A's_fluxes}). One can check that the vielbein $\,{\mathcal{V}_{\mathbb{M}}}^{\underline{\mathbb{N}}}\,$ satisfies the normalisation conditions \cite{deWit:2007mt}  
\beq
\begin{array}{rcll}
{\mathcal{V}_{\mathbb{M}}}^{\mathcal{IJ}} \, {\mathcal{V}_{\mathbb{N}}}_{\mathcal{IJ}}  - {\mathcal{V}_{\mathbb{M}}}_{\mathcal{IJ}} \, {\mathcal{V}_{\mathbb{N}}}^{\mathcal{IJ}} & = & i \, \Omega_{\mathbb{M} \mathbb{N}} & , \\[2mm]
\Omega^{\mathbb{M} \mathbb{N}} \, {\mathcal{V}_{\mathbb{M}}}^{\mathcal{IJ}} \, {\mathcal{V}_{\mathbb{N}}}_{\mathcal{KL}} & = & i \, {\delta^{\mathcal{IJ}}}_{\mathcal{KL}} & ,\\[2mm]
\Omega^{\mathbb{M} \mathbb{N}} \, {\mathcal{V}_{\mathbb{M}}}^{\mathcal{IJ}} \, {\mathcal{V}_{\mathbb{N}}}^{\mathcal{KL}} & = & 0 & .
\end{array}
\eeq

\section{Type II fluxes and the embedding tensor $\,f_{\a MNP}$}
\label{App:fluxes}

In this appendix, we summarise the identification between embedding tensor components $\,f_{\a MNP}\,$ (alternatively $\, \Lambda_{\a ABC}\,$ as explained in section~\ref{sec:typeII_models}) and type II flux backgrounds for the $\,\cN=1\,$ supergravity theory inside box --5-- in figure~\ref{fig:truncations}. 
\begin{table}[h!]
\renewcommand{\arraystretch}{1.25}
\begin{center}
\scalebox{0.87}[0.87]{
\begin{tabular}{ | c || c | c | c | c | c |}
\hline
couplings & SO($6,6$) & SO($2,2$) & type IIB & type IIA & fluxes\\
\hline
\hline
$1 $& $ -f_{+ \bar{a}\bar{b}\bar{c}} $  & $ - \Lambda_{+333} $& $ {F}_{ ijk} $& $F_{aibjck}$ & $  a_0 $\\
\hline
$U $& $f_{+ \bar{a}\bar{b}\bar{k}}$  &  $ \Lambda_{+334} $& ${F}_{ ij c} $& $F_{aibj}$ & $   a_1 $\\
\hline
$U^2 $& $ -f_{+ \bar{a}\bar{j}\bar{k}}$  & $ - \Lambda_{+344} $& ${F}_{i b c} $& $F_{ai}$ & $  a_2 $\\
\hline
$U^3 $& $f_{+ \bar{i}\bar{j}\bar{k}}$  & $ \Lambda_{+444} $& ${F}_{a b c} $& $F_{0}$ & $  a_3 $\\
\hline
\hline
$S $& $ -f_{- \bar{a}\bar{b}\bar{c}} $  &  $ - \Lambda_{-333} $& $ {H}_{ijk} $& $ {H}_{ijk} $  & $  - b_0$\\
\hline
$S \, U $& $f_{- \bar{a}\bar{b}\bar{k}}$  & $ \Lambda_{-334}$ & $ {H}_{ij c} $& ${\omega}^{c}_{ij}$ & $  - b_1 $\\
\hline
$S \, U^2 $& $ -f_{- \bar{a}\bar{j}\bar{k}}$   & $ - \Lambda_{-344} $& $ {H}_{ i b c}$ & $ {Q}_{ i }^{ b c}$  & $ - b_2 $\\
\hline
$S \, U^3 $& $f_{- \bar{i}\bar{j}\bar{k}}$   &  $ \Lambda_{-444} $& $ {H}_{a b c} $& $ {R}^{a b c} $ & $ - b_3 $\\
\hline
\hline
$T $& $f_{+ \bar{a}\bar{b}k}$   & $ \Lambda_{+233} $& $  Q^{a b}_k $&$ H_{a b k} $ & $  c_0 $\\
\hline
$T \, U $& $f_{+ \bar{a}\bar{j} k}=f_{+ \bar{i}\bar{b} k}\,\,\,,\,\,\,f_{+ a\bar{b}\bar{c}}$  &  $ \Lambda_{+234} \,\,\,,\,\,\, \Lambda_{+133} $& $ Q ^{a j}_k = Q^{i b}_k \,\,\,,\,\,\, Q^{b c}_a $& $ \omega^{j}_{k a} = \omega^{i}_{b k} \,\,\,,\,\,\, \omega_{b c}^a $  & $c_1 \,\,\,,\,\,\, \tilde {c}_1 $\\
\hline
$T \, U^2 $& $f_{+ \bar{i}\bar{b}c}=f_{+ \bar{a}\bar{j}c}\,\,\,,\,\,\,f_{+ \bar{i}\bar{j}k}$   &  $ \Lambda_{+134} \,\,\,,\,\,\, \Lambda_{+244} $& $ Q ^{ib}_c = Q^{a j}_c \,\,\,,\,\,\, Q^{ij}_k $& $ Q ^{ci}_b = Q^{j c}_a \,\,\,,\,\,\, Q^{ij}_k $ & $c_2 \,\,\,,\,\,\,\tilde{c}_2 $\\
\hline
$T \, U^3 $& $f_{+ \bar{i}\bar{j} c}$  &  $ \Lambda_{+144} $& $  Q^{ij}_{c} $& $  R^{ijc} $ & $c_3 $\\
\hline
\hline
$S \, T $& $f_{- \bar{a}\bar{b}k}$   &  $ \Lambda_{-233} $& $  P^{a b}_k $& & $  - d_0 $\\
\hline
$S \, T \, U $& $f_{- \bar{a}\bar{j} k}=f_{- \bar{i}\bar{b} k}\,\,\,,\,\,\,f_{- a\bar{b}\bar{c}}$   &  $ \Lambda_{-234} \,\,\,,\,\,\, \Lambda_{-133} $& $ P ^{a j}_k = P^{i b}_k \,\,\,,\,\,\, P^{b c}_a $&  & $- d_1 \,\,\,,\,\,\, - \tilde {d}_1 $\\
\hline
$S \, T \, U^2 $& $f_{- \bar{i}\bar{b}c}=f_{- \bar{a}\bar{j}c}\,\,\,,\,\,\,f_{- \bar{i}\bar{j}k}$   &  $ \Lambda_{-134} \,\,\,,\,\,\, \Lambda_{-244} $& $ P ^{ib}_c = P^{a j}_c \,\,\,,\,\,\, P^{ij}_k $&  & $ - d_2 \,\,\,,\,\,\, - \tilde{d}_2 $\\
\hline
$S \, T \, U^3 $& $f_{- \bar{i}\bar{j} c}$  &  $ \Lambda_{-144} $ & $  P^{ij}_{c} $&  & $ - d_3 $\\
\hline
\end{tabular}
}
\end{center}
\caption{Mapping between unprimed fluxes, embedding tensor components and couplings in the flux-induced superpotential. We have made the index splitting $\,M=\{a,i,\bar{a},\bar{i}\}\,$ for $\,\textrm{SO}(6,6)\,$ light-cone coordinates, not to be confused with the same indices appearing in appendix~\ref{App:indices}.}
\label{table:unprimed_fluxes}
\end{table}
\noindent This identification\footnote{Notice that refs~\cite{Dibitetto:2010rg,Dibitetto:2011gm} use light-cone coordinates for $\,\textrm{SO}(6,6)\,$ fundamental indices. In this basis, the metric takes the form 
$\,
\eta_{MN} = 
{\scriptsize{
\left( 
\begin{array}{cc}
0 & \mathds{1}_{6} \\
\mathds{1}_{6} & 0 
\end{array}
\right)}}
\,$
which is related to that in (\ref{eta_Lorentzian}) through an $\,\textrm{SO}(12)\,$ rotation of the form 
\beq
U=\frac{1}{\sqrt{2}} \,\left( 
\begin{array}{cc}
-\mathds{1}_{6} & \mathds{1}_{6} \\
\mathds{1}_{6} & \mathds{1}_{6} 
\end{array}
\right) \ .
\eeq
}
was originally proposed in ref.~\cite{Dibitetto:2010rg} and further developed in ref.~\cite{Dibitetto:2011gm}. We include it here to make the paper as self-consistent as possible\footnote{The minus sign affecting magnetic fluxes in the last column of tables \ref{table:unprimed_fluxes} and \ref{table:primed_fluxes} stems from the different choice of the SL($2$) complexified vielbein $L_{\alpha}$ in (\ref{LLstar}) compared to that of ref.~\cite{Dibitetto:2011gm}.}.

\begin{table}[h!]
\renewcommand{\arraystretch}{1.25}
\begin{center}
\scalebox{0.87}[0.87]{
\begin{tabular}{ | c || c | c |c | c | c |}
\hline
couplings & SO($6,6$) & SO($2,2$) & type IIB &  type IIA & fluxes\\
\hline
\hline
$T^3 \, U^3 $& $ -f_{+ abc} $ & $ - \Lambda_{+111} $& $ {F'}^{ijk} $&  & $  a_0' $\\
\hline
$T^3 \, U^2 $&  $f_{+ abk}$ &  $ \Lambda_{+112} $& ${F'}^{ ij c} $& &$   a_1' $\\
\hline
$T^3 \, U $& $ -f_{+ ajk}$  & $ - \Lambda_{+122} $& ${F'}^{i b c} $& &$  a_2' $\\
\hline
$ T^3 $& $f_{+ ijk}$ & $ \Lambda_{+222} $& ${F'}^{a b c} $& &$  a_3' $\\
\hline
\hline
$S \, T^3 \, U^3 $& $ -f_{- abc} $  & $ - \Lambda_{-111} $& $ {H'}^{ ijk} $& &$  - b_0'$\\
\hline
$S \, T^3 \, U^2 $& $f_{- abk}$  & $ \Lambda_{-112} $& $ {H'}^{i jc} $& &$ - b_1' $\\
\hline
$S \, T^3 \, U $& $ -f_{- ajk}$  & $ - \Lambda_{-122} $& $ {H'}^{ i b c} $& & $ - b_2' $\\
\hline
$S  \, T^3 $& $f_{- ijk}$   & $ \Lambda_{-222}$ & $ {H'}^{a b c} $& &$ - b_3' $\\
\hline
\hline
$T^2 \, U^3 $& $f_{+ ab\bar{k}}$ & $ \Lambda_{+114} $& $  {Q'}_{a b}^k $& &$  c_0' $\\
\hline
$T^2 \, U^2 $& $f_{+ aj\bar{k}}=f_{+ ib\bar{k}}\,\,\,,\,\,\,f_{+ \bar{a}bc}$ & $  \Lambda_{+124} \,\,\,,\,\,\, \Lambda_{+113} $ & $ {Q'}_{a j}^k = {Q'}_{i b}^k \,\,\,,\,\,\, {Q'}_{b c}^a $& &$c_1' \,\,\,,\,\,\, \tilde{c}_1' $\\
\hline
$T^2 \, U $& $f_{+ ib\bar{c}}=f_{+ aj\bar{c}}\,\,\,,\,\,\,f_{+ ij\bar{k}}$  & $ \Lambda_{+123} \,\,\,,\,\,\, \Lambda_{+224} $& $ {Q'}_{ib}^c = {Q'}_{a j}^c \,\,\,,\,\,\, {Q'}_{ij}^k $& &$c_2' \,\,\,,\,\,\,\tilde{c}_2' $\\
\hline
$T^2 $& $f_{+ ij\bar{c}}$  & $ \Lambda_{+223} $& $  {Q'}_{ij}^{c} $& &$c_3' $\\
\hline
\hline
$S \, T^2 \, U^3$& $f_{- ab\bar{k}}$  & $ \Lambda_{-114} $& $  {P'}_{a b}^k $& &$ - d_0' $\\
\hline
$S \, T^2 \, U^2 $& $f_{- aj\bar{k}}=f_{- ib\bar{k}}\,\,\,,\,\,\,f_{- \bar{a}bc}$ & $ \Lambda_{-124} \,\,\,,\,\,\, \Lambda_{-113} $& $ {P'}_{a j}^k = {P'}_{i b}^k \,\,\,,\,\,\, {P'}_{b c}^a $& &$ - d_1' \,\,\,,\,\,\, - \tilde {d}_1' $\\
\hline
$S \, T^2 \, U $& $f_{- ib\bar{c}}=f_{- aj\bar{c}}\,\,\,,\,\,\,f_{- ij\bar{k}}$  &$ \Lambda_{-123} \,\,\,,\,\,\, \Lambda_{-224} $ & $ {P'}_{ib}^c = {P'}_{a j}^c \,\,\,,\,\,\, {P'}_{ij}^k $& &$ - d_2' \,\,\,,\,\,\, - \tilde{d}_2' $\\
\hline
$S \, T^2  $& $f_{-ij\bar{c}}$ & $ \Lambda_{-223} $ & $  {P'}_{ij}^{c} $& &$ - d_3' $\\
\hline
\end{tabular}
}
\end{center}
\caption{Mapping between primed fluxes, embedding tensor components and couplings in the flux-induced superpotential. We have made the index splitting $\,M=\{a,i,\bar{a},\bar{i}\}\,$ for $\,\textrm{SO}(6,6)\,$ light-cone coordinates, not to be confused with the same indices appearing in appendix~\ref{App:indices}.}
\label{table:primed_fluxes}
\end{table}

Irrespective of their IIA or IIB string theory interpretation, the above set of fluxes generates the following $\,\cN=1\,$ flux-induced superpotential 
\beq 
\label{W_fluxes} 
W = (P_{F} +
P_{H} \, S ) + 3 \, T \, (P_{Q} + P_{P} \, S ) + 3 \, T^2 \, (P_{Q'}
+ P_{P'} \, S ) + T^3 \, (P_{F'} + P_{H'} \, S ) \ , 
\eeq
involving the three complex moduli $S$, $T$ and $U$ surviving the truncation from maximal supergravity to the theory inside box --5-- of figure~\ref{fig:truncations}. However, just by a simple inspection of tables \ref{table:unprimed_fluxes} and \ref{table:primed_fluxes}, it is clearly more convenient to adopt the terminology of the type IIB string theory when it comes to associate embedding tensor components to fluxes. In this picture, the superpotential in (\ref{W_fluxes}) contains flux-induced polynomials depending on both electric and magnetic pairs -- schematically $\,(e,m)\,$ -- of gauge $(F_{3},H_{3})$ fluxes and non-geometric $(Q,P)$ fluxes,
\beq
\begin{array}{lcll}
\label{Poly_unprim}
P_{F} = a_0 - 3 \, a_1 \, U + 3 \, a_2 \, U^2 - a_3 \, U^3 & \hspace{5mm},\hspace{5mm} & P_{H} = b_0 - 3 \, b_1 \, U + 3 \, b_2 \, U^2 - b_3 \, U^3 & ,  \\[2mm]
P_{Q} = c_0 + C_{1} \, U - C_{2} \, U^2 - c_3 \, U^3 & \hspace{5mm},\hspace{5mm} & P_{P} = d_0 + D_{1} \, U - D_{2} \, U^2 - d_3 \, U^3 & ,
\end{array}
\eeq
as well as those induced by their less known primed counterparts $\,(F'_{3},H'_{3})\,$ and $\,(Q',P')\,$ fluxes,
\beq
\begin{array}{lcll}
\label{Poly_prim}
P_{F'} = a_3' + 3 \, a_2' \, U + 3 \, a_1' \, U^2 + a_0' \, U^3 & \hspace{3mm},\hspace{3mm} &P_{H'} = b_3' + 3 \, b_2' \, U + 3 \, b_1' \, U^2 + b_0' \, U^3 & ,  \\[2mm]
P_{Q'} = -c_3' +  C'_{2} \, U + C'_{1} \, U^2 - c_0' \, U^3 & \hspace{3mm},\hspace{3mm} & P_{P'} = -d_3' + D'_{2} \, U + D'_{1} \, U^2 - d_0' \, U^3 & .
\end{array}
\eeq

\noindent For the sake of clarity, we have introduced the flux combinations $\,C_{i} \equiv 2 \, c_i - \tilde{c}_{i}\,$, $\,D_{i} \equiv 2 \, d_i - \tilde{d}_{i}\,$, $\,C'_{i} \equiv 2 \, c'_i - \tilde{c}'_{i}\,$ and $\,D'_{i} \equiv 2 \, d'_i - \tilde{d}'_{i}\,$ entering the superpotential (\ref{W_fluxes}), and hence also the scalar potential.

%
%

%

\providecommand{\href}[2]{#2}\begingroup\raggedright\endgroup

\end{document}